\documentclass[12pt]{article}
\pdfoutput=1

\usepackage[usenames,dvipsnames,svgnames,table,footnote]{xcolor} 
\usepackage[obeyspaces,hyphens,spaces]{url}
\usepackage{jcapmod}
\usepackage{verbatim}
\usepackage{graphics}
\usepackage{epsfig}
\usepackage{booktabs}
\usepackage{booktabs}
\usepackage[english]{babel}
\usepackage{amsmath, amssymb, amsbsy, amstext, amsthm, simplewick}
\usepackage{hyperref}
\usepackage{graphicx}
\usepackage{amsfonts}
\usepackage{upgreek}
\usepackage{framed}
\usepackage{tensor}
\usepackage{pifont}
\usepackage{latexsym, mathrsfs}
\usepackage{array}
\usepackage{hyperref}
\usepackage{xspace}
\usepackage{longtable}
\usepackage{multirow}
\usepackage{cases}
\usepackage{empheq}

\usepackage{bm}
\usepackage{bbm}
\usepackage{float}
\usepackage{afterpage}
\usepackage{setspace}
\usepackage{caption}
\usepackage[load-configurations=astronomy, range-units=brackets, range-phrase=-, per-mode=reciprocal, mode=math]{siunitx}
\usepackage{threeparttable}
\usepackage{subfigure}
\usepackage{tcolorbox}
\allowdisplaybreaks 

\usepackage{braket}

\newcommand{\be}{\begin{equation}}
\newcommand{\ee}{\end{equation}}
\newcommand{\bea}{\begin{eqnarray}}
\newcommand{\eea}{\end{eqnarray}}
\newcommand{\bml}{\begin{subequations}}
\newcommand{\eml}{\end{subequations}}
\newcommand{\bfig}{\begin{figure}}
\newcommand{\efig}{\end{figure}}

\newcommand{\slashed}{\hspace{-1.1ex}/}

\newcommand{\bmat}{\begin{pmatrix}}
\newcommand{\emat}{\end{pmatrix}}

\usepackage{ragged2e}

\usepackage{hhline}
\usepackage{array}
\newcolumntype{P}[1]{>{\centering\arraybackslash}p{#1}}
\usepackage{booktabs}
\usepackage{tikz}
\usetikzlibrary{calc,shadings,patterns,tikzmark,fadings}
\usepackage{cleveref} 
\Crefname{equation}{Eq.}{Eqs.}
\Crefname{section}{Sec.}{Secs.}
\Crefname{figure}{Fig.}{Figs.}
\Crefname{table}{Table}{Tables}

\definecolor{Blue}{rgb}{0.25, 0.41, 0.88}
\definecolor{Red}{rgb}{0.92,0.,0.}
\definecolor{darkorange}{rgb}{1.0,0.549,0.}
\definecolor{cobalt}{RGB}{44, 98, 120}
\definecolor{Mathematica1}{rgb}{0.368417, 0.506779, 0.709798}
\definecolor{Mathematica2}{rgb}{0.880722, 0.611041, 0.142051}
\definecolor{Mathematica3}{rgb}{0.560181, 0.691569, 0.194885}
\definecolor{Mathematica4}{rgb}{0.922526, 0.385626, 0.209179}
\definecolor{Mathematica5}{rgb}{0.528488, 0.470624, 0.701351}
\definecolor{Mathematica6}{rgb}{0.772079, 0.431554, 0.102387}
\definecolor{Mathematica7}{rgb}{0.363898, 0.618501, 0.782349}
\definecolor{Mathematica8}{rgb}{1, 0.75, 0}
\definecolor{Mathematica9}{rgb}{0.647624, 0.37816, 0.614037}
\definecolor{plotBlue}{RGB}{94, 130, 181}
\definecolor{plotRed}{RGB}{233, 85, 54}
\definecolor{plotGreen}{RGB}{142, 176, 50}
\definecolor{plotPurple}{RGB}{135, 120, 178}

\newcolumntype{C}[1]{>{\centering\let\newline\\\arraybackslash\hspace{0pt}}m{#1}}

\def\x{{\bf x}}

\makeatletter
\newlength{\apb@width}
\newcommand{\autoparbox}[2][c]{\settowidth{\apb@width}{#2}\parbox[#1]{\apb@width}{#2}}

\makeatother

\makeatletter
\newsavebox\myboxA
\newsavebox\myboxB
\newlength\mylenA

\newcommand*\xoverline[2][0.75]{
	\sbox{\myboxA}{$\m@th#2$}%
	\setbox\myboxB\null
	\ht\myboxB=\ht\myboxA%
	\dp\myboxB=\dp\myboxA%
	\wd\myboxB=#1\wd\myboxA
	\sbox\myboxB{$\m@th\overline{\copy\myboxB}$}
	\setlength\mylenA{\the\wd\myboxA}
	\addtolength\mylenA{-\the\wd\myboxB}%
	ifdim\wd\myboxB<\wd\myboxA%
	\rlap{\hskip 0.5\mylenA\usebox\myboxB}{\usebox\myboxA}%
	\else
	\hskip -0.5\mylenA\rlap{\usebox\myboxA}{\hskip 0.5\mylenA\usebox\myboxB}%
	\fi}
\makeatother

\numberwithin{equation}{section}
\numberwithin{figure}{section}
\numberwithin{table}{section}

\def\beq{\begin{equation}}
\def\eeq{\end{equation}}

\def\bea{\begin{eqnarray}}
\def\eea{\end{eqnarray}}

\def\beq{\begin{equation}}
\def\eeq{\end{equation}}
\def\bea{\begin{eqnarray}}
\def\eea{\end{eqnarray}}

\numberwithin{equation}{section}
\def\beq{\begin{equation}}
\def\eeq{\end{equation}}

\def\bea{\begin{eqnarray}}
\def\eea{\end{eqnarray}}

\DeclareRobustCommand{\SkipTocEntry}[4]{}

\usepackage{colortbl}
\definecolor{blue2}{cmyk}{1, 0.1, 0.1, 0.1}

\definecolor{pyBlue}{RGB}{31, 119, 180}
\definecolor{pyRed}{RGB}{214, 39, 40}
\definecolor{pyGreen}{RGB}{44, 160, 44}
\definecolor{pyBlue2}{RGB}{0, 111, 237}
\definecolor{pyRed2}{RGB}{224, 52, 36}

\newcolumntype{P}[1]{>{\centering\arraybackslash}p{#1}}
\newcolumntype{M}[1]{>{\centering\arraybackslash}m{#1}}

\begin{document}

\tcbset{colframe=black,arc=0mm,box align=center,halign=left,valign=center,top=-10pt}

\renewcommand{\thefootnote}{\fnsymbol{footnote}}

\pagenumbering{roman}
\begin{titlepage}
	\baselineskip=2.5pt \thispagestyle{empty}
	
	
\begin{center}
 ~{\Huge 
	 {\fontsize{22}{22} \textcolor{Sepia}{\bf\sffamily Entanglement negativity in de Sitter biverse from Stringy Axionic Bell pair: An analysis using Bunch-Davies vacuum}}}\\  \vspace{0.25cm}
\end{center}

		\begin{center}
	
	{\fontsize{14}{18}\selectfont Sayantan Choudhury${}^{\textcolor{Sepia}{1}}$\footnote{{\sffamily \textit{ Corresponding author, E-mail}} : {\ttfamily sayantan\_ccsp@sgtuniversity.org,  sayanphysicsisi@gmail.com}}}${{}^{,}}$
	\footnote{{\sffamily \textit{ NOTE: This project is the part of the non-profit virtual international research consortium ``Quantum Aspects of Space-Time \& Matter" (QASTM)} }. }${{}^{,}}$.~

\end{center}


\begin{center}
	
{ 
	
	\textit{${}^{1}$Centre For Cosmology and Science Popularization (CCSP),
SGT University, Gurugram, Delhi-NCR, Haryana- 122505, India.}
	}
\end{center}

\vspace{1.2cm}
\hrule \vspace{0.3cm}
\begin{center}
\noindent {\bf Abstract}\\
\end{center}

In this work, we study the signatures of quantum entanglement by computing entanglement negativity between two causally unrelated regions in $3+1$ dimensional global de Sitter space. We investigate a bipartite quantum field theoretic setup for this purpose, driven by an axionic Bell pair resulting from Type IIB string compactification on a Calabi-Yau three fold. We take into account a spherical surface that divides the spatial slice of the global de Sitter space into exterior and interior causally unrelated sub regions.  For the computational purpose we use the simplest possible initial choice of quantum vacuum,  which is Bunch-Davies state.  The quantitative quantum information theoretic measure for entanglement negativity turns out be consistent with the results obtained for entanglement entropy,  even we have to say it is better than that from quantum information theoretic point of view.  We design the problem in a hyperbolic open chart where one of the causally unrelated observers remains constrained and the scale dependence enters to the corresponding quantum information theoretic entanglement measure for axionic Bell pair. 
We find from our analysis that in the large scales initially maximally entangled Bunch-Davies state turns out to be strongly entangled or weakly entangled depending on the axionic decay constant and the supersymmetry breaking scale.  We also find that at the small scales the initial entanglement can be perfectly recovered.  We also discuss the possibility of having a biverse picture,  which is a mini version of the multiverse in the present theoretical set up.
Last but not the least,  we provide the necessary criteria for generating non vanishing quantum entanglement measures within the framework of quantum field theory of global de Sitter space as well as well as in primordial cosmology due to the axion derived from string theory.

\vskip10pt
\hrule
\vskip10pt

\text{Keywords:~~De-Sitter vacua,  Quantum Entanglement, Cosmology of Theories beyond}\\
 \text{ the SM, Quantum Information Theory aspects of Gravity,  String Cosmology.}

\end{titlepage}

\newpage
\setcounter{tocdepth}{2}

\tableofcontents

\newpage
	
	\clearpage
	\pagenumbering{arabic}
	\setcounter{page}{1}
	
	\renewcommand{\thefootnote}{\arabic{footnote}}

\clearpage

\section{Introduction}

In the present day research,  different quantum information theoretic measures of quantum entanglement is a remarkable probe in theoretical physics which helps us to distinguish the various type of long range correlated quantum mechanical states.
In this connection,  study of the explicit role of the long range quantum correlations in the framework of quantum field theory is extremely significant,  which is a fascinating topic of research itself.  See refs \cite{Maldacena:2012xp,Casini:2009sr,Amico:2007ag,Laflorencie:2015eck,Plenio:2007zz,Cerf:1996nb,Calabrese:2004eu,Martin-Martinez:2012chf,Nambu:2008my,Nambu:2011ae,VerSteeg:2007xs,Fischler:2013fba,Iizuka:2014rua,Choudhury:2017bou,Choudhury:2017qyl,Choudhury:2018ppd,Kanno:2014bma,Kanno:2014ifa,Kanno:2014lma,Albrecht:2018prr,Kanno:2017wpw,Kanno:2016qcc,Kanno:2015ewa,Colas:2022kfu,Burgess:2022nwu,Martin:2021znx,Martin:2021xml,Grain:2020wro,Martin:2018zbe,Martin:2015qta,Horodecki:2009zz} for more details.
The key ingredient of this study is the initial quantum mechanical vacuum states,  which are Chernikov-Tagirov,  Bunch-Davies, Hartle-Hawking,  $\alpha$ and Motta-Allen vacua \cite{Allen:1985ux,Adhikari:2021ked,Banerjee:2021lqu,Choudhury:2021tuu,Choudhury:2020yaa,Choudhury:2018ppd,Choudhury:2017qyl,Choudhury:2017glj,Choudhury:2017bou,Kanno:2014lma}.   Quantum entanglement is treated as one of the remarkable outcomes of the foundational theoretical aspects of quantum mechanics.  The prime reason of this thought is,  a local measurement in quantum mechanics may instantaneously put a significant impact of the outcome of the measurement beyond the physical light cone.  This is theoretically interpreted as Einstein-Podolsky-Rosen (EPR) paradox where the concept of causality violation is explicitly demonstrated \cite{Einstein:1935rr,Bell:1964kc,Aspect:1981zz,Aspect:1982fx,Horodecki:2009zz}.

Amongst various types of information theoretic measures entanglement entropy is considered to be a very useful quantitative as well as qualitative probe of quantum entanglement and this concept is commonly used in the framework of condensed matter physics,  quantum information theory and high energy physics.  But in this connection it is important to note that the technical computation of entanglement entropy is very difficult to perform in the context of quantum field theory.  Ryu and Takayanagi in refs.  \cite{Ryu:2006bv,Ryu:2006ef} first did the theoretically consistent computation of entanglement entropy for a strongly coupled quantum field theory having a gravitational dual counterpart using the underlying physical principles of AdS/CFT (or holographic gravitational dual prescription) \cite{Maldacena:1997re}.

In order to properly understand the direct implications of the previously proposed AdS/CFT (or holographic gravitational dual prescription) in the technical computation of entanglement entropy in presence of standard Bunch-Davies quantum vacuum state within the framework of quantum field theory of de Sitter space,  Maldacena and Pimentel prescribed an extremely useful computational strategy in ref.  \cite{Maldacena:2012xp} using free massive scalar quantum field.  After this work the proposed methodology was generalised for the same problem in presence of non standard $\alpha$ vacua in refs  \cite{Chernikov:1968zm,Bunch:1978yq,Hartle:1983ai,Iizuka:2014rua,Kanno:2014lma}.  Further in refs. \cite{Choudhury:2017bou,Choudhury:2017qyl,Choudhury:2018ppd},  these underlying physical concepts of the computation of entanglement entropy was applied within the framework of axion quantum field theory,   described in terms a specific type of effective interaction potential originated from Type II string theory compactification \cite{Panda:2010uq,Svrcek:2006yi,Beasley:2005iu} in presence of both standard Bunch-Davies and non standard $\alpha$ quantum vacua.  The underlying concept of quantum entanglement could applicable to the framework of cosmology,  specifically beyond the Hubble horizon scale.  Most importantly,  in this prescription if a particle pair is created in a casually connected Hubble horizon scale then it is naturally expected to be dissociated as an outcome of the de Sitter cosmological expansion.  In this particular case,  the findings from this prescription pointing towards the fact that two causally unrelated patches in the de Sitter cosmological space-time has to be entangled quantum mechanically.   As a result,  the corresponding observable quantum vacuum fluctuations associated with our own universe is entangled with the other part of the open patch of global de Sitter space,  which is commonly identified to be the {\it multiverse} in the present context of discussion.  To demonstrate this underlying physical picture the reduced density matrix formalism play a very significant role,  using which it was explicitly obtained that the quantum mechanical entanglement directly put impact on the cosmological power spectrum on the large scales and it is quantitatively similar to or larger than the curvature radius scale connected to the current issue \cite{Kanno:2014ifa}.

It is hugely expected that the present and the upcoming observational probes may detect the imprints and explicit effects of quantum mechanical entanglement in cosmological paradigm,  which we strongly believe will going to be extremely useful to understand a lot of unknown mysterious fascinating facts of our own universe.  In the past,  there are lot of efforts have been made to study the impacts of quantum mechanical entanglement in the theoretical framework of cosmology.  See refs.  \cite{Kanno:2014ifa,Choudhury:2018ppd,Kanno:2022kve,Kanno:2017wpw,Soda:2017yzu,Kanno:2015ewa,Adil:2022rgt,Bolis:2019fmq,Holman:2019spa,Albrecht:2014aga,Lello:2013qza,Lello:2013bva} for more details on these aspects.

The motivation of the background physical thought of the present paper are appended below point-wise:
\begin{enumerate}
\item It was first discovered in a number of refs.\cite{Garriga:2012qp,Garriga:2013pga,Frob:2014zka} to support the claim that the physical frame of the bubble nucleation process is observer dependent and is actually dictated by the rest frame of the observer. The bubble nucleation process being dependent on the observer is hence only expected.
The detailed study of such type of observer dependence in the framework of quantum mechanical entanglement is one of the prime motivations of the present work.  If one can able to address this issue in detail,  then various unexplored features of quantum cosmology can be explored with proper physical understanding.

\item Second, the explicit function of the Bunch Davies vacuum can be investigated in the context of quantum entanglement originating from an axion field embedded in an open patch of the global de Sitter space. This will be crucial because it allows one to directly examine whether it is possible to find the signs of quantum entanglement in the current {\it multiverse}  \cite{Kanno:2014ifa,Sato:1981gv,Vilenkin:1983xq,Linde:1986fc,Linde:1986fd,Bousso:2000xa,Susskind:2003kw} motivated theoretical set up.using current or future observational instruments.

\item 

Thirdly,  it was noted in the refs. \cite{Maldacena:2015bha,Choudhury:2016cso,Choudhury:2016pfr,Choudhury:2017bou,Choudhury:2017qyl,Choudhury:2018ppd} that the string theory-originated axion can be viewed as the ideal component to build the Bell pair, which is necessary to violate Bell's inequality within the context of primordial cosmology. Within the widely accepted framework of primordial cosmology, it is virtually difficult to break Bell's inequality, and it is for this reason that string theory and the axion play the most important roles.  It is significant to note that, in this context, it is nearly impossible to test the explicit role of quantum mechanical entanglement in observational probes without violating Bell's inequality in the context of primordial cosmology, which is necessary to break the degeneracy in the shape cosmological two-point function, also known as the primordial power spectrum obtained from various effective potentials from fundamental physical principles.  The current framework with the string theory-originated axion provides a perfect setup that is pointing towards a new physics coming from a non-standard cosmology because the generation of Bell's inequality violating pairs, also known as the Bell pair, is practically impossible within the framework of the standard primordial cosmological paradigm.  It is explicitly pointed in refs.\cite{Maldacena:2015bha,Choudhury:2016cso,Choudhury:2016pfr} that the prime signature of the Bell's inequality violation in non standard primordial cosmology is coming from the existence of one point function from axion,  which is absent in the well known standard cosmological paradigm.  This result has a great impact of producing quantum entanglement in the present scenario.  Also our derived results can be extremely useful to directly verify along with the observational probes the applicability as well as the justifiability of the string theory originated axion to address the issue of quantum entanglement in the present set up.  See refs. \cite{Espinosa-Portales:2022yok,Ando:2020kdz,Martin:2017zxs,Martin:2016nrr,Choudhury:2021mht,Maldacena:2015bha,Choudhury:2016cso,Choudhury:2016pfr,Choudhury:2017bou,Choudhury:2017qyl,Choudhury:2018ppd,Kanno:2017teu,Kanno:2017dci} for more details on the Bell's inequality and its violation in various contexts including cosmology.

\item Finally,  the motivation comes from the {\it multiverse} prescription appearing in the string theory landscape scenario,  which states that our universe may not be the single entity of the space-time but the part of a bigger size {\it multiverse} \cite{Kanno:2014ifa,Sato:1981gv,Vilenkin:1983xq,Linde:1986fc,Linde:1986fd,Bousso:2000xa,Susskind:2003kw}.  This is obviously a fascinating fact which one needs to study in detail.  However,  in the past the underlying physical concept of the {\it multiverse} has been criticised in various contexts as a hypothetical philosophical concept which is the outcome of complete theoretical imagination and cannot be tested via cosmological observation.  But the actual truth is fay beyond all of these criticism,  which is the detectable signatures can be obtained from the {\it multiverse} set up in presence of quantum mechanical entanglement.  It really helps us to produce at least two causally separated de sitter universe,  commonly known as the de Sitter bubbles in the present context.  But the numbers are not restricted in two and most importantly this is actually the starting point of {\it multiverse} which allows many more causally unrelated de Sitter bubbles.   

\end{enumerate}

Next,  we explicitly write down the underlying assumptions of the present work,  which are appended below point-wise:

\begin{enumerate}
\item To model the present {\it multiverse} motivated scenario \cite{Kanno:2014ifa,Sato:1981gv,Vilenkin:1983xq,Linde:1986fc,Linde:1986fd,Bousso:2000xa,Susskind:2003kw} ,  we consider two causally separated patches on the global de Sitter space by assuming that initially they are in a maximally entangled pure quantum mechanical state.   this particular set up we identify as the biverse picture which is the mini version of the original {\it multiverse} picture.

\item  In this theoretical set up further we introduce two observers whose actual purpose is to determine the quantum entanglement of a de Sitter universe.  We also assume that one of the observers is placed inside the a de Sitter bubble and want to determine how the signatures of quantum mechanical entanglement with the other de Sitter bubble can be visualized by the inside observer. Now the issue is that,  the inside observer can't able to see outside region of their own de Sitter bubble.  For this reason instead of using the total density matrix,  one needs to take the partial trace over the causally unrelated outside region,  which finally give rise to the reduced density matrix of the system in the present context. But as an outcome some information will be lost during this process and to describe the quantum mechanical state of the observer instead of using a pure state one needs to explicitly use a mixed state.  This scenario is completely different for an observer who is sitting on the other causally separated de Sitter bubble because in his frame of reference the quantum aspects is described by the pure quantum states.  For this reason we need study the effects and outcomes of the quantum entanglement between pure and mixed states in the present context of discussion.

\item To perform the computation for a single global de Sitter space picture as well in the case of the entangled two de Sitter space,  which is the biverse picture we need to properly understand the factorization of the Hilbert space in terms of the individual constituents which span the subspaces by forming complete orthonormal sets.  The geometrical structure of the global de Sitter space demands that in the hyperbolic open chart one can symmetrically factorize in the left region and in a right region,  which we have tagged as region {\bf L} and {\bf R} during performing the technical computation in this paper.  Since we have symmetrically factorize the total Hilbert space in the region {\bf L} and {\bf R},  it is viable to formulate the reduced description either in terms of the degrees of freedom appearing in the region {\bf L} or region {\bf R}.  In our computation we take the partial trace over the degrees of freedom appearing in the region {\bf R},  which forms a reduced density matrix in the region {\bf L}.

\item If the Hilbert space factorization performed correctly then one can further blindly trust the mode decomposition in the region {\bf L} and {\bf R} using which we have constructed the Bunch Davies quantum vacuum states in terms of harmonic oscillator modes which forms a complete orthonormal basis states.  Estimation of the quantum entanglement in terms of various quantum information theoretic measures are also based on this mode decomposition and hence one can rely on the tracing out the unwanted information from the bipartite quantum field theoretic set up under consideration to construct the reduced density matrix.  

\item  To study the outcome of quantum mechanical entanglement we have blindly trusted the above mentioned factorization for the second de Sitter space which is spanned by the well known Bunch Davies vacuum state.  For the other,  which is the first de Sitter bubble we have assumed that the detailed factorization structure is not explicitly needed for the computation we are interested to perform in this paper.  This is because of the fact that we need to take the partial trace operation over the first de Sitter vacuum which will remove all unwanted degrees of freedom from the quantum information theoretic measures of entanglement we want to compute to confirm the existence of quantum entanglement in the biverse picture.  Here one can do the computation in other way as well where one needs to take the partial trace operation with respect to the all degrees of freedom appearing in second de Sitter vacuum and properly factorize the first de Sitter vacuum in terms of {\bf L} and {\bf R} modes. Due to having the symmetrical structure of the set up that we are considering it is expected that for both the cases we will have the same physical outcome at the end of the computation. We just have taken the first possibility as a choice.

\item During the biverse construction we have first constructed the maximally entangled state which is one of the key ingredients to violate the Bell's CHSH inequality within the present framework.  This is a very challenging to construct within the framework of de Sitter space time as well as in primordial cosmology set up.  We have constructed the maximally entangled state which suffice the purpose.  But this could not be possible just having usual interaction in the scalar field theory embedded in the global patch of the de Sitter space.  The 
Bell's CHSH inequality violating pair is created in the region {\bf L} and {\bf R} with the help of specific type of interaction in the string theory originated axion driven scalar field theory.  The structure of the interaction is controlled by the time dependent axion decay constant and the supersymmetry breaking scale.  We have assumed that Bell pair is created and originated from the string theory originated axionic scalar field theory.  It might be possible to create such pair from some other theoretical sector as well which we have not considered in this paper for the time being.  

\item In the present context the global coordinates can be treated as the closed slicing from the point of view of FLRW cosmology.  One usually do primordial cosmology,  particularly inflation  \cite{Baumann:2009ds,Baumann:2009ni,Baumann:2014nda,Choudhury:2017cos,Choudhury:2015hvr,Choudhury:2015pqa,Choudhury:2014hja,Choudhury:2014sua,Choudhury:2014kma,Choudhury:2014uxa,Choudhury:2014sxa,Choudhury:2013woa,Choudhury:2013jya,Choudhury:2013zna,Choudhury:2012whm,Choudhury:2012ib,Choudhury:2012yh,Choudhury:2011jt,Choudhury:2011sq,Ali:2010jx,Panda:2010pj,Deshamukhya:2009wc,Ali:2008ij,Panda:2007ie,Panda:2006mw,Panda:2005sg,Chingangbam:2004ng,Choudhury:2003vr,Mazumdar:2001mm,Sami:2022qyo,Agarwal:2017wxo,Geng:2017mic,Ando:2020fjm,Pattison:2019hef,Hardwick:2018zry,Vennin:2015eaa,Vennin:2015egh,Vennin:2014lfa,Martin:2013tda} in the planar coordinate slicing of the de Sitter space.  By performing coordinate coordinate transformation one can transform the global to static and then static to the flat slicing in the planer patch of de Sitter space.  For this reason whatever results we have obtained in this paper for the global patch of the de Sitter space can be directly translated in the planer patch of the de Sitter space,   which further implies that our derived results hold good for primordial cosmological set up constructed out of the effective interaction studied for the string theory originated axion scalar field theory.  This is obviously an interesting aspect from our computation which shows the applicability of our derived results in the vast theoretical framework.

\item  The von Neumann and Renyi entropies \cite{Akhtar:2019qdn,Choudhury:2017qyl,Choudhury:2017bou} are the good quantum information theoretic measures of quantum entanglement in the context of bipartite quantum field theory designed in terms of pure state. But with the help of mixed state if we want to understand the features of an arbitrary bipartite quantum field theoretic system then instead of using entropic measures,  negativity or its logarithmic version would be perfect information theoretic measure of quantum entanglement.  Using this specific measure we compute the imprints of the quantum entanglement from the point of view of two causally separated observers in the open chart of global de Sitter space.

\end{enumerate}

    \begin{figure*}[htb]
    	\centering
    	{
    		\includegraphics[width=15.5cm,height=12.5cm] {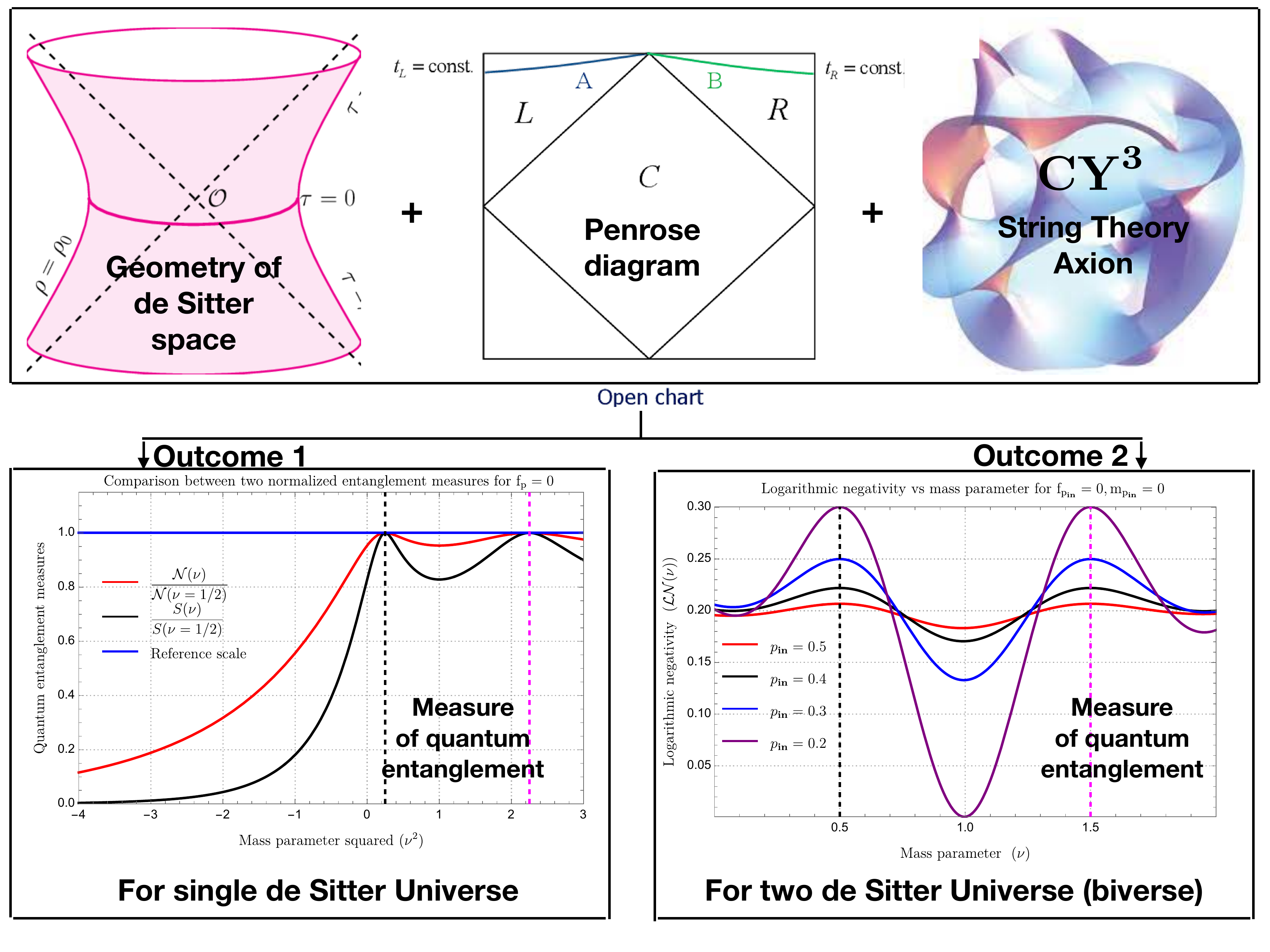}
    	}
    	\caption[Optional caption for list of figures]{Representative diagram of the overall quantum entanglement computation and all possible outcomes for single de Sitter universe and two de Sitter universe (biverse) obtained from our our analysis.} 
    	\label{Overall}
    \end{figure*}

In this work,  we compute the negativitity measure of quantum entanglement from an axionic effective potential which is obtained from Type IIB string theory compactification on a Calabi-Yau three fold in presence of NS5 brane.  Earlier this model have been studied in the framework of inflationary cosmology \cite{McAllister:2014mpa,McAllister:2008hb,Silverstein:2008sg,Flauger:2014ana}.  Since this axions can be treated as Bell pair,  we technically compute the expression for entanglement negativity from two causally unrelated patches of the open chart of the global de Sitter space.  From the quantitative results obtained from our computation we perform further a consistency check as well as the comparison of the results obtained in previous refs.\cite{Choudhury:2017bou,Choudhury:2017qyl,Choudhury:2018ppd}.  Furthermore,  to understand various unknown fascinating facts regarding the {\it multiverse} scenario we extend our computation in presence of axionic Bell pair in presence of de Sitter space along with the maximally entangled quantum Bunch Davies vacuum state.   We find from our analysis that in the large scales initially maximally entangled Bunch-Davies state turns out to be strongly entangled or weakly entangled depending on the axionic decay constant and the supersymmetry breaking scale.  We also find the at the small scales the initial entanglement can be perfectly recovered.
Last but not the least,  we provide the necessary criteria for generating non vanishing quantum entanglement measure within the framework of primordial cosmology due to the string axion.  In figure (\ref{Overall}) representative diagram of the overall quantum entanglement computation and all possible outcomes for single de Sitter universe and two de Sitter universe (biverse) obtained from our our analysis.

The organization of this paper is as follows:
\begin{itemize}
\item In \underline{\textcolor{purple}{\bf Section \ref{ka1}}} we briefly review the tools and techniques required to compute negativity and logarithmic negativity along with some easy demonstrations for the better understanding purpose.

\item In \underline{\textcolor{purple}{\bf Section \ref{ka2}}} we  mention the computational strategy for negativity in the hyperbolic open chart of the global de Sitter space.

\item In \underline{\textcolor{purple}{\bf Section \ref{ka3}}} we compute negativity by following aforementioned strategy for a specific effective potential,  Type IIB string compactification originated axion model. 

\item In \underline{\textcolor{purple}{\bf Section \ref{ka4}}} we compute entanglement in terms of negativity from two causally unrelated de Sitter bubbles,  which is determined from the point of view of two observers introduced during the calculation.  In this section we discuss the possibility of having biverse,  which is the mini version of the multiverse scenario.

\item Finally in \underline{\textcolor{purple}{\bf Section \ref{ka5}}} we summarize our all findings in this paper along with some future prospects of the computations performed in this paper.

\end{itemize}		    	

	\section{Basics of entanglement negativity and logarithmic negativity from Quantum Information Theory}	\label{ka2}
		\label{ka1}
Within the framework quantum mechanics,  particularly in quantum information theory	 to study the imprints and the underlying physical aspects of entanglement various measure have been proposed till date. 	 Entanglement negativity and its logarithmic version,  which is commonly known as logarithmic negativity are the very useful quantum information theoretic measures of quantum mechanical entanglement.  These measures are derived by making use of the positive partial transpose criterion for the separability.  The {\it Peres-Horodecki criterion} \cite{Peres:1996dw,Horodecki:1996nc,Simon:1999lfr,Horodecki:2009zz} is a prerequisite for the separation of the joint density matrix of two quantum mechanical systems $A$ and $B$. The phrase is sometimes referred to as the {\it PPT criterion},  which stands for positive partial transpose.  From the studies it was found that in the $ 2\times 2$ and $2\times 3$ dimensional cases the condition is sufficient.  Particularly this is used to decide the separability of mixed quantum states,  where the well known {\it Schmidt decomposition} does not apply.  It is important to note that,  in higher dimensions,  this specific concept gives inconclusive result,  and one should perform more advanced tests,  like {\it entanglement witnesses},  which describes a functional that distinguishes a specific entangled state from the separable ones.  In refs.  \cite{Peres:1996dw,Horodecki:1996nc,Simon:1999lfr,Horodecki:2009zz}, i t was explicitly shown that this measures are entanglement monotone and hence treated to be a proper measure of quantum entanglement.  In the following we give a brief overview on the topic of entanglement negativity and logarithmic negativity.
		 
		 Now let's take a look at a quantum system that can be described by $A$ and $B$. The direct products of the {\it Hilbert spaces} of the subsystems $A$ and $B$ define the corresponding total {\it Hilbert space} of the system  i.e.  ${\cal H}={\cal H}_A\otimes {\cal H}_B$.  Here ${\cal H}$ represents the  {\it Hilbert space} of the total system,   ${\cal H}_A$ and ${\cal H}_B$ represent the {\it Hilbert spaces} of the subsystems $A$ and $B$ respectively.

Further consider a pure quantum mechanical state,  by applying the {\it Schmidt decomposition} one can write:
\be |\Psi\rangle=\sum_{m} \sqrt{\lambda_m} |m\rangle_A \otimes |m\rangle_B, \ee
where $\lambda_m$ corresponds to the observed probability of the any general pure $m$-th state and satisfy the following constraint condition:
\be \sum_m \lambda_m=1,\ee
which physically implies that the total observed probability of the process,  which is obtained by summing over all possible pure states has to be conserved in the present context of discussion.

Now,  it is important to note if we are interested in the mixed states then to technically describe the behaviour of the subsystem inclusion of the concept of reduced density matrix is extremely important and this is actually described by the reduced density operator of the subsystem.  In the present description we consider two subsystems $A$ and $B$,  which are equally likely and both of them carry same weight in the present construction.  For this reason to describe the subsystem either we will talk about the description in terms of $A$ or $B$.  Let us say for the time being that we are interested to know about the underlying physics of the subsystem $A$ which can be obtained by taking the partial trace operation on the information of the subsystem $B$.  This process will finally give rise to the following reduced density matrix of the subsystem $A$,  which is given by:
\be \rho_A={\rm Tr}_B  |\Psi\rangle\langle\Psi|=\sum_m \lambda_m  |m\rangle_A {}_A\langle m|.\ee
Further,  utilizing this reduced density matrix of the subsystem $A$ one can explicitly compute the expression for the von Neumann entropy,  which is given by:
\be S=-{\rm Tr} \left[\rho_A \ln \rho_A\right]=-\sum_m \lambda_m\ln \lambda_m. \ee
Technically the above mentioned equation was introduced to describe the first entanglement measure in quantum information theory.  In the specific situation,  when there is no quantum entanglement,  then in terms of the probability we can write:
\be \lambda_m =
\left\{
	\begin{array}{ll}
		1  \quad\quad\quad\quad\quad\quad & \mbox{if } m=1  \\
		0 & \mbox{if } m\neq 1
	\end{array}
\right. \ee
In the above mentioned both the cases the measure of entanglement entropy vanish explicitly.  If such a situation appears in a particular physical systems,  then we can comment that there would be no quantum mechanical entanglement for that type of systems.  On the other hand,  if we find some physical systems where the probability lie within the window $0<\lambda_m<1 \forall m$ then the corresponding entanglement entropy measure is non-zero and it is treated to be good measure within the framework of quantum information theory.  But one word of caution is that,  the entanglement entropy measure sometimes gives inconclusive result.  For an example,  if we are interested to describe the strictly classical correlation in terms of mixed state then the entanglement entropy measure gives non-zero result, which is itself a surprising fact. In this specific type of situation entanglement entropy measure within the framework of quantum information theory fails to distinguish between the impacts of quantum mechanical correlations and classical correlations. 

Then obviously a natural question comes to a physicists mind that what is the way out of this situation? Is it possible to define more powerful entanglement measure in the present framework? Answer to all of these questions are {\it Yes},  it is possible to define such measures which can conclusively distinguish the quantum and classical effects in the correlation functions computed from the underlying physical theory.  Based on the separability criterion entanglement negativity and logarithmic negativity measures are defined which actually serve the purpose. 

Let us now give a brief outline on the connection between separability and quantum entanglement,  and its usefulness in the present context of discussion.  A quantum mechanical state is considered to be separable iff the density matrix of the total system can be expressed in terms of a tensor or outer products of the individual density matrices belong to the each subsystems under consideration.  Technically this statement can be written as:

\be \rho=\sum_m \lambda_m \left(\rho^{A}_m\otimes\rho^{B}_m  \right)\quad\quad \quad {\rm where} \quad \lambda_m\geq 0\quad\forall m.\ee
Here $\rho$ represents the density matrix of the total system under consideration.  For the $m$-th state corresponding to the subsystems $A$ and $B$ the density matrices are defined as:
\be \rho^{A}_m:=|m\rangle_A {}_A\langle m|\quad \quad {\rm and}\quad\quad\rho^{B}_m:=|m\rangle_B {}_B\langle m|.\ee
Hence in terms of the information coming from the subsystems $A$ and $B$,  the density matrix for the total system can be recast as:		
\be \label{ww1}\rho=\sum_m \lambda_m \left(|m\rangle_A {}_A\langle m|\otimes|m\rangle_B {}_B\langle m|  \right)\quad\quad \quad {\rm where} \quad \lambda_m\geq 0\quad\forall m.\ee	
One can further generalize the structure of the density matrix by considering both the contributions from entangled and non-entangled states,  which is given by the following expression:	
		\be \rho=\sum_m\sum_n\sum_p\sum_q {\cal D}_{mnpq} \left(|m\rangle_A {}_A\langle n|\otimes|p\rangle_B {}_B\langle q|  \right)\quad  {\rm where} \quad  {\cal D}_{mnpq}\geq 0\quad\forall m,n,p,q,\ee	
	where ${\cal D}_{mnpq}$ represents a more general coefficient which capture both the effects from entangled and non-entangled quantum states.  Let's investigate a partial transposition operation with respect to the subsystem $A$ in light of the new generation definition of the total density matrix, which results in the following expression:
	\bea \rho^{T_A}&=&\sum_m\sum_n\sum_p\sum_q {\cal D}_{mnpq} \left(\left(|m\rangle_A {}_A\langle n|\right)^{T_A}\otimes|p\rangle_B {}_B\langle q|  \right)\nonumber\\
	&=&\sum_m\sum_n\sum_p\sum_q {\cal D}_{mnpq} \left(|n\rangle_A {}_A\langle m|\otimes|p\rangle_B {}_B\langle q|  \right).\eea	
	Following a partial transpose operation with regard to the subsystem $A$, we obtain the following results for the non-entangled quantum state:
	\bea \rho^{T_A}&=&\sum_m \lambda_m \left(\left(|m\rangle_A {}_A\langle m|\right)^{T_A}\otimes|m\rangle_B {}_B\langle m|  \right)\nonumber\\
	&=&\sum_m \lambda_m \left(|m\rangle_A {}_A\langle m|\otimes|m\rangle_B {}_B\langle m|  \right)\nonumber\\
	&=&\rho,\eea	
	which corresponds to the fact that the total density matrix remains unchanged.   Here since for the non-entangled state $\lambda_m\forall m$,  this directly implies that $\rho^{T_A}\geq 0$. This also confirms that if after performing a partial transpose operation on the total system has at least one negative eigenvalue then the system cannot be described in terms of the above mentioned form stated in eqn (\ref{ww1}) and consequently the underlying quantum state considered in this discussion has to be entangled.  This is actually very interesting outcome coming from the present computation in support of quantum entanglement.
	
	Utilizing these facts one step forward the definition of an quantum information theoretic entanglement measure,  {\it entanglement negativity} is proposed,  which is given by:
	\be {\cal N}=\sum_{\lambda_m<0}|\lambda_m|,\ee
	where in this definition summation over all negative eigenvalues are explicitly taken into account.  From this definition,  we can clearly see that when we have ${\cal N}=0$,  there is no quantum entanglement in the system.  Apart from having a very good physical thought behind the construction of the entanglement negativity measure,  unfortunately it turns out that this is not an additive measure and most importantly not at all suitable for the description of many body or multi-subsystem appearing within the framework of quantum field theory.  Hence,  to give a more physically applicable quantum information theoretic measure another powerful quantity has been introduced,  which is known as {\it logarithmic negativity} and is treated to be most improved version of the entanglement negativity measure in the present context of discussion.  
	
	Let us now discuss in detail that how one can technically define the {\it logarithmic negativity} by making use of the background physical facts discussed in the present context.   To define this quantity,  first of all let us introduce the trace norm of the partial transposed version of the total density matrix over the subsystem $A$,  which is described by the following expression:
	\bea ||\rho^{T_A}||&=&{\rm Tr}\sqrt{\left(\rho^{T_A}\right)^{\dagger}\rho^{T_A}}\nonumber\\
	&=&\sum_m |\lambda_m|\nonumber\\
	&=&\left(\sum_{\lambda_m>0}|\lambda_m|+\sum_{\lambda_m<0}|\lambda_m|\right)\nonumber\\
	&=&\left({\cal N}+\sum_{\lambda_m>0}|\lambda_m|\right)\nonumber\\
	&=& \left(2{\cal N}+1\right).\eea
	Here in the list line of the above expression to write down corresponding trace norm in terms of the quantum entanglement negativity measure we have utilized the following sets of useful constraints:
	\be {\rm Tr} \left(\rho\right)=1,  \quad {\rm Tr} \left(\rho^{T_A}\right)=1,  \quad \sum_m\lambda_m=1.\ee
	Then the logarithm of the trace norm of the of the total density matrix over the subsystem $A$ is identified to be logarithmic negativity,  which is given by the following expression:
	\be L{\cal N}=\ln\left(||\rho^{T_A}||\right)=\ln\left(2{\cal N}+1\right).\ee
	This implies the fact that,  when ${\cal N}\neq 0$,  then $L{\cal N}\neq 0$ and the corresponding quantum state is considered to be entangled. 
	
	Let us now consider a special case where we are dealing with the pure quantum mechanical state.  In this specific case our next objective is to explicitly compute the expression for the logarithmic negativity.  To serve this purpose we use the well known {\it Schmidt decomposition} technique for pure quantum state,  using which we can write:
	\be |\Psi\rangle=\sum_m \sqrt{\lambda_m}\left(|m\rangle_A\otimes|m\rangle_B\right).\ee
	Then using this representation of the pure quantum state one can further define the expression for the density matrix for the total system,  which is given by the following expression:
	\bea \rho &=& |\Psi\rangle \langle \Psi|=\sum_m\sum_n \sqrt{\lambda_m \lambda_n}\left(\left(|m\rangle_A\otimes|m\rangle_B\right)\left( {}_A\langle n|\otimes{}_B\langle n|\right)\right).\eea
	Further,  taking the partial transpose operation on the above mentioned density matrix of the total system over the subsystem $A$,  we get the following expression:
	\bea \label{pt}\rho^{T_A} &=& \sum_m\sum_n \sqrt{\lambda_m \lambda_n}\left(\left(|m\rangle_A\otimes|m\rangle_B\right)\left( {}_A\langle n|\otimes{}_B\langle n|\right)\right)^{T_A}\nonumber\\
	&=& \sum_m\sum_n \sqrt{\lambda_m \lambda_n}\left(\left(|n\rangle_A{}_A\langle m|\right)\otimes\left(|n\rangle_A{}_A\langle m|\right)\right)\nonumber\\
	&=&  \sum_m\sum_{n,m=n} \sqrt{\lambda_m \lambda_n}\left(\left(|n\rangle_A{}_A\langle m|\right)\otimes\left(|n\rangle_A{}_A\langle m|\right)\right)\nonumber\\
	&&\quad\quad\quad\quad+\sum_m\sum_{n,m\neq n} \sqrt{\lambda_m \lambda_n}\left(\left(|n\rangle_A{}_A\langle m|\right)\otimes\left(|n\rangle_A{}_A\langle m|\right)\right)\nonumber\\
&=&  \sum_m\lambda_m\left(\left(|m\rangle_A{}_A\langle m|\right)\otimes\left(|m\rangle_A{}_A\langle m|\right)\right)\nonumber\\
	&&\quad\quad\quad\quad+\sum_m\sum_{n,m\neq n} \sqrt{\lambda_m \lambda_n}\left(\left(|n\rangle_A{}_A\langle m|\right)\otimes\left(|n\rangle_A{}_A\langle m|\right)\right).\eea
	For the sake of simplicity,  to diagonalize the structure of the total density matrix we further introduce a new basis for the pure quantum states,  in which two new state vectors are defined by the following expressions:
	\bea |\Psi^{+}_{mn}\rangle: &=&\frac{1}{\sqrt{2}}\left[\left(|m\rangle_A\otimes|n\rangle_B\right)+\left(|n\rangle_A\otimes|m\rangle_B\right)\right]\quad\quad{\rm for}\quad\quad m<n,\\
	 |\Psi^{-}_{mn}\rangle: &=&\frac{1}{\sqrt{2}}\left[\left(|m\rangle_A\otimes|n\rangle_B\right)-\left(|n\rangle_A\otimes|m\rangle_B\right)\right]\quad\quad{\rm for}\quad\quad m<n.\eea
	 It is very easy to check the following two constraints are always satisfied by the new basis state vectors $|\Psi^{+}_{mn}\rangle$ and $|\Psi^{-}_{mn}\rangle$,  which are given by:
	 \bea \left({}_A\langle m|\otimes{}_B\langle m|\right)|\Psi^{+}_{mn}\rangle &=&\frac{1}{\sqrt{2}}\left[\underbrace{\left({}_A\langle m|\otimes{}_B\langle m|\right)\left(|m\rangle_A\otimes|n\rangle_B\right)}_{=0}+\underbrace{\left({}_A\langle m|\otimes{}_B\langle m|\right)\left(|n\rangle_A\otimes|m\rangle_B\right)}_{=0}\right]\nonumber\\
	 &=&0,\quad\quad\quad\\
	\left({}_A\langle m|\otimes{}_B\langle m|\right)|\Psi^{-}_{mn}\rangle &=&\frac{1}{\sqrt{2}}\left[\underbrace{\left({}_A\langle m|\otimes{}_B\langle m|\right)\left(|m\rangle_A\otimes|n\rangle_B\right)}_{=0}-\underbrace{\left({}_A\langle m|\otimes{}_B\langle m|\right)\left(|n\rangle_A\otimes|m\rangle_B\right)}_{=0}\right]\nonumber\\
	&=&0. \eea
	This implies the fact that both the new basis state vectors $|\Psi^{+}_{mn}\rangle$ and $|\Psi^{-}_{mn}\rangle$  are orthogonal to the state $|m\rangle_A\otimes|m\rangle_B$,  which is obviously a very helpful information for the further construction.  Then the partial transpose with respect to subsystem $A$ for the total density matrix as stated in eqn (\ref{pt}) can be written in the diagonalized basis in terms of the two state vectors $|\Psi^{+}_{mn}\rangle$ and $|\Psi^{-}_{mn}\rangle$ as:
	\be \rho^{T_A}=\sum_m\lambda_m |m\rangle_A{}_A\langle m|+\sum_{m}\sum_{n,m<n}\sqrt{\lambda_m\lambda_n}\left(|\Psi^{+}_{mn}\rangle\rangle \Psi^{+}_{mn}|-|\Psi^{-}_{mn}\rangle\rangle \Psi^{-}_{mn}|\right).\ee
	If we closely look into the above expression then here we have have the following three possibilities:
	\begin{enumerate}
	\item The part $|m\rangle_A{}_A\langle m|$ has the eigenvalues $\lambda_m$ for all values of $m$.  Here $\lambda_m>0$ for this construction.
	
	\item The part $|\Psi^{+}_{mn}\rangle\rangle \Psi^{+}_{mn}|$ has the eigenvalues $\lambda_m\lambda_n$ for all values of $m$ and $n$.  Here $\lambda_m>0$ and $\lambda_n>0$,  then $\lambda_m\lambda_n>0$ for this construction.
	
	\item The part $|\Psi^{-}_{mn}\rangle\rangle \Psi^{-}_{mn}|$ has the eigenvalues $-\lambda_m\lambda_n$ for all values of $m$ and $n$.  Here $\lambda_m>0$ and $\lambda_n>0$,  then $-\lambda_m\lambda_n<0$ for this construction.  So the negative eigenvalues appear in this computation,  which is clearly the indication of having quantum mechanical entanglement in the present framework.  It is important to note that,  if in these sets of eigenvalues at least two of the $\lambda_m \forall m>2\neq 0$,  then the negative eigenvalues always appear in this computation. 
	\end{enumerate}
	Then,  in the new diagonalized basis the trace norm can be further computed as:
	\be ||\rho^{T_A}||=\left(\sum_m \lambda_m + 2 \sum_m\sum_{n,m<n}\sqrt{\lambda_m\lambda_n}\right)=\sum_m\sum_n \sqrt{\lambda_m\lambda_n}=\left(\sum_m \lambda_m \right)^2.\ee
	Utilizing this derived result,  further the logarithmic negativity can be computed as:
	\be L{\cal N}=\ln\left(||\rho^{T_A}||\right)=2\ln\left(\sum_m \lambda_m\right).\ee
	In general for a given quantum mechanical state our job is to compute the expression for the entanglement negativity or logarithmic negativity analytically or numerically depending on the physical system from which we want to estimate the final result.  In the next section,  using the methodology and the definitions presented in this section we will compute the expressions for the mentioned quantum information theoretic entanglement measures using string theory originated axion model placed in the open chart of the hyperbolic global de Sitter space-time.
	\section{Computational strategy of negativity between two causally unrelated open charts of global de Sitter space with axion}
	\label{ka2} 
	In this section,  our prime objective is to give a overall flavour of the computation of the reduced density matrix for the open chart of global de Sitter space,   which is necessarily needed to compute the expressions for the entanglement negativity and logarithmic negativity in the present context of discussion.  See refs.  \cite{Maldacena:2012xp,Iizuka:2014rua,Kanno:2014lma,Choudhury:2017bou,Choudhury:2017qyl,Choudhury:2018ppd} for the more details on the formalism,  where this method have been used earlier to compute the expression for the entanglement entropy from the von Neumann and Renyi measures.  
	 \subsection{Quantum structure of open chart}
    \label{ka2a}
    To compute the reduced density matrix as well as the mentioned quantum information theoretic measures we need to first of all understand the underlying quantum field theoretic set up in detail.  To serve the purpose we need to first of all understand the Hilbert space structure of the open chart of global de Sitter space.  This would be also helpful to implement the two causally unrelated patches of the open charts in terms of the quantum mechanical Hilbert space.  In this construction let us first of all consider a closed surface $\displaystyle \Sigma$ in a space like hypersurface.   As an outcome of this fact,  this hypersurface can be visualized in terms of an exterior and an interior region.  For further simplification purpose we identify the interior region as {\bf RI},  which is tagged with ${\bf L}$ region the associated Penrose diagram presented in figure~(\ref{penrose}).  Similarly,  the exterior region is identified as {\bf RII},  which is the ${\bf R}$ region in the corresponding diagram.  By looking at the geometrical structure of the two causally unrelated patches of the open chart it is expected that the representative quantum field theory has to be bipartite,  which will suffice the purpose for the quantum mechanical construction of the corresponding set up.  To describe this set up let us consider the fact that the Hilbert space of the total system is described by ${\bf \cal H}$.  Then for the bipartite quantum set up the total Hilbert space ${\bf \cal H}$ can be decomposed in terms of the tensor products of two Hilbert spaces associated with the tagged ${\bf L}$ and ${\bf R}$ region,  as described above.  Technically this is written as,  ${\bf \cal H}={\bf \cal H}_{\bf L}\otimes {\bf \cal H}_{\bf R}.$  Here ${\bf \cal H}_{\bf L}$ and ${\bf \cal H}_{\bf R}$ represent the corresponding Hilbert spaces of the subsystems ${\bf L}$ and ${\bf R}$ respectively.  This is perfectly consistent with the local description of quantum field theory.  However,  in the more generalized description,  it might not always true that in the underlying quantum theory the total Hilbert space can be factorized very easily and can be written in terms of tensor products of the subsystem Hilbert spaces.  This is because of the fact that the underlying physical theory might not be bipartite in nature.  Dealing with this type of physical systems at the level of computation is extremely difficult to handle.  To handle such systems at the level of analysis sometimes the concept of short distance lattice cut-off is introduced within the framework of quantum field theory,  which actually perfectly allows the bipartite construction in the computational set up.  Additionally,  it is important to note that,  in this construction the localized modes play significant role to construct the building blocks of tagged ${\bf L}$ and ${\bf R}$ regions respectively.   Consequently,  this quantum mechanical formulation allows us to construct the expression for the reduced density matrix of the region ${\bf L}$ by tracing out all the information of the region ${\bf R}$.  Similar argument holds for the construction of density matrix for the region ${\bf R}$.  In our computation,  we assume that the region ${\bf R}$ and the region ${\bf L}$ are the replica or identical copy of each other.  For this reason constructing any one of the reduced density matrix is sufficient enough to describe the subsystem in the present context of discussion.  Here the reduced density matrix can be written as:
	\bea \rho_{\bf L}={\rm Tr}_{\bf R}\left(|\Psi_{\bf BD}\rangle \langle \Psi_{\bf BD}|\right).\eea
	Here the underlying initial quantum vacuum structure is characterized in terms of the state vector $|\Psi_{\bf BD}\rangle$,  commonly known as the {\it Bunch-Davies vacuum}.  Now from the above mentioned expression it is quite obvious that if we can able to construct the mathematical structure of the  {\it Bunch-Davies vacuum} state vector in the mentioned Hilbert space,  then we can immediately compute the expression for the reduced density matrix,  using which one can further derive the expressions for entanglement entropy measures,  von Neumann and Renyi,  entanglement negativity and logarithmic negativity.  Now the prime point at this stage is to fix the structure of the initial quantum vacuum state vector one need to know the geometrical structure of the background space time,  which is open chart of global de Sitter space and also about the explicit matter content which is embedded in the mentioned space time.  In our case,  we will embed string theory originated axion field as a scalar matter on the mentioned space time,  which will going to help us to explicitly compute the mode functions in open chart.  Once the mode functions are computed as a function of momentum and various quantum numbers,  then the specific structure of the {\it Bunch-Davies vacuum} state vector can be fixed from the present computational algorithm.  In the next subsection we elaborately discuss the geometrical structure of the open chart of the global de Sitter space time,  which we believe will going to extremely helpful to underlying technical construction of the physical problem that we are studying in this paper.

    \subsection{Geometric structure of open chart}
    \label{ka2b}
    \begin{figure*}[htb]
    	\centering
    	{
    		\includegraphics[width=15.5cm,height=16.5cm] {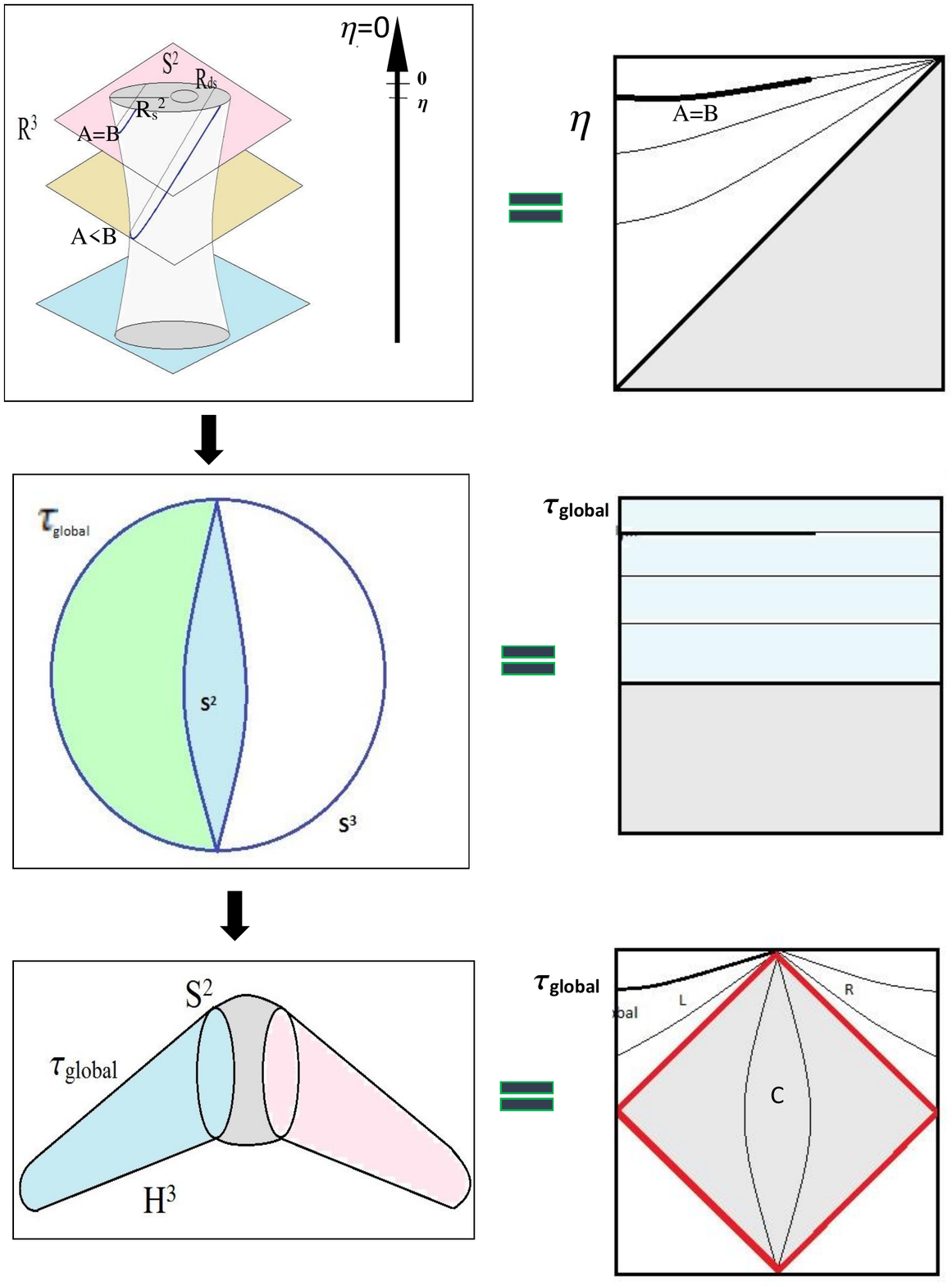}
    	}
    	\caption[Optional caption for list of figures]{Representative diagram and penrose diagram for the geometrical construction of the bipartite system. } 
    	\label{penrose}
    \end{figure*}
    In this subsection we are going to discuss about the geometrical  structure of open chart in global de Sitter space time.  Let us consider a surface of two sphere (${ \bf S}^2$) whose geometry is described by the following equation:
    \bea \sum^{3}_{i=1}x^2_{i}=R^2,\eea
    where $R$ is the radius of the two sphere and $x_{i}\forall i=1,2,3$ represent three orthogonal coordinates.  For the practical purpose here we need to consider that the radius of the two sphere is sufficiently large enough compared to the horizon size.   Now since the horizon size is described by the conformal time scale,  then here we need $R\gg \tau$,  where $\tau$ represents the conformal time.   In practical purposes this can be achieved by taking the late time limit $\tau\rightarrow 0$.   Here it is important to note that the surface of the two sphere that we are discussing in this section is invariant under the ${\rm \bf SO(1,3)}$ transformation.  In this construction the actual symmetry group is 
  ${\rm \bf SO(1,4)}$,  which is known as the isometry group of de Sitter space and ${\rm \bf SO(1,3)}$ forms a subgroup of this bigger isometry group.  In the representative penrose diagram this particular geometry is appearing in the tagged left region  as shown in figure~(\ref{penrose}).  In our computation we choose equal conformal time slice on three sphere ${\bf S}^3$,  which is appearing in the context of the global coordinates of de Sitter space.  Most importantly,  it is noted that the surface of two sphere is treated to be the equator of the three sphere in this geometrical construction.  This conformal map on the boundary of $3+1$ dimensional global de Sitter space effectively produce divergent contributions,  which can be easily regulated by taking another conformal reverse map from three sphere to two sphere via a global time surface.  
    
    Further,  we consider the following $1+4$ dimensional Euclidean global hyperbolic geometry which is defined as:
    \bea {\bf Euclidean~global~hyperbolic~geometry:\Longrightarrow}\quad\quad\quad\sum^{5}_{j=1}Y^2_{j}= \frac{1}{H^2},\eea
  where the corresponding $1+4$ dimensional coordinates are given by the following equation:  
    \bea
           \label{hypg}
  \displaystyle  Y_{j}&=&\displaystyle \frac{1}{H}\times\left\{\begin{array}{ll}
 \displaystyle  \cos\tau_{\bf E}~\sin\sigma_{\bf E}~\hat{\bf n}_{j}~~~~~~~~~~~~ &
                                                           \mbox{\small {\textcolor{black}{\bf  for $j=1,2,3$}}}  
                                                          \\ 
          \displaystyle \sin\tau_{\bf E} & \mbox{\small { \textcolor{black}{\bf for $j=4$}}}\\ 
                    \displaystyle \cos\tau_{\bf E}~\cos\sigma_{\bf E} & \mbox{\small { \textcolor{black}{\bf for $j=5$}}}.~~~~~~~~
                                                                    \end{array}
                                                          \right.
                                                          \eea 
     Here $\hat{\bf n}_{j}\forall j=1,2,3$
    are the three components which is in ${\bf R}^{3}$.  The corresponding Euclidean metric is  is given by:
    \bea {\bf Lorentzian~metric:\Longrightarrow}\quad\quad\quad ds^2_{\bf E}= \frac{1}{H^{2}}\left\{d\tau^2_{\bf E}+\cos^2\tau_{\bf E}\left(d\sigma^2_{\bf E}+\sin^2\sigma_{\bf E}~d\Omega^2_{\bf 2}\right)\right\},
    \eea
    where $d\Omega^2_{\bf 2}$ is the line element in two sphere. Now we take analytic continuation in the fifth coordinate,  which gives:
    \bea Y_{5}=\frac{1}{H}\cos\tau_{\bf E}~\cos\sigma_{\bf E}\xrightarrow[]{\bf Analytic~continuation} X_{0}=iY_{5}=\frac{i}{H}\cos\tau_{\bf E}~\cos\sigma_{\bf E}\eea
  and consider the coordinate redefinition of the part,  which give rise to:
   \bea {\bf Redefined~coordinates:\Longrightarrow}\quad\quad\quad X_{k}=Y_{k}~~~~~~~~~{ \forall k=1,2,3,4}.\eea
    As a result,  the corresponding Lorentzian geometry is described by the following representative equation:
    \bea {\bf Lorentzian~geometry:\Longrightarrow}\quad\quad\quad  \frac{1}{H^{2}}=\sum^{4}_{\mu=0}X^2_{\mu}=\left(-X^2_0+\sum^{4}_{j=1}X^2_j\right).\eea
    Since we have transformed the coordinate in Lorentzian signature,  the corresponding Lorentzian geometry for the different regions in the representative penrose diagram is described by the following coordinates:
    \bea
               \label{penr2}
      \displaystyle \textcolor{black}{\bf Region-R}&:\Longrightarrow&\displaystyle\left\{\begin{array}{ll}
     \displaystyle \tau_{\bf E}=\frac{\pi}{2}-it_{\bf R}~~~~~~~~~~~~ &
                                                               \mbox{\small {\textcolor{black}{\bf for $t_{\bf R}\geq 0$}}}  
                                                              \\ 
              \displaystyle \sigma_{\bf E}=-ir_{\bf R} & \mbox{\small { \textcolor{black}{\bf for $r_{\bf R}\geq 0$}}}.~~~~~~~~
                                                                        \end{array}
                                                              \right.\\
\label{penr3}
      \displaystyle \textcolor{black}{\bf Region-C}&:\Longrightarrow&\displaystyle\left\{\begin{array}{ll}
     \displaystyle \tau_{\bf E}=t_{\bf C}~~~~~~~~~~~~~~~~~~ &
                                                               \mbox{\small {\textcolor{black}{\bf for $\displaystyle-\frac{\pi}{2}\leq t_{\bf C}\leq \frac{\pi}{2}$}}}  
                                                              \\ 
              \displaystyle \sigma_{\bf E}=\frac{\pi}{2}-ir_{\bf C} & \mbox{\small { \textcolor{black}{\bf for $-\infty<r_{\bf C}< \infty$}}}.~~~~~~~~~~~~~~
                                                                        \end{array}
                                                              \right. \\
\label{penr4}
      \displaystyle \textcolor{black}{\bf Region-L}&:\Longrightarrow&\displaystyle\left\{\begin{array}{ll}
     \displaystyle \tau_{\bf E}=-\frac{\pi}{2}+it_{\bf L}~~~~~~~~~~~~ &
                                                               \mbox{\small { \textcolor{black}{\bf for $t_{\bf L}\geq 0$}}}  
                                                              \\ 
              \displaystyle \sigma_{\bf E}=-ir_{\bf L} & \mbox{\small { \textcolor{black}{\bf for $r_{\bf L}\geq 0$}}}.~~~~~~~~
                                                                        \end{array}
                                                              \right.                                         \eea
 Finally,  the representative Lorentzian metric for the three different regions,  as mentioned above can be written as  \cite{Maldacena:2012xp,Iizuka:2014rua,Kanno:2014lma,Choudhury:2017bou,Choudhury:2017qyl,Choudhury:2018ppd}:
  \bea
                 \label{r2z}
        \displaystyle \textcolor{black}{\bf Region-R}&:\Longrightarrow&\displaystyle\left\{\begin{array}{ll}
       \displaystyle ds^2_{\bf R}=\frac{1}{H^{2}}\left[-dt^2_{\bf R}+\sinh^2t_{\bf R}\left(dr^2_{\bf R}+\sinh^2r_{\bf R}~d\Omega^2_{\bf 2}\right)\right], 
                                                                          \end{array}
                                                                \right.\\
  \label{r3}
        \displaystyle \textcolor{black}{\bf Region-C}&:\Longrightarrow&\displaystyle\left\{\begin{array}{ll}
       \displaystyle  ds^2_{\bf C}=\frac{1}{H^{2}}\left[dt^2_{\bf C}+\cos^2t_{\bf C}\left(-dr^2_{\bf C}+\cosh^2r_{\bf C}~d\Omega^2_{\bf 2}\right)\right], \end{array}
                                                                \right. \\
  \label{r4}
        \displaystyle \textcolor{black}{\bf Region-L}&:\Longrightarrow&\displaystyle\left\{\begin{array}{ll}
       \displaystyle  ds^2_{\bf L}=\frac{1}{H^{2}}\left[-dt^2_{\bf L}+\sinh^2t_{\bf L}\left(dr^2_{\bf L}+\sinh^2r_{\bf L}~d\Omega^2_{\bf 2}\right)\right].  \end{array}
                                                                \right.                                         \eea 
    \begin{figure*}[htb]
    \centering
    \subfigure[For ${\bf L}$ and ${\bf R}$.]{
        \includegraphics[width=14.2cm,height=9cm] {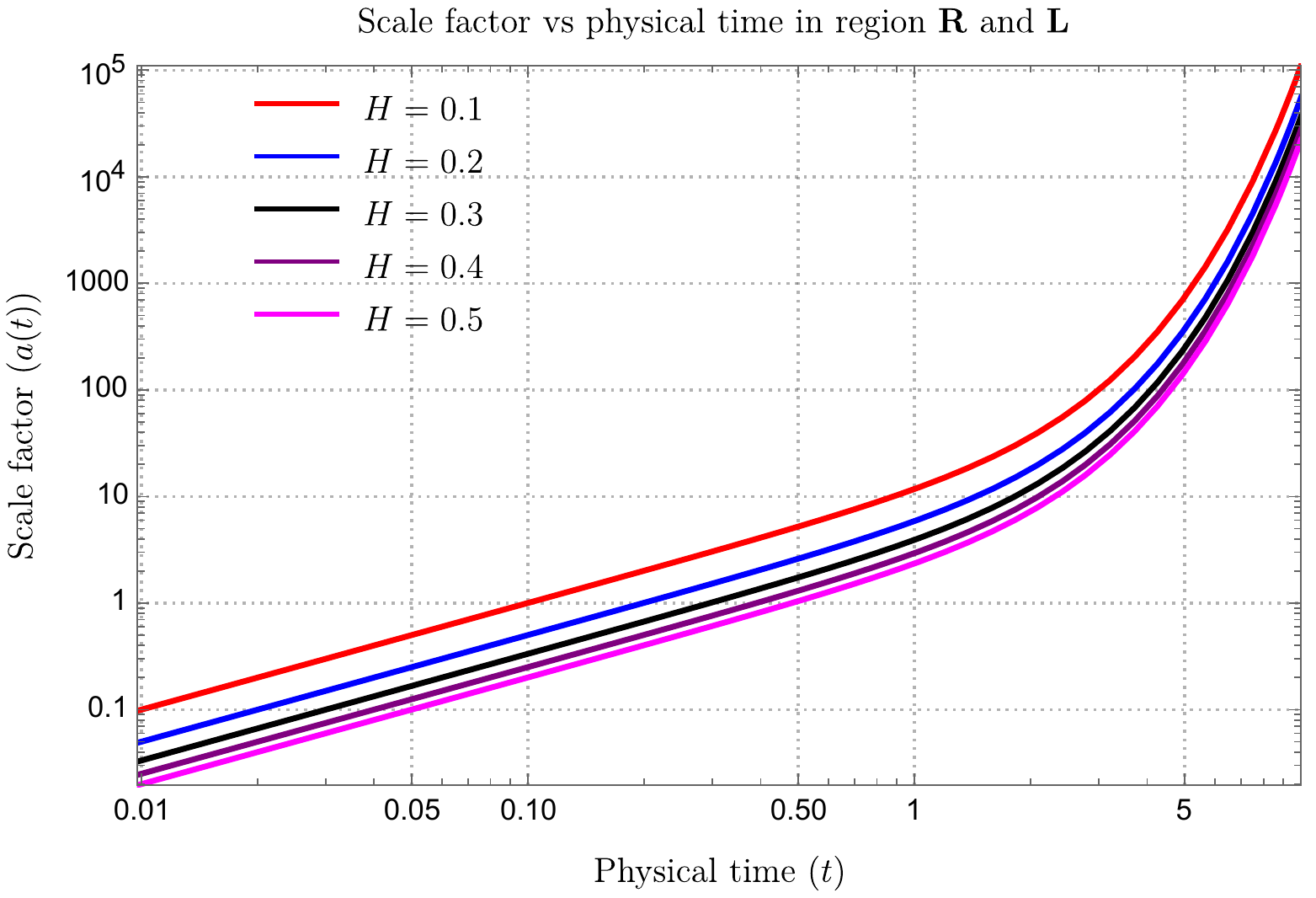}
        \label{s1}
    }
    \subfigure[For ${\bf C}$.]{
        \includegraphics[width=14.2cm,height=9cm] {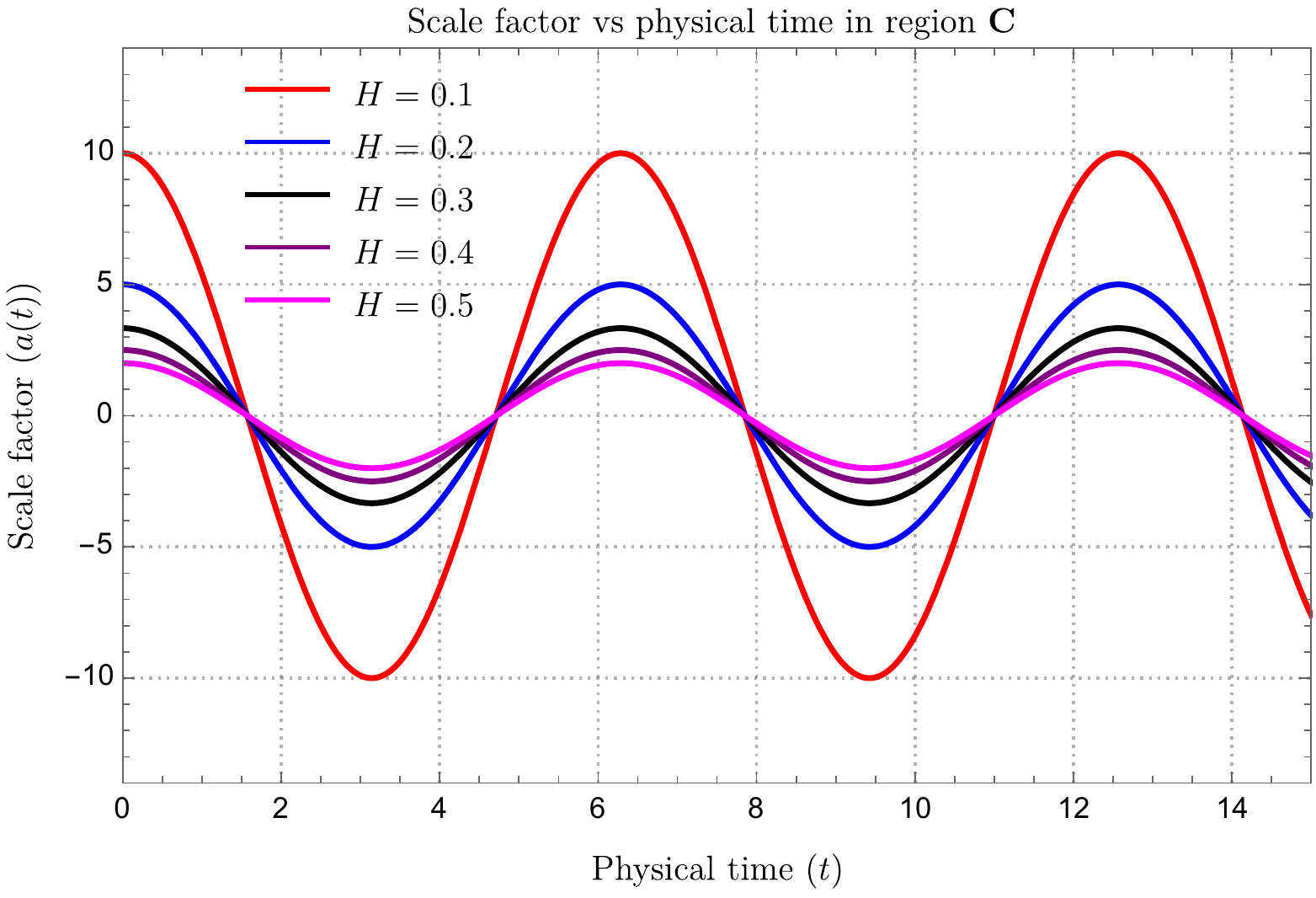}
        \label{s2}
       }
    \caption[Optional caption for list of figures]{Graphical behaviour of the scale factor ($a(t)$) with respect to physical time ($t$) in the region {\bf L},  {\bf R} and {\bf C} of the penrose diagram.}  
    \label{scale}
    \end{figure*}  
 Here the scale factor $a(t)$ in the three regions in terms of physical time variable $t$ can be written as:
   \bea
                 \label{f1}
        \displaystyle \textcolor{black}{\bf Region-R}&:\Longrightarrow&\displaystyle\left\{\begin{array}{ll}
       \displaystyle a(t_{\bf R})=\frac{1}{H}\sinh t_{\bf R},
                                                                          \end{array}
                                                                \right.\\
  \label{f2}
        \displaystyle \textcolor{black}{\bf Region-C}&:\Longrightarrow&\displaystyle\left\{\begin{array}{ll}
       \displaystyle  a(t_{\bf C})=\frac{1}{H}\cos t_{\bf C}, \end{array}
                                                                \right. \\
  \label{f3}
        \displaystyle \textcolor{black}{\bf Region-L}&:\Longrightarrow&\displaystyle\left\{\begin{array}{ll}
       \displaystyle  a(t_{\bf L})=\frac{1}{H}\sinh t_{\bf L}  \end{array}
                                                                \right.                                         \eea 
                                                                Here $H$ is Hubble parameter which is treated to be a constant in the above mentioned three regions of consideration.  In figure (\ref{scale}(a)) and figure (\ref{scale}(b)) we have plotted the behaviour of the scale factor $a(t)$ with respect to the physical time scale $t$ in regions {\bf R}, {\bf R} and {\bf C} respectively.  From the plots its clearly observed that in the case of region {\bf L} and {\bf R} the behaviour is same and show growing behaviour with respect to time.  This is expected from the geometrical construction global de Sitter space.  On the other hand,  for the region {\bf C} it shows periodic oscillatory behaviour.  In all of these plots we have used different values of the Hubble constants in dimensionless units.

   In the next subsection,  we will compute the expression for the wave function in presence of string theory originated axion effective interactions either in the region {\bf L} or {\bf R} of the open chart of the hyperbolic slices of global de Sitter space.  According to our construction of the geometrical set up the region {\bf L} and {\bf R} are replica of each other and exactly symmetric,  which can easily be observed from the representative pensore diagram as shown in figure~(\ref{penrose}).   For this particular reason we derive the wave function in both the regions,  but taking the partial trace over the contributions of the region {\bf R},  we use only the effective contributions from the region {\bf L} during the construction of reduced density matrix.  The similar trick can be applicable to the region {\bf R} as well.

    \subsection{Mode function and wave function of axion in an open chart}
    \label{ka2c}

In this part, we compute the equations for the global de Sitter space time hyperbolic open chart's hyperbolic open mode functions, associated wave functions, and quantum number and momentum dependent mode functions. We use the string theory-derived axion effective potential, which manifests as a scalar field in the matter sector, to make it happen. This outcome will be very helpful in computing the decreased density matrix expression as well as the entanglement negativity and its logarithmic counterpart in the following subsection. It is crucial to emphasise that in the current discussion, we are interested in the axion monodromy model.     
     
 In the present construction the axion field is generated from $RR$ sector of the {\bf Type IIB} string theory set up,  where the effective potential and the related interaction is originated as an outcome of the compactification on a Calabi-Yau three fold in presence of $NS5$ brane.  For the details from the string theory construction of the present set up see the refs. \cite{Panda:2010uq,Svrcek:2006yi,Beasley:2005iu} for the better understanding on the subject and the model we are considering in our paper.  We have mentioned clearly in the introductory part of this paper that,  axion would be the best candidate to study the effects of CHSH version of Bell's inequality violation.  For this reason the computational outcomes of this section will going to help us to study the non-local effects of quantum mechanical entanglement,  which is necessarily required to violate Bell's inequality from the quantum information theoretic point of view.
 One important fact we need to explicitly point out that,  in the context of bipartite quantum field theoretic set up explicit violation of Bell's CHSH inequality necessarily requires the existence of quantum mechanical entanglement in the prescribed theoretical set up.  But the converse fact is not always necessarily true.  Regarding this exception see ref. \cite{Horodecki:2009zz} which will give a clear underlying physical idea of the mentioned fact.  In this description the specific theoretical structure of the quantum mechanical state play very significant role.  Because within the framework quantum information theory bipartite system is characterised by a specific type of quantum state,  commonly known as {\it Werner} type of states which produce non zero quantum entanglement measure and this is the most important underlying concept to connect with the Bell's CHSH inequality violation \cite{Horodecki:2009zz}.  But the underlying behaviour of such states are not very clear in the present framework and without knowing the details it is very complicated to pursue the present analysis.

    \begin{figure*}[htb]
    \centering
    \subfigure[For $b<0$]{
        \includegraphics[width=14.2cm,height=9cm] {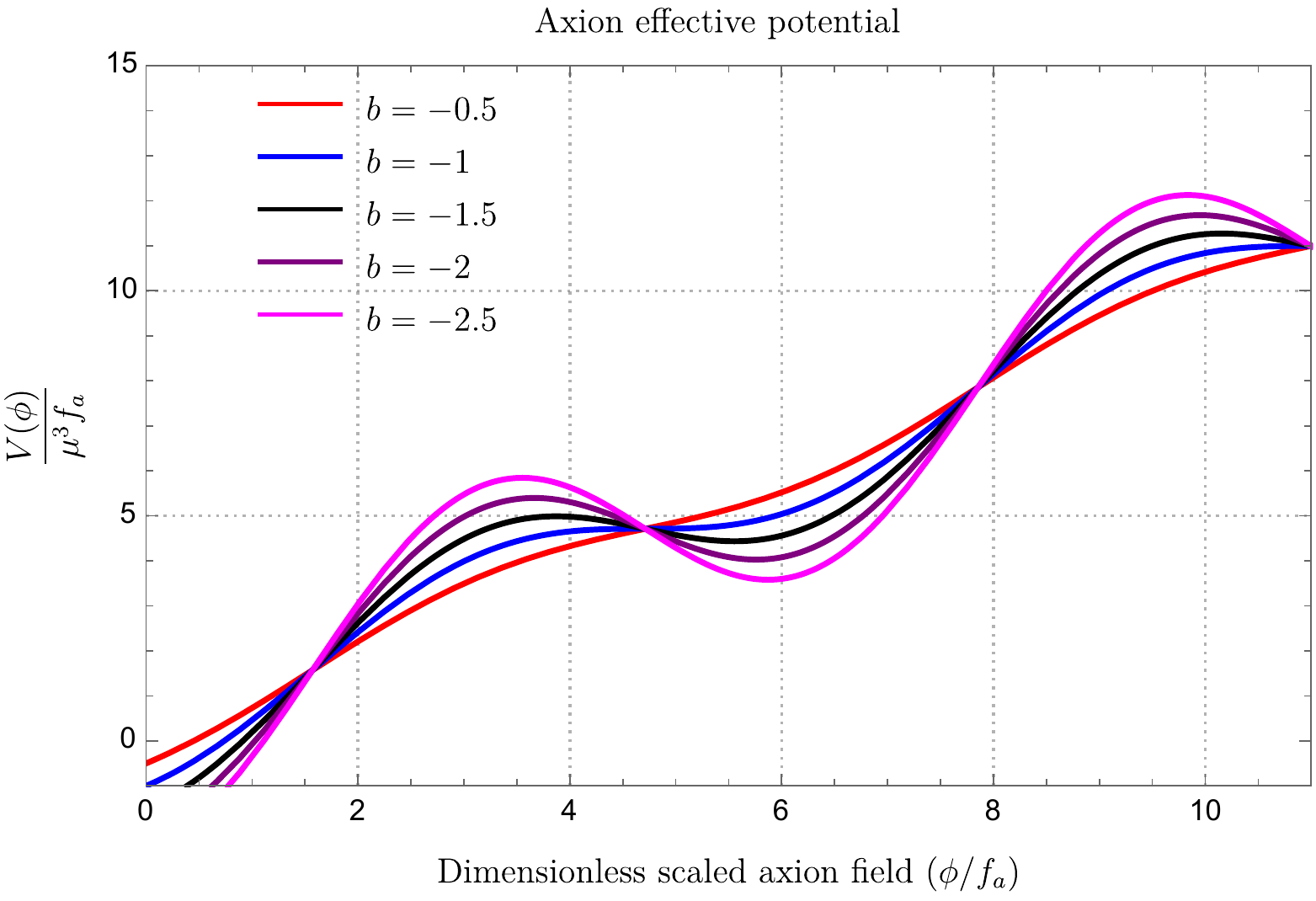}
        \label{fig1}
    }
    \subfigure[For $b>0$]{
        \includegraphics[width=14.2cm,height=9cm] {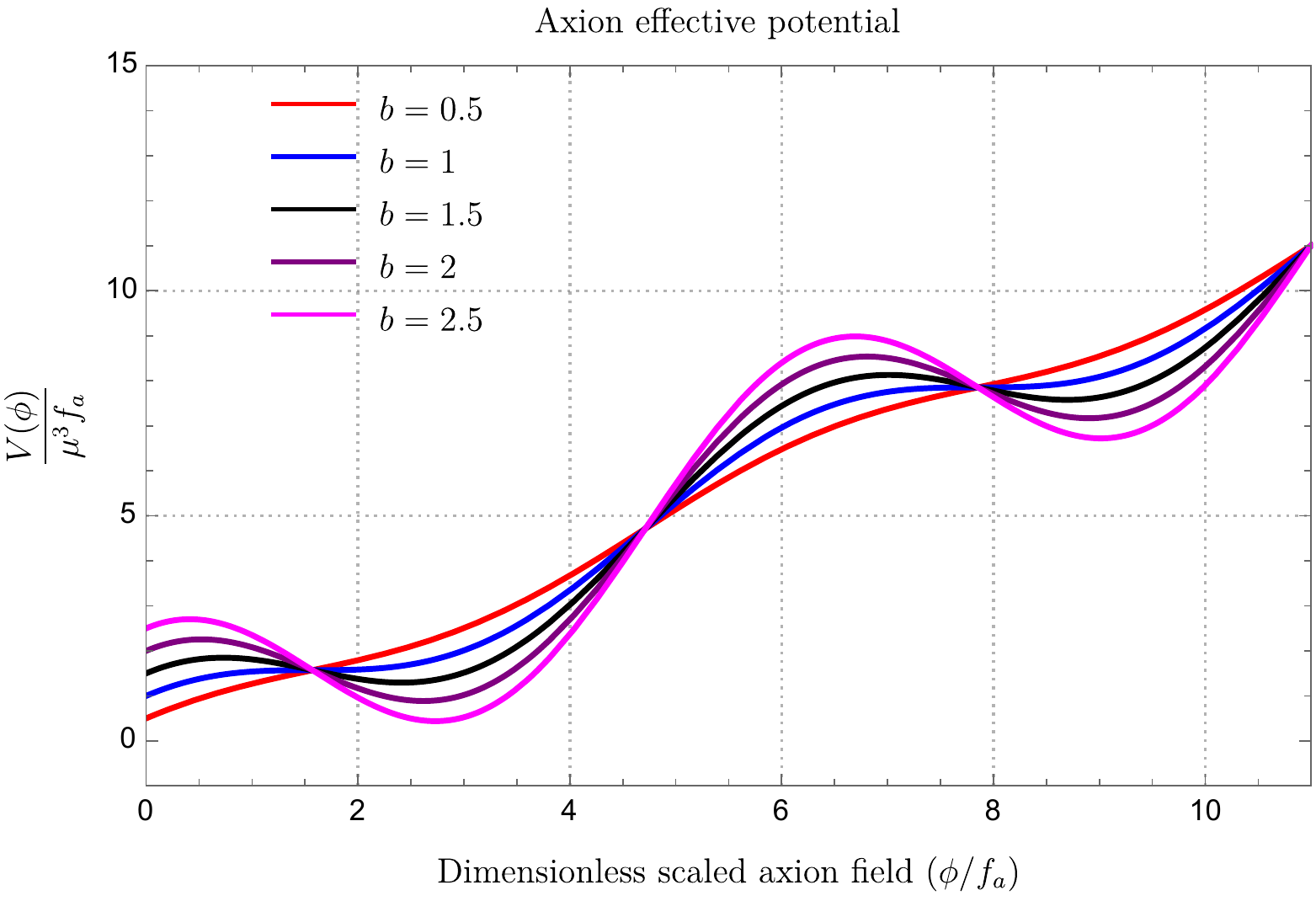}
        \label{fig2}
       }
    \caption[Optional caption for list of figures]{Graphical behaviour of the axion effective potential with respect to the field obtained from Type IIB String Theory compactification for stringy parameter $b<0$ and $b>0$.} 
    \label{pote1}
    \end{figure*}  
        Let us start with the following canonical
        effective action for axion field minimally coupled with the gravitational sector by space-time metric in $1+3$ dimensions:
         \be\label{axeff}   {\bf Effective~action:}~~~~~S= \int d^{4}x \sqrt{-g}\left[-\frac{1}{2}(\partial \phi)^2 -V(\phi)\right],\ee
       This action is derived from string compactification and is extremely useful to study the quntum information theoretic measures related to quantum entanglement.  Here $\phi$ is the dimensionful axionic field,  which is described by the following effective potential \cite{Panda:2010uq,Svrcek:2006yi,Beasley:2005iu} :
    \bea\label{axion} {\bf Effective~potential:}~~~~~V(\phi)&=&\mu^3\phi+\Lambda^4_{G}\cos\left(\frac{\phi}{f_{a}}\right)\nonumber\\
    &=&\mu^3f_{a}\left[\left(\frac{\phi}{f_{a}}\right)+b\cos\left(\frac{\phi}{f_{a}}\right)\right].~~~~~~~~~~\eea

    \begin{figure*}[htb]
    \centering
    \subfigure[For~conformal~time ]{
        \includegraphics[width=14.2cm,height=9cm] {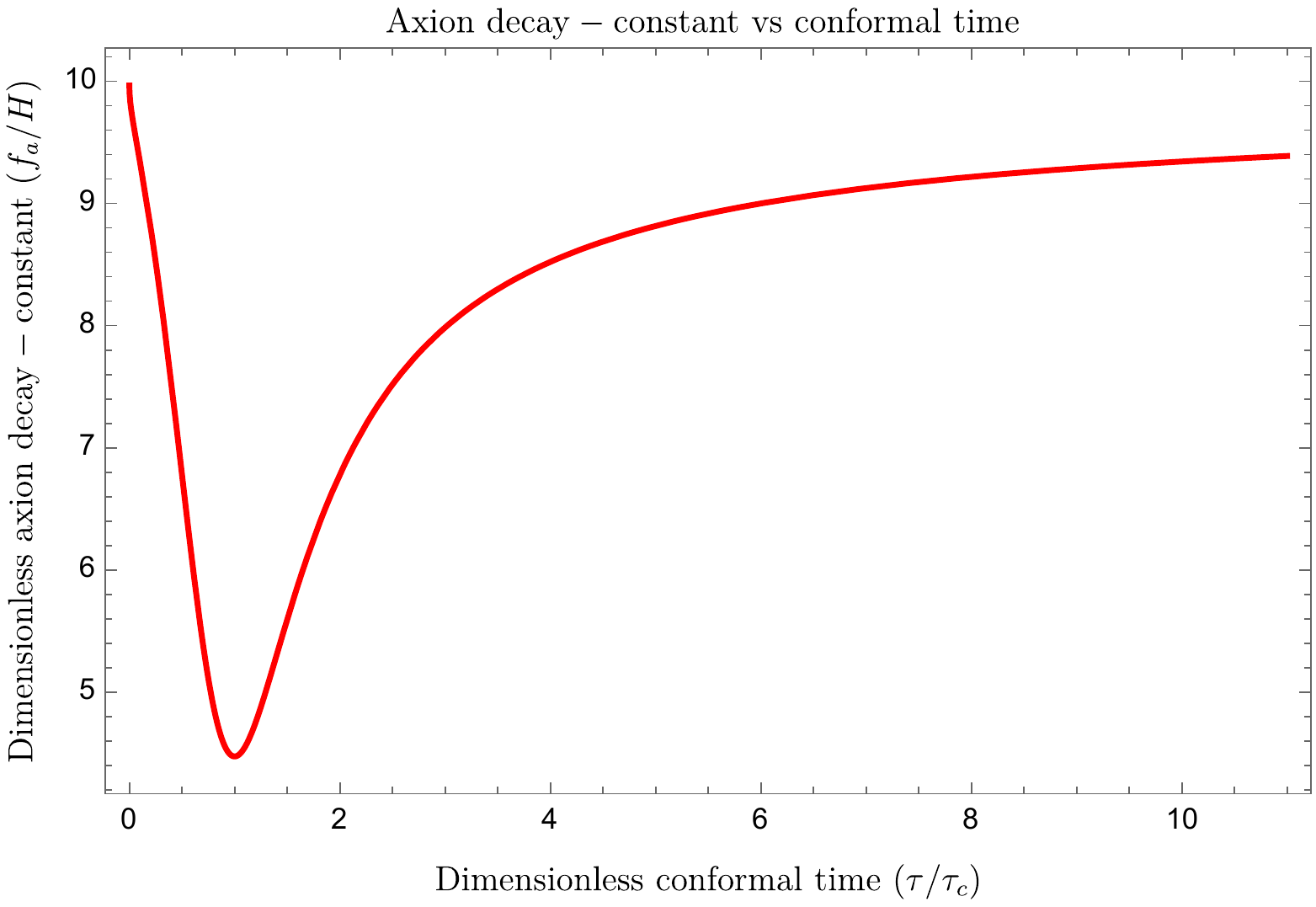}
        \label{fig3a}
    }
    \subfigure[For~physical~time]{
        \includegraphics[width=14.2cm,height=9cm] {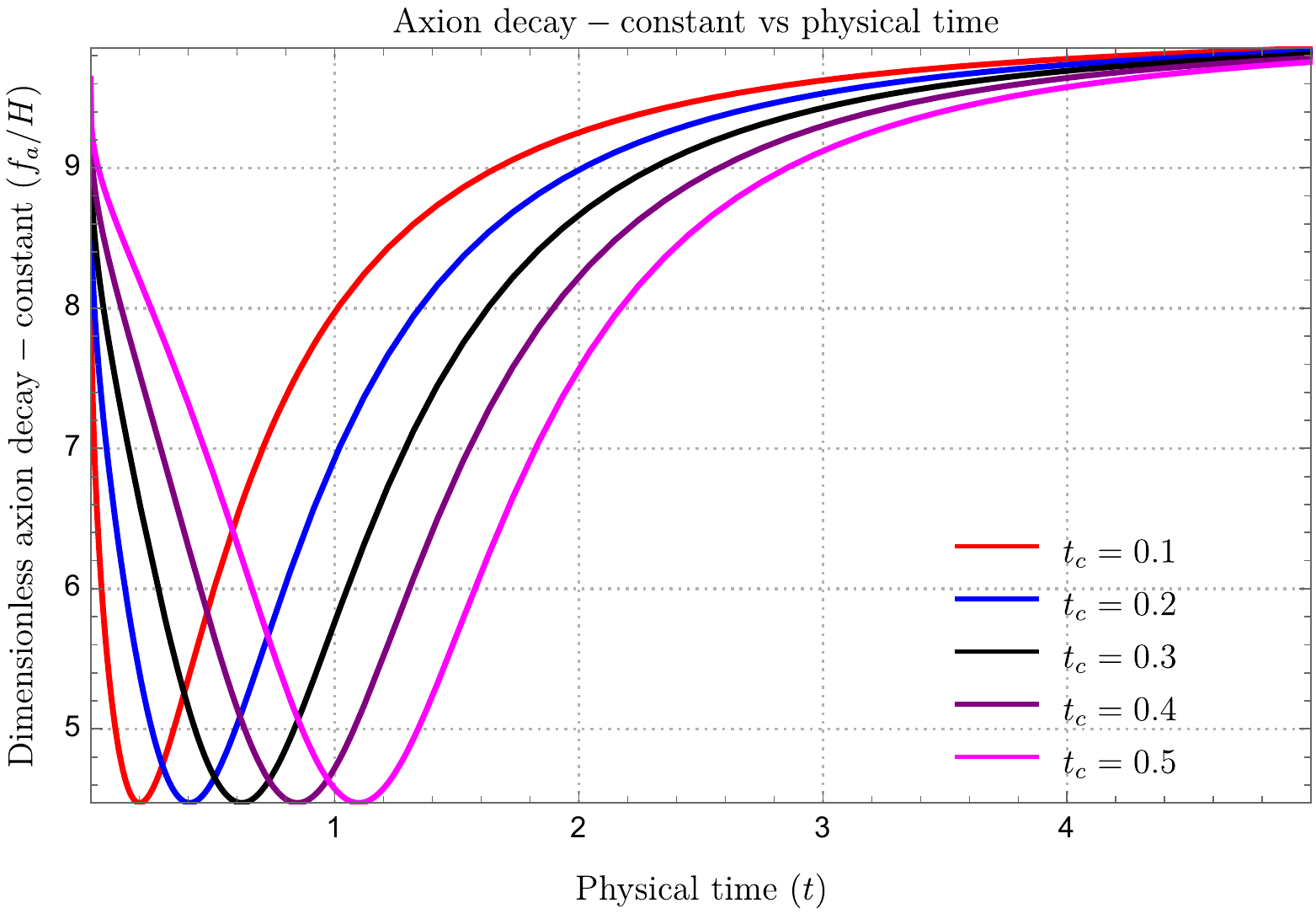}
        \label{fig3b}
       }
    \caption[Optional caption for list of figures]{Schematic behaviour of the dimensionless axion decay constant with the conformal and physical time scale.} 
    \label{fig3}
    \end{figure*}  
    In figure (\ref{pote1}(a)) and figure (\ref{pote1}(b)) we have plotted the behaviour of the dimensionless axion effective potential $V(\phi)/\mu^3 f_a$ with respect to the dimensionless axion field variable $\phi/f_a$ for the stringy parameter $b>0$ and $b>0$.  The precise definition of the stringy parameter $b$ is written in the later part of this paper explicitly.  Both the behaviour are useful to understand the contribution of the linear and non-perterbative periodic part in the total axion effective potential.

  In this effective potential $\mu^3$ represents the coupling parameter of linear interaction which is associated with the underlying theoretical scale,  can be expressed as:
  \bea {\bf Scale~of~effective~potential:}~~~~~\mu^3=\frac{1}{f_a \alpha^{'2}g_s}\exp(4A_0)+\frac{R^2 m^4_{SUSY}}{f_a \alpha^{'}L^4}\exp(2A_0),~~~~~\eea
  where $\exp(A_0)$ represents the warp factor of the lower portion of the Klebanov-Strassler throat geometry,   $R$ is the radius which stabilized the $5$ brane and antibrane in the corresponding string theory construction,   the only mass scale involved in this construction is $m_{SUSY}$,  which actually represents the underlying supersymmetry breaking scale in this specific set up,  the Regge slope is $\alpha^{'}$ which is proportional to the inverse string tension,  string coupling is $g_s$ and finally $L^6$ represents the world volume.  Here first part of the effective potential breaks the shift symmetry and the rest of the part preserves the symmetry $\phi\rightarrow \phi+2\pi f_a$.  Here $f_a$ quantifies the axionic decay parameter,  which is in general conformal time dependent ($\tau$) and we have chosen the following useful profile:
                     \bea {\bf Axion~decay~constant~profile:}~~~~~f_a=\sqrt{100-\frac{80}{1+\left(\ln\frac{\tau}{\tau_c}\right)^2}}~H,\eea
                         which was used in refs.~\cite{Maldacena:2015bha,Choudhury:2016cso,Choudhury:2016pfr} to validate Bell's CHSH inequality violation in early universe cosmology.  Here $H$ is the Hubble parameter,  and $\tau_c$ is the characteristic time scale at which we have $f_a=2\sqrt{5} H$,  which is almost a constant in the background geometrical set up we are considering.
 Now since the whole problem we are going to solve in terms of the physical time variable $t$ in both the regions {\bf R} and {\bf L} it is necessarily to find out the relationship between the conformal time scale $\tau$ and the physical time scale $t$.  Here we find the following connecting relationship:
 \bea  {\bf Dimensionless~conformal~time~scale:}~~~~~\frac{\tau}{\tau_c}=1+\frac{\displaystyle \ln\left[\frac{{\rm tanh}\left(\frac{t}{2}\right)}{{\rm tanh}\left(\frac{t_c}{2}\right)}\right]}{\displaystyle\ln\left[{\rm tanh}\left(\frac{t_c}{2}\right)\right]},\eea  
 where $\tau_c$ is given by the following expression:
 \bea {\bf Characteristic~time~scale:}~~~~~\tau_c=H\ln\left[{\rm tanh}\left(\frac{t_c}{2}\right)\right].\eea  
 For this reason the given profile of the axion decay parameter can be further written in terms of the physical time $t$ as:
                      \bea {\bf Axion~decay~constant~profile:}~~~~~f_a=\sqrt{100-\frac{80}{1+\left(\ln\left[\frac{ \ln\left[{\rm tanh}\left(\frac{t}{2}\right)\right]}{\ln\left[{\rm tanh}\left(\frac{t_c}{2}\right)\right]}\right]\right)^2}}~H,\nonumber\\
                      &&\eea
                      where $t_c$ represents the characteristic time scale in the physical time scale which basically $\tau_c$ in the conformal time scale.  In figure (\ref{fig3}(a)) and figure (\ref{fig3}(b)) we have plotted the behaviour of the dimensionless axion decay parameter $f_a/H$ with respect to conformal as well as the physical time scale.  Both the plots almost depict similar feature in both the time scales.  For this reason the chosen profile is extremely useful for our analysis as it is not changing by changing the definition of the associated time scales.
  
   Additionally,  to write down the effective potential in an simplest form we further introduce a new dimensionless quantity,  $b$,  which is defined as:
\bea {\bf New~dimensionless~parameter:}~~~~~ b= \frac{\Lambda^4_{G}}{\mu^3 f_{a}}.\eea
To define the new quantity $b$,  a characteristic scale has been introduced,  $\Lambda_{G}$,  which is given by:
\bea {\bf Characteristic~scale:}~~~~~ \Lambda_{G}=\sqrt{\frac{m_{SUSY} L^3}{ \sqrt{\alpha^{'}}g_{s}}}~\underbrace{\exp\left(-cS_{inst}\right)}_{\bf Instantonic~decay }.\eea
Here $S_{inst}$ represents the instantonic action which finally give rise present structure of the effective potential within the framework of string theory,  the instanton coupling parameter $c\sim{\cal O}(1)$ which is actually treated to be constant term in this computation.  Finally one can able fix the form of the warp factor in terms of all the stringy parameters,  which is given by:
\bea  {\bf Warp~factor:}~~~~~ \exp(A_0)=\left(\frac{\Lambda_G}{m_{SUSY}}\right)^2\frac{L}{R}\sqrt{\alpha^{'}g_s}=\frac{L^{4}}{m_{SUSY}R}\sqrt{\frac{\alpha^{'}}{g_s}}~\underbrace{\exp\left(-cS_{inst}\right)}_{\bf Instantonic~decay },~~~~~~\eea
which further corresponds to the following expression for the coupling parameter $\mu^3$,  which is given by:
  \bea &&{\bf Scale~in~terms~of~instantonic~decay:}~~~~~\nonumber\\
  \mu^3 &=& \frac{g^2_s}{f_a }\left(\frac{\Lambda_G}{m_{SUSY}}\right)^8\left(\frac{L}{R}\right)^4+\frac{ \alpha^{'}g^2_sR^2 m^4_{SUSY}}{f_aL^4}\left(\frac{\Lambda_G}{m_{SUSY}}\right)^4\left(\frac{L}{R}\right)^2\nonumber\\
  &=&\frac{1}{f_a g^{3}_s} \frac{L^{16}}{m^{4}_{SUSY}R^{4}}~\underbrace{\exp\left(-4cS_{inst}\right)}_{\bf Faster~decay }+\frac{m^2_{SUSY}L^{4}}{f_a g_s }~\underbrace{\exp\left(-2cS_{inst}\right)}_{\bf Slower~decay },\eea

  which is obviously a necessary input to fix the corresponding overall scale of effective potential derived from string theory.  Last but not the least,  one can further able to compute the expression for the string scale associated with the problem,  in terms of the other stringy parameters as:
    \begin{figure*}[ht]
    \centering
   {
        \includegraphics[width=14.2cm,height=8cm] {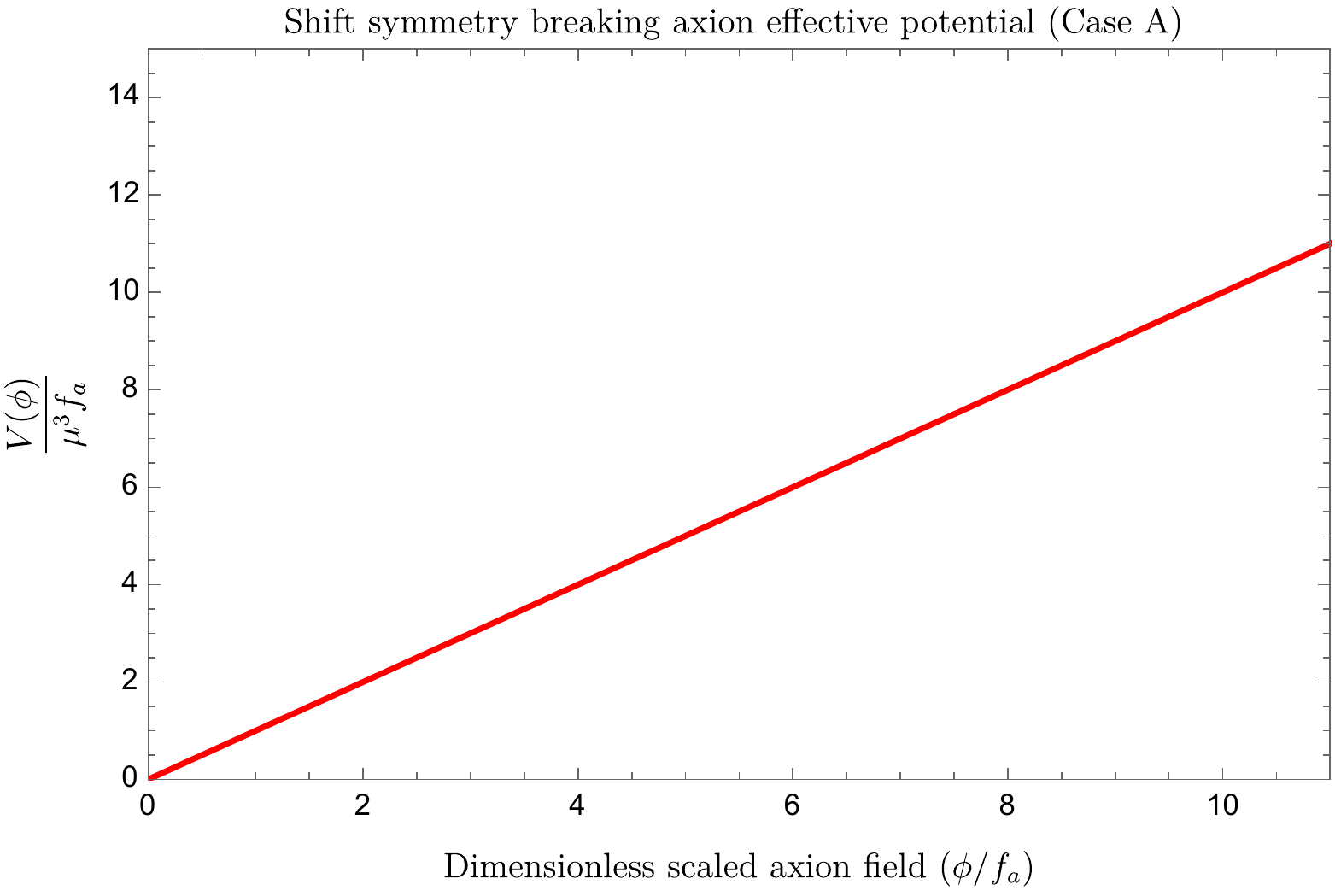}
    }
    \caption[Optional caption for list of figures]{Schematic behaviour of the approximated axion effective potential obtained from Case A.} 
    \label{fig4}
    \end{figure*}  
  \bea {\bf String~scale:}~~~~~M_s=\frac{1}{\sqrt{\alpha^{'}}} \exp(A_0)=\left(\frac{\Lambda_G}{m_{SUSY}}\right)^2\frac{L}{R}\sqrt{g_s}=\frac{L^{4}}{m_{SUSY}R\sqrt{g_s}}~\underbrace{\exp\left(-cS_{inst}\right)}_{\bf Instantonic~decay }\nonumber\\
  \eea
    \begin{figure*}[htb]
    \centering
    \subfigure[For $b<0$]{
        \includegraphics[width=14.2cm,height=9cm] {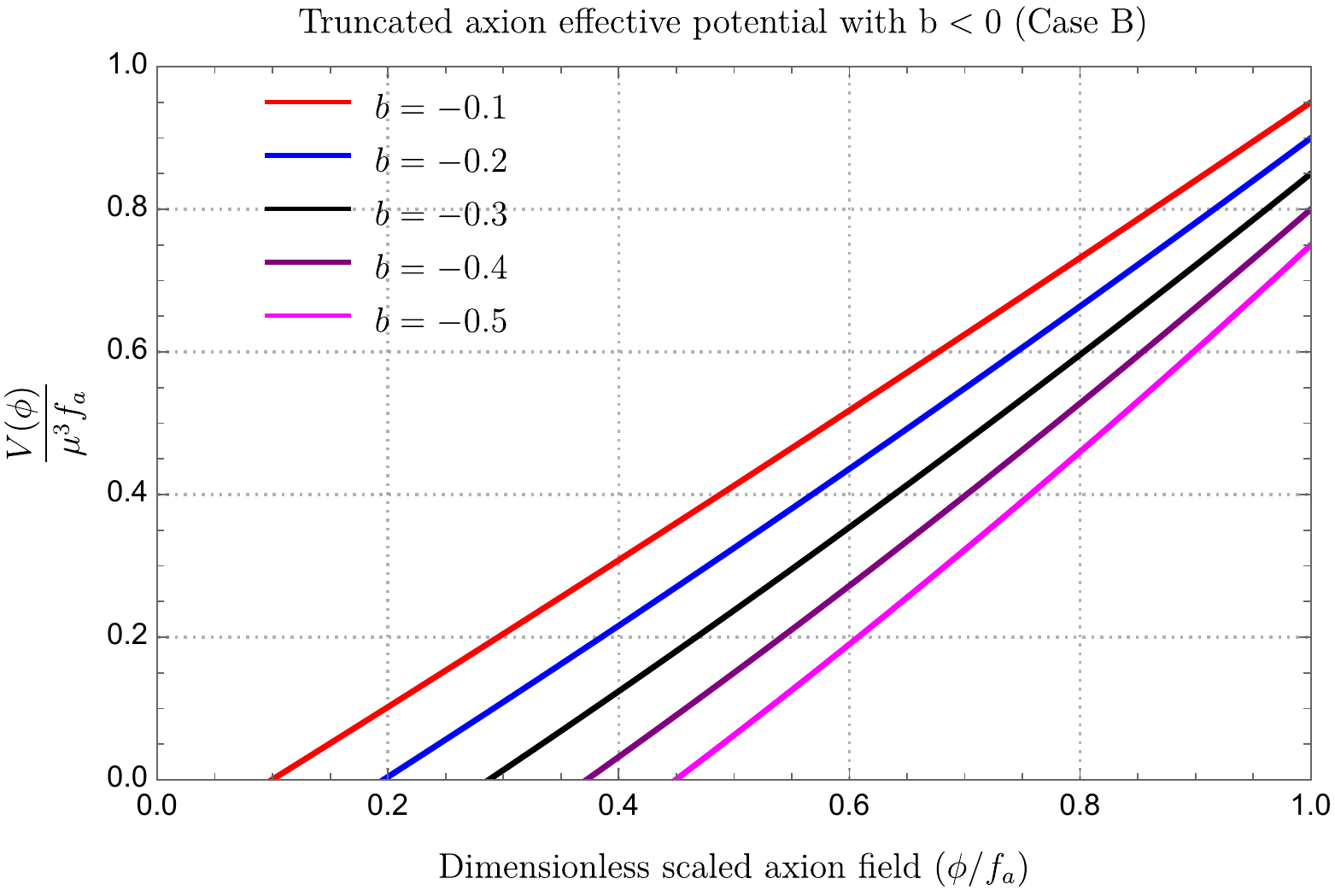}
        \label{fig5}
    }
    \subfigure[For $b>0$]{
        \includegraphics[width=14.2cm,height=9cm] {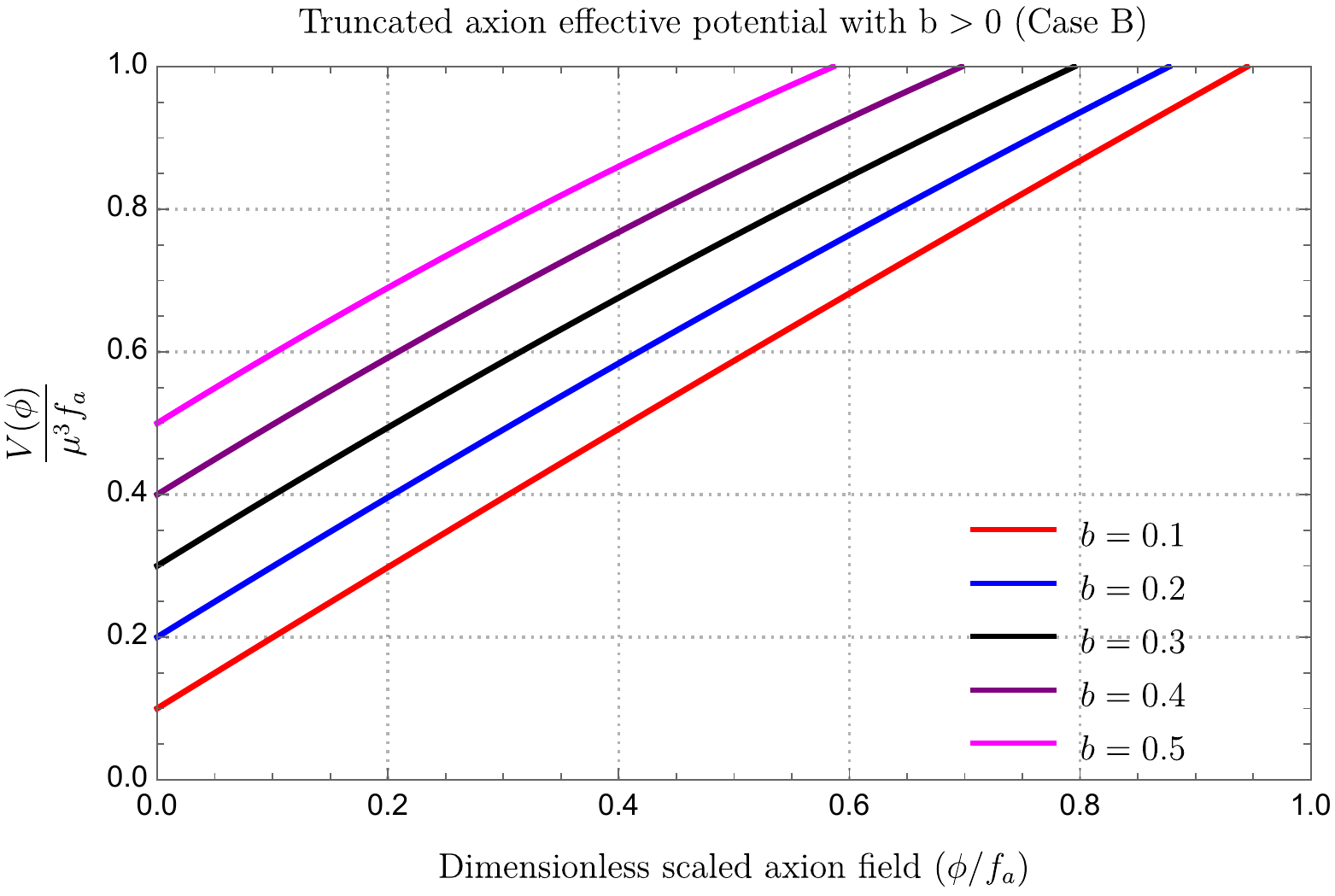}
        \label{fig6}
       }
    \caption[Optional caption for list of figures]{Schematic behaviour of the truncated axion effective potential obtained from Case B with stringy parameter $b<0$ and $b>0$.} 
    \label{pote2}
    \end{figure*}  
  The representative action along with the effective potential is a very important input for the rest of the computation performed in this paper.  

Here the following two physical approximation can be used to simplify the problem in a very simpler language:
    \begin{enumerate}
    \item \underline{\textcolor{black}{\bf Case~A:}}\\
    This is the specific situation where we only consider the part of the effective potential which breaks the shift symmetry $\phi\rightarrow \phi+2\pi f_a$,  which is given by:
        \bea\label{af1}  {\bf Shift~symmetry~breaking~effective~potential:}~~V(\phi)\approx \mu^3f_a\left(\frac{\phi}{f_{a}}\right).\eea
        In the present computational purpose,  particularly in the field equation the above mentioned potential contributes as a source term in terms of $\mu^3$ which basically fix the overall energy scale in terms of the stringy model parameters.  In figure~(\ref{fig4}),  the corresponding shift symmetry breaking part of the axion effective potential is plotted,  which shows linear behaviour.

    \item \underline{\textcolor{black}{\bf Case~B:}}\\
    In this specific case we consider the small field limiting approximation in which the dimensionless field variable $\displaystyle \frac{\phi}{f_a}\ll 1$.   For this reason one can approximate the shift symmetry $\phi\rightarrow \phi+2\pi f_a$ preserving non-perturbative contribution as:
     \bea \displaystyle \cos\left(\frac{\phi}{f_{a}}\right)\approx 1-\frac{1}{2}\left(\frac{\phi}{f_{a}}\right)^2,\eea where due to having the truncation at the quadratic order term at the end t.he previously mentioned shift symmetry is broken.  In prince one can take the full non-perturbative term, but at the level of eqn of motion and later deling with such terms are extremely difficult in the field theory language. In the present computational purpose the higher order terms can be neglected due to having small field limit without loosing any generality,  on the cost of breaking the shift symmetry.  Consequently in the present case the effective potential can be approximated as:
        \bea\label{af2}  {\bf Truncated~effective~potential:}~~~~~V(\phi)&\approx&\Lambda^4_{G}+\mu^3f_a\left(\frac{\phi}{f_{a}}\right)-\frac{\Lambda^4_{G}}{2}\left(\frac{\phi}{f_{a}}\right)^2\nonumber\\
        &=&\mu^3f_{a}\left(b+\left(\frac{\phi}{f_{a}}\right)\right)-\frac{m^2_{eff}}{2}\left(\frac{\phi}{f_{a}}\right)^2\nonumber\\
        &=&\mu^3f_{a}\left(b+\left(\frac{\phi}{f_{a}}\right)-\frac{b}{2}\left(\frac{\phi}{f_{a}}\right)^2\right),~~~~~~~~~~\eea
      where we define the effective mass in terms of stringy parameters by the following expression:
                         \bea 
                {\bf Effective~mass:}~~~~~          m^2_{eff}&=&\mu^3 b f_{a}=\Lambda^4_{G}=\left(\frac{m_{SUSY} L^3}{ \sqrt{\alpha^{'}}g_{s}}\right)^2~\underbrace{\exp\left(-4cS_{inst}\right)}_{\bf Instantonic~decay }.~~~~~~~~~~  \eea
                In figure~(\ref{pote2}(a)) and figure~(\ref{pote2}(b)),  the corresponding trucated axion effective potential is plotted,  which shows deviation from the linear behaviour for both the signatures of stringy parameter $b<0$ and $b>0$ for small field limit where $\phi\ll f_a$ approximation works perfectly well.  
                     
        \end{enumerate}

   We are now interested in the field equations of axion which can be obtained by varying the effective action stated in equation (\ref{axeff}) with respect to the axion field itself,  give rise to the following expressions for the above mentioned two cases:
    \bea \underline{\textcolor{black}{\bf For ~Case~A}:}~~~~~~~~~~~~~~~~~~~~~~~~~~~~~~\left[\frac{1}{a^3(t)}\partial_{t}\left(a^3(t)\partial_{t}\right)-\frac{1}{H^2a^2(t)}\hat{\bf L}^2_{\bf H^3}\right]\phi&=&\mu^3,\\
    \underline{\textcolor{black}{\bf For ~Case~B}:}~~~~~~~~~~~~~~~\left(\left[\frac{1}{a^3(t)}\partial_{t}\left(a^3(t)\partial_{t}\right)-\frac{1}{H^2a^2(t)}\hat{\bf L}^2_{\bf H^3}\right]+m^2_{eff}\right)\phi&=&\mu^3=\frac{m^2_{eff}f_a}{b},\nonumber\\
    \eea
 where the scale factor $a(t)$ for global de Sitter space is given by the following expression:
 \bea
                   \label{r2zzaa}
          \displaystyle \textcolor{black}{\bf Scale~factor}~~~~~~~~~~~~~~a(t)&=&\displaystyle\frac{1}{H}\sinh t~~~~{\rm where}~~~t=\Bigg(t_{\bf R}({\rm in}~ {\bf R}),t_{\bf L}({\rm in}~ {\bf L})\Bigg).\eea
 In this context we introduce a Laplacian operator $\hat{\bf L}^2_{\bf H^3}$ in hyperbolic slice ${\bf H^3}$ which is defined as:
          \bea \textcolor{black}{\bf Laplacian~operator:}~~~~\hat{\bf L}^2_{\bf H^3}&=&\frac{1}{\sinh^2r}\left[\partial_{r}\left(\sinh^2r~\partial_{r}\right)+\frac{1}{\sin\theta}\partial_{\theta}\left(\sin\theta~\partial_{\theta}\right)+\frac{1}{\sin^2\theta}\partial^2_{\Phi}\right].\nonumber\\
          &&\eea     
              which has the following properties:
              \begin{enumerate}
              \item Laplacian operator $\hat{\bf L}^2_{\bf H^3}$ satisfies the following eigenvalue equation:
              \bea\textcolor{black}{\bf Eigenvalue~equation:}~~~~ \hat{\bf L}^2_{\bf H^3}{\rm\cal Y}_{plm}(r,\theta,\Phi)&=&\lambda_p{\rm\cal Y}_{plm}(r,\theta,\Phi),    \eea
              with quantum number $p$ dependent eigenvalue: 
              \bea \textcolor{black}{\bf Eigenvalue:}~~~~\lambda_p=-(1+p^2).\eea
        \item  Eigenfunction of the  Laplacian operator $\hat{\bf L}^2_{\bf H^3}$ is ${\rm\cal Y}_{plm}(r,\theta,\Phi)$ which is defined as:
             \bea   \textcolor{black}{\bf Eigenfunction:}~~~~{\cal Y}_{plm}(r,\theta,\Phi)&=&\frac{\Gamma\left(ip+l+1\right)}{\Gamma\left(ip+1\right)}~\frac{p}{\sqrt{\sinh r}}~{\cal P}^{-\left(l+\frac{1}{2}\right)}_{\left(ip-\frac{1}{2}\right)}\left(\cosh r\right)Y_{lm}(\theta,\Phi),\nonumber\\
             &&\eea         
             where $p$, $l$ and $m$ are three quantum numbers associated with the above mentioned eigenfunction.  Here $Y_{lm}(\theta,\Phi)$ is the well known spherical harmonics which is dependent on two quantum numbers $l$ and $m$ and on two angular coordinates as it defined in ${\bf S}^2$ and last but not the least the radial solution is characterized by the function,  ${\cal P}^{-\left(l+\frac{1}{2}\right)}_{\left(ip-\frac{1}{2}\right)}\left(\cosh r\right)$,  which is the well known associated Legendre polynomial in this context.
              \end{enumerate}
              After quantization the classical solution obtained from the field equation is promoted in terms of the quantum operator and by following the well known canonical quantization technique the corresponding quantum operator can be written in terms of the creation and annihilation operators along with basis Bunch Davies mode function,  which is nothing but the classical counterpart of the solution of the field equation.  The total quantum solution for the axion field operator for both the {\bf Case A} and {\bf Case B} can be written in the following compact form:         
    \bea\label{total} \textcolor{black}{\bf Quantum~Mode~function:}~~~~\widehat{ \phi}(t,r,\theta,\Phi)&=&\int^{\infty}_{0} dp \sum_{\sigma=\pm 1}\sum^{p-1}_{l=0}\sum^{+l}_{m=-l}\left[a_{\sigma plm}{\cal U}_{\sigma plm}(t,r,\theta,\Phi)\right.\nonumber\\ &&\left.~~~~~+a^{\dagger}_{\sigma plm}{\cal U}^{*}_{\sigma plm}(t,r,\theta,\Phi)\right]\forall ~~t=(t_{\bf R},t_{\bf L}).\nonumber\\
    &&\eea   
In this context the Bunch-Davies vacuum is defined by the following expression:
\bea  \textcolor{black}{\bf Bunch-Davies~vacuum:}~~a_{\sigma p l m}|{\bf BD}\rangle&=&0~~~~~~~~~~ \forall \sigma=(+1,-1);0<p<\infty;\nonumber\\
&&~~~~~~~~~~~~l=0,\cdots,p-1,m=-l,\cdots,+l.~~~~~~~~~~~ \eea
Here ${\cal U}_{\sigma plm}(t,r,\theta,\Phi)$ represents the classical solution of the field equation for the axion for both the {\bf Case A} and {\bf Case B} which forms the complete basis.  After quantization this basis functions,  which is sometimes referred as the mode functions are tagged by the three quantum numbers,  $p$, $l$ and $m$,  which are appearing as an outcome of the canonical quantization of the modes in the present context of discussion.   The solution of the mode functions can be obtained by solving the corresponding axion field equations,  which are basically solving partial differential equations using the well known method of separation of variables for the {\bf Case A} and {\bf Case B}.  This finally give rise to the following expression for both the mentioned possibilities:
  \bea \textcolor{black}{\bf Bunch-Davies~Mode~function:}~~{\cal U}_{\sigma plm}(t,r,\theta,\Phi)&=&\frac{1}{a(t)}\chi_{p,\sigma}(t){\cal Y}_{plm}(r,\theta,\Phi)\nonumber\\
  &=&\frac{H}{\sinh t}\chi_{p,\sigma}(t){\cal Y}_{plm}(r,\theta,\Phi).~~~~~~~~\eea 
  Here it is important to note that the time dependent part of the mode function $\chi_{p,\sigma}(t)$ only works for the positive frequencies and hence forms a complete set in the present theoretical set up.  This part of the solution is dependent on the momentum $p$ which is actually the wave number and in the quantum mechanical picture it is playing the role of a quantum number as clearly mentioned earlier.  Particularly this time dependent part of the wave function is extremely significant for the present discussion as we are interested in the dynamical behaviour of the mode function in the {\bf R} and {\bf L} region of the open chart of global de Sitter space time.  This particular part will going to control the behaviour of the quantum entanglement measure in the mentioned space time.  If we can able to extract the hidden features from the time dependent part of the field equations from the {\bf Case A} and {\bf Case B} then half of the computational job is done.  Now since in both the cases we are dealing with inhomogeneous second order differential equations the total solution can be written as the sum of complementary part ($\chi^{(c)}_{p,\sigma}(t)$) and particular integral part ($\chi^{(p)}_{p,\sigma}(t)$) i.e.
          \bea \textcolor{black}{\bf Total~time~dependent~solution:}~~\chi_{p,\sigma}(t)=\underbrace{\chi^{(C)}_{p,\sigma}(t)}_{\bf Complementary~part}+\underbrace{\chi^{(P)}_{p,\sigma}(t)}_{\bf Particular~integral~part}.\nonumber\\
          \eea
         Here the complementary part ($\chi^{(c)}_{p,\sigma}(t)$) of the time dependent solution of the mode function satisfy the homogeneous part of the field equation for the {\bf Case A} and {\bf Case B}  can be written as:
              \bea
                                 \label{com}&&\textcolor{black}{\bf Complementary~part~of~the~field~equation:}~~\nonumber\\
                        \displaystyle 0&=&\displaystyle\left\{\begin{array}{ll}
                       \displaystyle \left[\partial^2_t +3\coth t~ \partial_t+\frac{(1+p^2)}{\sinh^2t}\right]\chi^{(C)}_{p,\sigma}(t)~~~~~~~~~~~~~~~~~ &
                                                                                 \mbox{\small {\textcolor{black}{\bf for Case A}}}  
                                                                                \\ \\
                                \displaystyle \left[\partial^2_t +3\coth t~ \partial_t+\frac{(1+p^2)}{\sinh^2t}+\frac{m^2_{eff}}{H^2}\right]\chi^{(C)}_{p,\sigma}(t) & \mbox{\small {\textcolor{black}{\bf for Case B}}}.~~~~~~~~
                                                                                          \end{array}
                                                                                \right.\eea
                                                                                The solution of the above equations for the {\bf Case A} and {\bf Case B} combiningly can be written as:
 \bea
\label{sol}&&\textcolor{black}{\bf Complementary~solution:}~~\nonumber\\
                         \displaystyle \nonumber&&\\ \chi^{(c)}_{p,\sigma}(t)&=&\displaystyle\left\{\begin{array}{ll}
                        \displaystyle \left\{\frac{1}{2\sinh\pi p}\left[\frac{\left(e^{\pi p}-i\sigma~e^{-i\pi\nu}\right)}{\Gamma\left(\nu+\frac{1}{2}+ip\right)}{\cal P}^{ip}_{\left(\nu-\frac{1}{2}\right)}(\cosh t_{\bf R})\displaystyle\right.\right.\\
                        \left.\left.  \displaystyle~~~ ~~~~~~~~~~~-\frac{\left(e^{-\pi p}-i\sigma~e^{-i\pi\nu}\right)}{\Gamma\left(\nu+\frac{1}{2}-ip\right)}{\cal P}^{-ip}_{\left(\nu-\frac{1}{2}\right)}(\cosh t_{\bf R})\right]\right\}_{\sigma=\pm 1}~~ &
                                                                          ~~~ ~~~~~~~~~~~~~~~~~~~~~ \mbox{\small {\textcolor{black}{\bf for R}}}  
                                                                                 \\ 
                                 \displaystyle \left\{\frac{\sigma}{2\sinh\pi p}\left[\frac{\left(e^{\pi p}-i\sigma~e^{-i\pi\nu}\right)}{\Gamma\left(\nu+\frac{1}{2}+ip\right)}{\cal P}^{ip}_{\left(\nu-\frac{1}{2}\right)}(\cosh t_{\bf L})\displaystyle\displaystyle\right.\right.\\
                        \left.\left.  \displaystyle~~~ ~~~~~~~~~~~-\frac{\left(e^{-\pi p}-i\sigma~e^{-i\pi\nu}\right)}{\Gamma\left(\nu+\frac{1}{2}-ip\right)}{\cal P}^{-ip}_{\left(\nu-\frac{1}{2}\right)}(\cosh t_{\bf L})\right]\right\}_{\sigma=\pm 1} & ~~~ ~~~~~~~~~~~~~~~~~~~~~~ \mbox{\small {\textcolor{black}{\bf for L}}},~~
                                                                                           \end{array}                          \right.                                                                                           \eea 
where 
we introduce a new parameter $\nu$,  which is known as mass parameter are defined for {\bf Case A} and {\bf Case B} as:
   \bea
                                 \label{nu}
                        \displaystyle \textcolor{black}{\bf Mass~parameter:}~~\nu&=&\displaystyle\left\{\begin{array}{ll}
                       \displaystyle \frac{3}{2}~~~~~~~~~~~~~~~~~ &
                                                                                 \mbox{\small {\textcolor{black}{\bf for Case A}}}  
                                                                                \\ 
                                \displaystyle \sqrt{\frac{9}{4}-\frac{m^2_{eff}}{H^2}}=\sqrt{\frac{9}{4}-\frac{\mu^3 b}{f_a H^2}}=\sqrt{\frac{9}{4}-\frac{\Lambda^4_G}{f^2_a H^2}} ~~~~~~~~~& \mbox{\small {\textcolor{black}{\bf for Case B}}}.~~~~~~~~
                                                                                          \end{array}
                                                                                \right.\eea
 Here the solution has following properties:                                                                               
\begin{enumerate}
\item Here $\sigma=\pm 1$ for {\bf R} and {\bf L} regions. 
 
\item In the {\bf Case B} if we consider $m_{eff}\ll H$,  the mass parameter is approximated as $\nu=3/2$ which is exactly the {\bf Case A}.

\item In the {\bf Case B} if we consider $m_{eff}=\sqrt{2}H$,  then the mass parameter is given by $\nu=1/2$ and this is the case of conformally coupling. 

\item In the {\bf Case B} if we consider $m_{eff}<\sqrt{2}H$,  the mass parameter is lying within the range,  $1/2<\nu<3/2$,  which is the low mass region.

\item In the {\bf Case B} if we consider $m_{eff}=H$,  then the mass parameter is given by,  $\nu=5/2$,  which is the intermediate mass region.

\item In the {\bf Case B} if we consider $m_{eff}\gg H$,  then the mass parameter is given by,  $\displaystyle \nu=i\sqrt{\frac{m^2_{eff}}{H^2}
-\frac{9}{4}}\approx i\frac{m_{eff}}{H}$,  which is the high mass region.

\item  In the {\bf Case B} if we consider $\sqrt{2}H<m_{eff}<3H/2$, then the mass parameter $\nu$ is lying within the range,  $0<\nu<1/2$.

\item Complementary part of the solution satisfy,  $\chi^{(C)}_{p,\sigma}(t)=\chi^{(C)}_{-p,\sigma}(t)$.

\item  Here one can define the following Klien-Gordon inner product in terms of the complementary part of the time dependent fields equation:
\bea {\bf Klien-Gordon~product:}~~~~\left(\left(\chi^{(C)}_{p,\sigma}(t),\chi^{(C)}_{p,\sigma^{'}}(t)\right)\right)_{\bf KG}={\cal N}_{p\sigma}\delta_{\sigma\sigma^{'}}, \eea
where ${\cal N}_{p\sigma}$ is the normalization constant,  which is given by:
\bea
                                 \label{norm}
                        \displaystyle {\bf Normalization:}~~~~{\cal N}_{p\sigma}&=&\displaystyle
                                \displaystyle \frac{4}{\pi}\frac{\left[\cosh\pi p-\sigma\cos\left(\nu-\frac{1}{2}\right)\right]}{|\Gamma\left(\nu+\frac{1}{2}+ip\right)|^2}\forall \sigma=\pm 1.\eea
                                This will going to be extremely useful to further fix the overall normalization factor of the complementary part of the time dependent contribution of the corresponding mode function in the present context of discussion.   

\end{enumerate}

On the other hand the particular integral part satisfy the following time dependent part of the field equation:
   \bea
                                 \label{com2}&& \textcolor{black}{\bf Particular~part~of~the~field~equation:}~~\nonumber\\
                        \displaystyle &&\displaystyle\left\{\begin{array}{ll}
                       \displaystyle \left[\partial^2_t +3\coth t~ \partial_t+\frac{(1+p^2)}{\sinh^2t}\right]\chi^{(P)}_{p,\sigma}(t)=\mu^3~~~~~~~~~~~~~~~~~ &
                                                                                 \mbox{\small {\textcolor{black}{\bf for Case A}}}  
                                                                                \\ 
                                \displaystyle \left[\partial^2_t +3\coth t~ \partial_t+\frac{(1+p^2)}{\sinh^2t}+\frac{m^2_{eff}}{H^2}\right]\chi^{(P)}_{p,\sigma}(t)=\frac{m^2_{eff}f_a}{b}~~~~~~~~ & \mbox{\small {\textcolor{black}{\bf for Case B}}}.~~~~~~~~
                                                                                          \end{array}
                                                                                \right.\eea
Here in both the cases we have inhomogeneous differential equations and the inhomogeneous contributions in both the cases playing the role of source terms in the present context of discussion.  Since we have chosen a specific time dependent profile of the axion decay parameter to prepare Bell CHSH inequality violating pair for the {\bf Case B} the source term in this particular case will be time dependent in a very specific fashion.

It is a well known fact that apart from having any type of structure of the source contribution one can able to solve the inhomogeneous differential equation using the Green's function method.  This leads to the following solution of the particular integral part for both the cases considered for our analysis in this paper:
\bea
                                 \label{com3} \textcolor{black}{\bf Particular~solution:}~~
                        \displaystyle  \chi^{(P)}_{p,\sigma}(t)&=&\displaystyle\left\{\begin{array}{ll}
                       \displaystyle\int dt^{'}~G_{\sigma}(t,t^{'})~\mu^3~~~~~~~~~~~~~~~~~ &
                                                                                 \mbox{\small {\textcolor{black}{\bf for Case A}}}  
                                                                                \\ 
                                \displaystyle\int dt^{'}~G_{\sigma}(t,t^{'})~\frac{m^2_{eff}f_a(t^{'})}{b}~~~~~~~~ & \mbox{\small {\textcolor{black}{\bf for Case B}}}.~~~~~~~~
                                                                                          \end{array}
                                                                                \right.\eea
   where $G_{\sigma}(t,t^{'})$ is the Green's function for axion field, which is given by the following general expression:
   \bea\label{green} \textcolor{black}{\bf Green's~function:}~~G_{\sigma}(t,t^{'})&=&\sinh^2 t\sum^{\infty}_{n=0}\frac{1}{\left(p^2-p^2_{n}\right)}\chi^{(C)}_{p_{n},\sigma}(t)\chi^{(C)}_{p_{n},\sigma}(t^{'})\quad\quad{\rm where}\quad\sigma=\pm 1.\nonumber\\
   &&\eea 
   
   Further,  we use the following new notations to express the total solution of the time dependent part of the field equation:
\bea 
                                   \label{bbvx}
                          \displaystyle \textcolor{black}{\bf New~notation:}~~{\cal P}^{q}&=& {\cal P}^{ip}_{\left(\nu-\frac{1}{2}\right)}(\cosh t_{q}),\quad{\cal P}^{{q},n}=
                                  \displaystyle {\cal P}^{ip_n}_{\left(\nu-\frac{1}{2}\right)}(\cosh t_{q}) \quad\quad {\rm where}\quad q=\left({\bf R},{\bf L}\right).\nonumber\\
                                  &&
                           \eea

                                                 As a result, the solution's entire time-dependent component can be reduced to the following:
   \bea
                                   \label{cxcxx}  
                          \displaystyle \boxed{\boxed{\chi_{p,\sigma}(t)=\sum_{q={\bf R},{\bf L}}\left\{\underbrace{\frac{1}{{\cal N}_{p}}\left[\alpha^{\sigma}_{q}~{\cal P}^{q}+\beta^{\sigma}_{q}~{\cal P}^{q*}\right]}_{\textcolor{black}{\bf Complementary~solution}}+\underbrace{\sum^{\infty}_{n=0}\frac{1}{{\cal N}_{p_n}\left(p^2-p^2_n\right)}\left[\bar{\alpha}^{\sigma}_{q,n}~\bar{\cal P}^{q,n}+\bar{\beta}^{\sigma}_{q,n}~\bar{\cal P}^{*q,n}\right]}_{\textcolor{black}{\bf Particular~solution}}\right\}\forall \sigma=\pm 1}},\nonumber\\
                          &&\eea 
                          where we use two new redefined symbol for the further simplification purpose in the present computation:
                                                   \bea
                                 \label{com5}                              
                        \displaystyle \overline{\cal P}^{q,n}&=& \sinh^2t~  {\cal P}^{q,n}\times\displaystyle\left\{\begin{array}{ll}
                       \displaystyle\int dt^{'}~\chi^{(C)}_{p_n,\sigma,q}(t^{'})~\mu^3~~~~~~~~~~~~~~~~~ &
                                                                                 \mbox{\small {\textcolor{black}{\bf for Case A}}}  
                                                                                \\ 
                                \displaystyle\int dt^{'}~\chi^{(C)}_{p_n,\sigma,q}(t^{'})~\frac{m^2_{eff}f_a(t^{'})}{b}~~~~~~~~ & \mbox{\small {\textcolor{black}{\bf for Case B}}}.~~~~~~~~
                                                                                          \end{array}
                                                                                \right.\\
         {\cal N}_{p}&=&2\sinh \pi p ~\sqrt{{\cal N}_{p\sigma}}=4\sinh \pi p ~\sqrt{ \frac{\left[\cosh\pi p-\sigma\cos\left(\nu-\frac{1}{2}\right)\right]}{\pi|\Gamma\left(\nu+\frac{1}{2}+ip\right)|^2}}\forall \sigma=\pm 1,q={(\bf R},{\bf L})~~~~~~~~~~~~\\
         {\cal N}_{p_n}&=&2\sinh \pi p_n ~\sqrt{{\cal N}_{p_n\sigma}}=4\sinh \pi p_n ~\sqrt{ \frac{\left[\cosh\pi p_n-\sigma\cos\left(\nu-\frac{1}{2}\right)\right]}{\pi|\Gamma\left(\nu+\frac{1}{2}+ip_n\right)|^2}}\forall \sigma=\pm 1,q={(\bf R},{\bf L})~~~~~~~~~~~~\eea                                                                      
  In the above mentioned solution as mentioned in equation~(\ref{cxcxx}),  we define few expansion coefficients,  which are given by:
   \bea 
   \label{dex1}&&{\bf Expansion~coefficients:}~~\nonumber\\
   \alpha^{\sigma}_{\bf R}&=&\frac{1}{\sigma}\alpha^{\sigma}_{\bf L}=\frac{\left(e^{\pi p}-i\sigma e^{-i\pi\nu}\right)}{\Gamma\left(\nu+\frac{1}{2}+ip\right)},\alpha^{\sigma}_{{\bf R},n}=\frac{1}{\sigma}\alpha^{\sigma}_{{\bf L},n}=\frac{\left(e^{\pi p_n}-i\sigma e^{-i\pi\nu}\right)}{\Gamma\left(\nu+\frac{1}{2}+ip_n\right)}\\
                  \beta^{\sigma}_{\bf R}&=& \frac{1}{\sigma}\beta^{\sigma}_{\bf L}= -\frac{\left(e^{-\pi p}-i\sigma e^{-i\pi\nu}\right)}{\Gamma\left(\nu+\frac{1}{2}-ip\right)}., \beta^{\sigma}_{{\bf R},n}= \frac{1}{\sigma}\beta^{\sigma}_{{\bf L},n}= -\frac{\left(e^{-\pi p_n}-i\sigma e^{-i\pi\nu}\right)}{\Gamma\left(\nu+\frac{1}{2}-ip_n\right)}.   ~~~~~~~~
 \eea 
                                                                                                                                      
 Further the solution written in equation~(\ref{cxcxx}) can be written in matrix notation for the further simplification:
 \bea\label{xmatrix} \boxed{\boxed{{\bf Matrix~solution:}~~~~{\bf \chi}^{I}=\underbrace{\frac{1}{{\cal N}_p}{\cal  M}^{I}_{J}{\cal P}^{J}}_{\bf Complementary~solution}+\underbrace{\sum^{\infty}_{n=0}\frac{1}{{\cal N}_{p,(n)}}\left({\cal  M}_{(n)}\right)^{I}_{J}{\cal P}^{J}_{(n)}}_{\bf Particular~solution}}}\eea 
 where we define two square matrices for the complementary and particular part as:
  \bea {\cal M}^{I}_{J}&=&\left(\begin{array}{ccc} \alpha^{\sigma}_{q} &~~~ \beta^{\sigma}_{q} \\ \beta^{\sigma^{*}}_{q} &~~~ \alpha^{\sigma^{*}}_{q}  \end{array}\right),~~
  \left({\cal M}_{(n)}\right)^{I}_{J}=\left(\begin{array}{ccc} \bar{\alpha}^{\sigma}_{q,n} &~~~ \bar{\beta}^{\sigma}_{q,n} \\ \bar{\beta}^{\sigma^{*}}_{q,n} &~~~ \bar{\alpha}^{\sigma^{*}}_{q,n}  \end{array}\right),~~\sigma=\pm 1, q=({\bf R},{\bf L}), (I,J)=1,2,3,4.~~~~~~~~~~\eea
 and similarly the useful column matrices can be expressed as:
 \bea  {\cal P}^{J}_{(n)}=\left(\begin{array}{ccc} {\cal P}^{q,n} \\ {\cal P}^{{q^*},n}\\
          \end{array}\right),~~
  \chi^{I}=\left(\begin{array}{ccc} \chi_{\sigma}(t) \\ \chi^{*}_{\sigma}(t),
   \end{array}\right),~~
   {\cal P}^{J}=\left(\begin{array}{ccc} {\cal P}^{q} \\ {\cal P}^{{q^*}},\\
      \end{array}\right)~~\sigma=\pm 1, q=({\bf R},{\bf L}), (I,J)=1,2,3,4.~~~~~~~~~~\eea
Also we introduce another new symbol for the normalization factor ${\cal N}_{p,(n)}$ as obtained for the particular part of the solution,  which is given by:
\bea {\cal N}_{p,(n)}&=&2\sinh \pi p_n ~\sqrt{{\cal N}_{p_n\sigma}} ~\left(p^2-p^2_n\right)\nonumber\\
&=&4\sinh \pi p_n ~\left(p^2-p^2_n\right)~\sqrt{ \frac{\left[\cosh\pi p_n-\sigma\cos\left(\nu-\frac{1}{2}\right)\right]}{\pi|\Gamma\left(\nu+\frac{1}{2}+ip_n\right)|^2}}\forall \sigma=\pm 1,q={(\bf R},{\bf L}).~~~~~\eea
Hence the Bunch-Davies mode function can be rewritten as:  
          \bea   \frac{H}{\sinh t}a_{I}\chi^{I}=\frac{H}{\sinh t}a_{I}\left[\frac{1}{{\cal N}_p}{\cal  M}^{I}_{J}{\cal P}^{J}+\sum^{\infty}_{n=0}\frac{1}{{\cal N}_{p,(n)}}\left({\cal  M}_{(n)}\right)^{I}_{J}{\cal P}^{J}_{(n)} \right],~~{\rm where}~~ a_{I}=(a_{\sigma},
           a^{\dagger}_{\sigma}).~~~~~~~~
           \eea                                  
Further we define:
           \bea\label{def1} b_{J} &=& a^{(C)}_{I}{\cal M}^{I}_{J},~~~ b_{J(n)} = a^{(P)}_{I(n)}\left({{\cal M}_{(n)}}\right)^{I}_{J},~~{\rm where}~~ a^{(C)}_{I}=(a^{(C)}_{\sigma},
                      a^{(C)\dagger}_{\sigma}), a^{(P)}_{I(n)}=(a^{(P)}_{\sigma,n},a^{(P)\dagger}_{\sigma,n}).~~~~~~~~~~~
           \eea
This implies that the operator ansatz is following:
 \bea a_{I}&=& \left[a^{(c)}_{I}+\sum^{\infty}_{n=0}a^{(p)}_{I(n)}\right], a^{(c)}_{I} = b_{J}\left({\cal M}^{-1}\right)^{I}_{J},~~~ a^{(p)}_{I(n)} = b_{J(n)}\left({\cal M}^{-1}_{(n)}\right)^{I}_{J},\eea 
  where inverse square matrices are defined as:
   \bea \left({\cal M}^{-1}\right)^{I}_{J}&=&\left(\begin{array}{ccc} \gamma_{\sigma q} &~~~ \delta_{\sigma q} \\ \delta^{*}_{\sigma q} &~~~ \gamma^{*}_{\sigma q}  \end{array}\right),
        ~~~~\left({\cal M}^{-1}_{(n)}\right)^{I}_{J}=\left(\begin{array}{ccc}\overline{\gamma}_{\sigma q,n} &~~~ \overline{\delta}_{\sigma q,n} \\ \overline{\delta}^{*}_{\sigma q,n} &~~~\overline{\gamma}^{*}_{\sigma q,n}  \end{array}\right),\eea
 The individual components of these matrices are given by:
           \bea 
           \label{r1}\gamma_{j\sigma}&=& \displaystyle     \displaystyle \frac{\Gamma\left(\nu+\frac{1}{2}+ip\right)~e^{\pi p+i\pi\left(\nu+\frac{1}{2}\right)}}{4\sinh\pi p}\left(\begin{array}{ccc} \frac{1}{\displaystyle e^{\pi p+i\pi \left(\nu+\frac{1}{2}\right)}+1} &~~~ \frac{1}{\displaystyle e^{\pi p+i\pi \left(\nu+\frac{1}{2}\right)}-1} \\ \frac{1}{\displaystyle e^{\pi p+i\pi \left(\nu+\frac{1}{2}\right)}+1} &~~~ -\frac{1}{\displaystyle e^{\pi p+i\pi \left(\nu+\frac{1}{2}\right)}-1}  \end{array}\right)~~~~~~~ \\
         \label{r2}\delta^{*}_{j\sigma}&=&
                                                                               \displaystyle \frac{\Gamma\left(\nu+\frac{1}{2}-ip\right)~e^{i\pi\left(\nu+\frac{1}{2}\right)}}{4\sinh\pi p}\left(\begin{array}{ccc} \frac{1}{\displaystyle e^{\pi p}+e^{i\pi \left(\nu+\frac{1}{2}\right)}} &~~~ -\frac{1}{\displaystyle e^{\pi p}-e^{i\pi \left(\nu+\frac{1}{2}\right)}} \\  \frac{1}{\displaystyle e^{\pi p}+e^{i\pi \left(\nu+\frac{1}{2}\right)}} &~~~ \frac{1}{\displaystyle e^{\pi p}-e^{\pi p+i\pi \left(\nu+\frac{1}{2}\right)}}  \end{array}\right)~~~~~~~     \\
              \label{r3}  \overline{\gamma}_{j\sigma,n}&=& \displaystyle \frac{\Gamma\left(\nu+\frac{1}{2}+ip_n\right)~e^{\pi p_n+i\pi\left(\nu+\frac{1}{2}\right)}}{4\sinh\pi p_n}\left(\begin{array}{ccc} \frac{1}{\displaystyle e^{\pi p_n+i\pi \left(\nu+\frac{1}{2}\right)}+1} &~~~ \frac{1}{\displaystyle e^{\pi p_n+i\pi \left(\nu+\frac{1}{2}\right)}-1} \\ \frac{1}{\displaystyle e^{\pi p_n+i\pi \left(\nu+\frac{1}{2}\right)}+1} &~~~ -\frac{1}{\displaystyle e^{\pi p_n+i\pi \left(\nu+\frac{1}{2}\right)}-1}  \end{array}\right)~~ \\
                   \label{r4}       \overline{\delta}^{*}_{j\sigma,n}&=& \displaystyle \frac{\Gamma\left(\nu+\frac{1}{2}-ip_n\right)~e^{i\pi\left(\nu+\frac{1}{2}\right)}}{4\sinh\pi p_n}\left(\begin{array}{ccc} \frac{1}{\displaystyle e^{\pi p_n}+e^{i\pi \left(\nu+\frac{1}{2}\right)}} &~~~ -\frac{1}{\displaystyle e^{\pi p_n}-e^{i\pi \left(\nu+\frac{1}{2}\right)}} \\  \frac{1}{\displaystyle e^{\pi p_n}+e^{i\pi \left(\nu+\frac{1}{2}\right)}} &~~~ \frac{1}{\displaystyle e^{\pi p_n}-e^{\pi p_n+i\pi \left(\nu+\frac{1}{2}\right)}}  \end{array}\right)~~~~~~~   \eea 
                   Additionally,  we have following two constraints:
 \bea a^{(C)}_{I}\underbrace{\left[\sum^{\infty}_{n=0}\frac{1}{{\cal N}_{p,(n)}}\left({\cal  M}_{(n)}\right)^{I}_{J}{\cal P}^{J}_{(n)}\right]}_{\bf Particular~solution}= 0,~~~~~~~ 
 a^{(P)}_{I(n)}\underbrace{\left[\frac{1}{{\cal N}_p}{\cal  M}^{I}_{J}{\cal P}^{J}\right]}_{\bf Complementary~solution}=0.\eea
Then in terms of the previously mentioned matrix elements the annihilation and creation operators are explicitly defined as:
  \bea\label{vb1} {\bf Annihilation~operator:}~~a_{\sigma}&=&\sum_{q={\bf R},{\bf L}}\left\{\left[\gamma_{q\sigma}b_{q}+\delta^{*}_{q\sigma}b^{\dagger}_{q}\right]+\sum^{\infty}_{n=0}\left[\overline{\gamma}_{q\sigma,n}\overline{b}_{q,n}+\overline{\delta}^{*}_{q\sigma,n}\overline{b}^{\dagger}_{q,n}\right]\right\}\forall \sigma=\pm 1,\\ \label{vb2}  {\bf Creation~operator:}~~a^{\dagger}_{\sigma}&=&\sum_{q={\bf R},{\bf L}}\left\{\left[\gamma^{*}_{q\sigma}b^{\dagger}_{q}+\delta_{q\sigma}b_{q}\right]+\sum^{\infty}_{n=0}\left[\overline{\gamma}^{*}_{q\sigma,n}\overline{b}^{\dagger}_{q,n}+\overline{\delta}_{q\sigma,n}\overline{b}_{q,n}\right]\right\}\forall \sigma=\pm 1.~~~~~~~~~~\eea   
 Now the Bunch-Davies quantum vacuum state can be written in terms of the tensor product of ${\bf R}$ and ${\bf L}$ vacua by making use of the following Bogoliubov transformation:
  \bea|{\bf BD}\rangle &=&\exp\left(\widehat{\cal K}\right)~\bigg(|{\bf R}\rangle \otimes|{\bf L}\rangle\bigg)~~~~~~~{\rm for}~~~~~~~{\cal H}_{\bf BD}:={\cal H}_{\bf R}\otimes {\cal H}_{\bf L},\eea 
  where the new quantum Bogoliubov operator $\hat{\cal K}$ can be expressed as:
  \bea {\bf Bogoliubov~operator~I:}~~\widehat{\cal K}&=&\Bigg(\underbrace{\frac{1}{2}\sum_{i,j={\bf R},{\bf L}}m_{ij}~b^{\dagger}_{i}~b^{\dagger}_{j}}_{\bf Complementary~part}+\underbrace{\frac{1}{2}\sum_{i,j={\bf R},{\bf L}}\sum^{\infty}_{n=0}\overline{m}_{ij,n}~\overline{b}^{\dagger}_{i,n}~\overline{b}^{\dagger}_{j,n}}_{\bf Particular~integral~part}\Bigg),~~~~~~~\eea
  where our objective is to determine the coefficients $m_{ij}$ and $\bar{m}_{ij,n}$ in this work. Also the ${\bf R}$ and ${\bf L}$ vacuum states are defined as:
  \bea   {\bf Factorization~of~states:}~~|{\bf R}\rangle&=&\bigg(|{\bf R}\rangle_{(C)}+\sum^{\infty}_{n=0}|{\bf R}\rangle_{(P),n}\bigg),~~~  |{\bf L}\rangle=\bigg( |{\bf L}\rangle_{(C)}+\sum^{\infty}_{n=0}|{\bf L}\rangle_{(P),n}\bigg),\nonumber\\
  \eea 
which satisfy the following constraints:
      \bea b_{{\bf L}}|{\bf L}\rangle_{(C)}&=&0,b_{\bf R}|{\bf R}\rangle_{(C)}=0,\\
      \overline{b}_{{\bf L},n}|{\bf L}\rangle_{(P)}&=& 0, \overline{b}_{{\bf R},n}|{\bf R}\rangle_{(P)}=0.~~~~~~~~~~~\eea
     which satisfy the following commutation algebra:
      \bea 
      \left[ b_i,b^{\dagger}_j\right]&=&\delta_{ij},~~~~ \left[ b_i,b_j\right]=0=
      \left[ b^{\dagger}_i,b^{\dagger}_j\right].~~~~~~~~~~~~~~~\\ 
            \left[ \overline{b}_{i,n},\overline{b}^{\dagger}_{j,m}\right]&=&\delta_{ij}{\delta}_{nm},~~~~\left[ \overline{b}_{i,n},\overline{b}_{j,m}\right]= 0=
            \left[ \overline{b}^{\dagger}_{i,m},\overline{b}^{\dagger}_{j,m}\right].~~~~~~~~~~~
            \eea
            This implies the following relation:
   \bea \Bigg(\underbrace{\left(m_{ij}\gamma_{j\sigma}+\delta^{*}_{i\sigma}\right)b^{\dagger}_{i}}_{\bf Complementary~part}+\underbrace{\sum^{\infty}_{n=0}\left(\overline{m}_{ij,n}\overline{\gamma}_{j\sigma,n}+\overline{\delta}^{*}_{i\sigma,n}\right)\overline{b}^{\dagger}_{i,n}}_{\bf Particular~integral~part}\Bigg)\left(|{\bf R}\rangle\otimes |{\bf L}\rangle\right)&=& 0.\eea
    which further correspond to the following conditions:
         \bea \left(m_{ij}\gamma_{j\sigma}+\delta^{*}_{i\sigma}\right)&=& 0,\\
 \left(\bar{m}_{ij,n}\bar{\gamma}_{j\sigma,n}+\bar{\delta}^{*}_{i\sigma,n}\right)&=& 0~~~~\forall n,\eea
         using which we define the following mass matrices,  which are given by: 
           \bea  m_{ij}&=& -\delta^{*}_{i\sigma}\left(\gamma^{-1}\right)_{\sigma j}\equiv\left(\begin{array}{ccc} m_{\bf RR} &~~~ m_{\bf RL} \\ m_{\bf LR} &~~~ m_{\bf LL}  \end{array}\right)
\displaystyle\approx \frac{e^{i\theta}\sqrt{2}~e^{-p\pi}}{\sqrt{\cosh 2\pi p+\cosh 2\pi \nu}}\left(\begin{array}{ccc} \cos\pi\nu &~~~ i\sinh p\pi \\ i\sinh p\pi &~~~ \cos\pi\nu  \end{array}\right).                                                                        ~~~~~~~~~~~\eea
           \bea  \bar{m}_{ij,n}=-\overline{\delta}^{*}_{i\sigma,n}\left(\overline{\gamma}^{-1}\right)_{\sigma j,n}\equiv\left(\begin{array}{ccc} \overline{m}_{{\bf RR},n} &~~~ \overline{m}_{{\bf RL},n} \\ \overline{m}_{{\bf LR},n} &~~~ \overline{m}_{{\bf LL},n}  \end{array}\right)\displaystyle\approx \frac{e^{i\theta}\sqrt{2}~e^{-p_n\pi}}{\sqrt{\cosh 2\pi p_n+\cosh 2\pi \nu}}\left(\begin{array}{ccc} \cos\pi\nu &~~~ i\sinh p_n\pi \\ i\sinh p_n\pi &~~~ \cos\pi\nu  \end{array}\right).                                                                      \nonumber\\
       \eea 
  The eigenvalues of the solutions are given by:
\bea
  \lambda_{\pm}&=&\bigg(m_{\bf RR}\pm m_{\bf RL}\bigg)=e^{i\theta}\frac{\sqrt{2}~e^{-p\pi}\left(\cos\pi\nu \pm i\sinh p\pi\right)}{\sqrt{\cosh 2\pi p+\cos 2\pi \nu}},\\
  \lambda_{\pm,n}&=& \bigg(\bar{m}_{{\bf RR},n}\pm \bar{m}_{{\bf RL},n}\bigg)=e^{i\theta}\frac{\sqrt{2}~e^{-p_n\pi}\left(\cos\pi\nu \pm i\sinh p_n\pi\right)}{\sqrt{\cosh 2\pi p_n+\cos 2\pi \nu}}.\eea
  But this extremely complicated to take the partial trace operation from the contributions obtained from ${\bf R}$ and ${\bf L}$.  For this reason it is unsuitable basis for our calculation.

 To perform the above mentioned operation we need to perform another Bogoliubov transformation in terms of the following suitable basis,  where the new quantum operators are defined as:
 \bea 
   c_{\bf R}&=& \bigg(u~b_{\bf R}+v~b^{\dagger}_{\bf R}\bigg),~~~C_{{\bf R},n}= \bigg(U_n~b_{{\bf R},n}+V_n~b^{\dagger}_{{\bf R},n}\bigg).\\
c_{\bf L}&=&\bigg( \overline{u}~b_{\bf L}+\overline{v}~b^{\dagger}_{\bf L}\bigg),~~~C_{{\bf L},n}= \bigg(\overline{U}_n~b_{{\bf L},n}+\overline{V}_n~b^{\dagger}_{{\bf L},n}\bigg),\eea   
   which satisfy the following normalization constraints on the mode function:
        \bea 
          \bigg( |u|^2-|v|^2\bigg)&=& 1,~~~\bigg(|U_n|^2-|V_n|^2\bigg)= 1.\\  
                     \bigg(|\bar{u}|^2-|\bar{v}|^2\bigg)&=& 1 ,~~~\bigg(|\bar{U}_n|^2-|\bar{V}_n|^2\bigg)= 1.\eea  
                      Using this information Bunch-Davies vacuum state can be expressed in the newly Bogoliubov transformed basis as:
 \bea |{\bf BD}\rangle &=&\frac{1}{{\cal N}_{p}}\exp\left(\widehat{\cal W}\right)\left(|{\bf R}^{'}\rangle\otimes |{\bf L}^{'}\rangle\right)\quad\quad {\rm where}\quad {\cal N}_{p}\approx\frac{1}{\displaystyle \sqrt{\left[1-\left(|\gamma_p|^2+\sum^{\infty}_{n=0}|\Gamma_{p,n}|^2\right)\right]}},\quad\quad\quad\eea  
 where $|{\bf R}^{'}\rangle$ and $ |{\bf L}^{'}\rangle$ are new basis operators in ${\cal H}_{\bf BD}:={\cal H}_{\bf R}\otimes {\cal H}_{\bf L}$.  Additionally,  we introduce a new quantum operator $\widehat{\cal W}$,  which is defined as:
   \bea  {\bf Bogoliubov~operator~II:}~~\widehat{\cal W}&=&\bigg(\underbrace{\gamma_{p}~c^{\dagger}_{\bf R}~c^{\dagger}_{\bf L}}_{\bf Complementary~part}+\underbrace{\sum^{\infty}_{n=0}\Gamma_{p,n}~C^{\dagger}_{{\bf R},n}~C^{\dagger}_{{\bf L},n}}_{\bf Particular~integral~part}\bigg),\quad\eea
     where our prime objective is to determine $\gamma_{p}$ and $\Gamma_{p,n}$ from the present calculation.
     
    The new oscillator algebra is given by: 
           \bea 
           \left[ c_i,c^{\dagger}_j\right]&=&\delta_{ij},~~~~ \left[ c_i,c_j\right]=0=
           \left[ c^{\dagger}_i,c^{\dagger}_j\right].~~~~~~~~~~~~~~~\\
                 \left[ C_{i,n},C^{\dagger}_{j,m}\right]&=&\delta_{ij}{\delta}_{nm},~~~~\left[ C_{i,n},C_{j,m}\right]= 0=
                 \left[ C^{\dagger}_{i,m},C^{\dagger}_{j,m}\right].~~~~~~~~~~~~~~~
                 \eea
   Here the new operators are defined as:
      \bea  c_{\bf R}|{\bf BD}\rangle &=&\gamma_{p}~c^{\dagger}_{\bf L}|{\bf BD}\rangle,~~~
     c_{\bf L}|{\bf BD}\rangle =\gamma_{p}~c^{\dagger}_{\bf R}|{\bf BD}\rangle,\\ 
   C_{{\bf R},n}|{\bf BD}\rangle &=&\Gamma_{p,n}~C^{\dagger}_{{\bf L},n}|{\bf BD}\rangle,~~~
          C_{{\bf L},n}|{\bf BD}\rangle =\Gamma_{p,n}~C^{\dagger}_{{\bf R},n}|{\bf BD}\rangle.\eea
   In the new basis we have the following expressions:
            \bea c_{J}&=& b_{I}{\cal G}^{I}_{J},~
            C_{J(n)}=\bar{b}_{J(n)}\left({\cal G}_{(n)}\right)^{I}_{J}~{\rm where}~~{\cal G}^{I}_{J}=\left(\begin{array}{ccc} U_q &~~~ V^{*}_q \\ V_q &~~~ U^{*}_q  \end{array}\right)
                    ,~
                    \left({\cal G}_{(n)}\right)^{I}_{J}=\left(\begin{array}{ccc} \overline{U}_{ q,n} &~~~ \overline{V}^{*}_{\sigma q,n} \\ \overline{V}_{ q,n} &~~~ \overline{U}^{*}_{ q,n}  \end{array}\right),\quad\quad\quad\quad\eea
where the components of the new matrices are given by:
                    \bea U_q &\equiv& {\rm \bf diag}\left(u,\overline{u}\right),~~V_q \equiv {\rm \bf diag}\left(v,\overline{v}\right),~~ \overline{U}_{q,n} \equiv {\rm \bf diag}\left(U_n,\overline{U}_n\right),~~\overline{V}_{q,n} \equiv {\rm \bf diag}\left(V_n,\overline{V}_n\right).~~\quad\quad\quad\eea
           Finally we derive the following sets of homogeneous algebraic equations: 
  \bea
  m_{\bf RR}u+v-\gamma_{p} m_{\bf RL}\overline{v}^{*}&=& 0,\\
   m_{\bf RR}\overline{u}+\overline{v}-\gamma_{p} m_{\bf RL}v^{*}&=& 0,\\ 
   m_{\bf RL}u-\gamma_{p} \overline{u}^{*}-\gamma_{p}m_{\bf RR}\overline{v}^{*}&=& 0,\\
   m_{\bf RL}\overline{u}-\gamma_{p} u^{*}-\gamma_{p}m_{\bf RR}v^{*}&=& 0,\\
    \bar{m}_{{\bf RR},n}U_n+V_n-\Gamma_{p,n}\overline{m}_{{\bf RL},n}\overline{V}^{*}_n&=& 0,\\
      \overline{m}_{{\bf RR},n}\overline{U}_n+\overline{V}_n-\Gamma_{p,n} \overline{m}_{{\bf RL},n}V^{*}_n&=& 0,\\ 
       \overline{m}_{{\bf RL},n}U_n-\Gamma_{p,n} \overline{U}^{*}_n-\Gamma_{p,n} \overline{m}_{{\bf RR},n}\overline{V}^{*}_n&=& 0,\\
       \overline{m}_{{\bf RL},n}\overline{U}_n-\Gamma_{p,n}  U^{*}_n-\Gamma_{p,n} \overline{m}_{{\bf RR},n}V^{*}_n&=& 0,~~~~~~~~~~\eea
       Here we have the following properties of the above mentioned equations:
       \begin{enumerate}
       \item {\bf \underline{Property I:}}  
  \bea m_{\bf RR}&=& m_{\bf LL}=m^{*}_{\bf RR}=\omega=\frac{\sqrt{2}~e^{-p\pi}\cos\pi\nu}{\sqrt{\cosh 2\pi p+\cos 2\pi \nu}},\\
       m_{\bf RL}&=& m_{\bf LR}=-m^{*}_{\bf RL}=\zeta=e^{i\frac{\pi}{2}}\frac{\sqrt{2}~e^{-p\pi}\sinh p\pi}{\sqrt{\cosh 2\pi p+\cos 2\pi \nu}},\eea
       
       \item {\bf \underline{Property II:}}  
                      \bea \bar{m}_{{\bf RR},n}&=& \bar{m}_{{\bf LL},n}=\bar{m}^{*}_{{\bf RR},n}=\omega_n=\frac{\sqrt{2}~e^{-p_n\pi}\cos\pi\nu}{\sqrt{\cosh 2\pi p_n+\cos 2\pi \nu}},\\
                     \bar{m}_{{\bf RL},n}&=& \bar{m}_{{\bf LR},n}=-\bar{m}^{*}_{{\bf RL},n}=\zeta_n=e^{i\frac{\pi}{2}}\frac{\sqrt{2}~e^{-p_n\pi}\sinh p_n\pi}{\sqrt{\cosh 2\pi p_n+\cos 2\pi \nu}}.\eea
 \item  {\bf \underline{Property III:}} \\
  If we have the following two conditions:
\bea \gamma^{*}_{p}=-\gamma_{p},~~~
   \Gamma^{*}_{p,n}=-\Gamma_{p,n},\eea 
then we can fix the following constraint:
\bea v^{*}=\overline{v},~~
   u^{*}=\overline{u},~~ V^{*}_n=\overline{V}_n,~~ 
      U^{*}_n=\overline{U}_n. \eea
      As a result,  we have two normalization conditions instead of two,  which are given by:  
\bea  \bigg(|u|^2-|v|^2\bigg)=1, \quad\quad \bigg(|U_n|^2-|V_n|^2\bigg)=1.\eea 
       \end{enumerate}  
         Finally,   we have the following expressions:
     \bea
         \gamma_{p}&=&\frac{1}{2m_{\bf RL}}\left[\left(1+m^2_{\bf RL}-m^2_{\bf RR}\right) 
      - \sqrt{\left(1+m^2_{\bf RL}-m^2_{\bf RR}\right)^2-4m^2_{\bf RL}}\right]\nonumber\\
         &=&i\frac{\sqrt{2}}{\sqrt{\cosh 2\pi p +\cos 2\pi \nu}+\sqrt{\cosh 2\pi p +\cos 2\pi \nu+2}}.~~~~~~~~~~~\\
          \Gamma_{p,n}&=&\frac{1}{2\bar{m}_{{\bf RL},n}}\left[\left(1+\bar{m}^2_{{\bf RL},n}-\bar{m}^2_{{\bf RR},n}\right)
        - \sqrt{\left(1+\bar{m}^2_{{\bf RL},n}-\bar{m}^2_{{\bf RR},n}\right)^2-4\bar{m}^2_{{\bf RL},n}}\right]\nonumber\\
          &=&i\frac{\sqrt{2}}{\sqrt{\cosh 2\pi p_n +\cos 2\pi \nu}+ \sqrt{\cosh 2\pi p_n +\cos 2\pi \nu+2}}.~~~~~~~~~~\eea   
          Here it is important to note that we have taken the negative signature in front of the square root contribution to strictly satisfy the constraint,  $|\gamma_p|<1$ and  $|\Gamma_{p,n}|<1$.  For the other signature,  which is appearing from other branch of solution these mentioned constraints are not satisfied at all and for this reason the other solutions are not physical in the present context of discussion.  Also the above equations satisfy the following normalization conditions:
 \bea  \bigg(|\overline{u}|^2-|\overline{v}|^2\bigg)=1, \quad\quad \bigg(|\overline{U}_n|^2-|\overline{V}_n|^2\bigg)=1.\eea 
 general solutions of these equations can be expressed as:
\bea \overline{u}&=&\frac{1-\gamma_p \zeta}{\sqrt{|1-\gamma_p \zeta|^{2}-|\omega|^{2}}}=u^{*}=u,\quad\quad\overline{v}=\frac{\omega}{\sqrt{|1-\gamma_p \zeta|^{2}-|\omega|^{2}}}=v^{*}=v,\\
 \overline{U}_n&=&\frac{1-\Gamma_{p_n} \zeta_n}{\sqrt{|1-\Gamma_{p_n} \zeta_n|^{2}-|\omega|^{2}}}={U}^{*}_n={U}_n\quad\quad\overline{V}_n=\frac{\omega_n}{\sqrt{|1-\Gamma_{p_n} \zeta_n|^{2}-|\omega|^{2}}}={V}^{*}_n={V}_n.\quad\quad\quad\eea
  Here we have two additional constraints which are satisfied during this computation:
  \bea \omega^{*}=\omega,\quad\quad \zeta^{*}=-\zeta,\quad\quad \gamma^{*}_p=-\gamma_p,\quad\quad\Gamma^{*}_{p,n}=-\Gamma_{p,n}.\eea
  These derived expressions will be used further to derive the rest of the results in this paper.

 \subsection{\textcolor{black}{Construction of reduced density matrix in open chart}}
    \label{ka2d}
    
   In this portion our job is construct the expression for reduced density operator corresponding to the previously defined Bunch Davies state.  After quantization the corresponding quantum state
   is characterized in terms of the three important quantum  numbers $p, l$ and $m$.  As we have already mentioned before that the contributions coming from region {\bf R} and region {\bf L} are exactly same and symmetric in nature.  Because of this reason we are going to derive the expression for the reduced density operator in the region {\bf L} by taking partial trace over all the contributions coming from region {\bf R}.  This leads to the following result for the corresponding density operator:
         \bea \rho_{{\bf L};p,l,m}&=&{\bf \rm Tr}_{\bf R}\bigg(|{\bf BD}\rangle \langle {\bf BD}|\bigg)\nonumber\\ 
         &=&
      \Bigg\{ \frac{\left(1-|\gamma_{p}|^2\right)}{\left(1+f_p\right)}\sum^{\infty}_{k=0}|\gamma_{p}|^{2k}|k;p,l,m\rangle_{{\bf L}^{'}}{}_{{\bf L}^{'}}\langle k;p,l,m|\nonumber\\
        &&\quad\quad\quad\quad\quad\quad\quad\quad+\frac{f^{2}_{p}}{\left(1+f_p\right)}\sum^{\infty}_{n=0}\sum^{\infty}_{r=0}|\Gamma_{p,n}|^{2r}|n,r;p,l,m\rangle_{{\bf L}^{'}}{}_{{\bf L}^{'}}\langle n,r;p,l,m|\Bigg\},~~~~~~~~~\eea
      where $\gamma_{p}$ and $\Gamma_{p,n}$ we have computed in the previous subsection and the new normalization factor ${\it f}_{p}$ is defined as:
      \bea {\it f}_{p}&=&\frac{1}{\displaystyle\left(\sum^{\infty}_{n=0}\frac{1}{1-|\Gamma_{p,n}|^2}\right)}.\eea 
In the definition of the density matrix the quantum mechanical states $|k;p,l,m\rangle_{{\bf L}^{'}}$ and $|n,r;p,l,m\rangle_{{\bf L}^{'}}$ can be further written in terms of creation operators in the new basis $|{\bf L}^{'}\rangle$ as:
      \bea |k;p,l,m\rangle_{{\bf L}^{'}}&=& \frac{1}{\sqrt{k!}}(c^{\dagger}_{\bf L})^{k}|{\bf L}^{'}\rangle,\\
       |n,r;p,l,m\rangle_{{\bf L}^{'}}&=& \frac{1}{\sqrt{r!}}(C^{\dagger}_{{\bf L},n})^{r}|{\bf L}^{'}\rangle.\eea
     In the new representative basis the reduced density operator takes the following diagonal form:
       \bea \rho_{\bf L}&=&\Bigg\{\underbrace{\left(1-|\gamma_{p}|^2\right){\bf diag}\Bigg(1,|\gamma_{p}|^2,|\gamma_{p}|^{4},|\gamma_{p}|^{6}\cdots\Bigg)}_{\bf Complementary~part}\nonumber\\
       &&\quad\quad\quad\quad\quad\quad +\underbrace{f^{2}_{p}\sum^{\infty}_{n=0}{\bf diag}\Bigg(1,|\Gamma_{p,n}|^2,|\Gamma_{p,n}|^{4},|\Gamma_{p,n}|^{6}\cdots\Bigg)}_{\bf Particular~integral~part}\Bigg\},~~~~~~~~~\eea
   Here it is important to note that:
              \bea \frac{1}{\left(1+f_p\right)}{\bf Tr}\left[\left(1-|\gamma_{p}|^2\right){\bf diag}\left(1,|\gamma_{p}|^2,|\gamma_{p}|^{4},|\gamma_{p}|^{6}\cdots\right)\right]&=&\frac{\left(1-|\gamma_{p}|^2\right)}{\left(1+f_p\right)}\sum^{\infty}_{k=0}|\gamma_{p}|^{2k}=\frac{1}{\left(1+f_p\right)},~~~~~~~~~~~~~~\\   \frac{f^2_p}{\left(1+f_p\right)}{\bf Tr}\left[\sum^{\infty}_{n=0}{\bf diag}\left(1,|\Gamma_{p,n}|^2,|\Gamma_{p,n}|^{4},|\Gamma_{p,n}|^{6}\cdots\right)\right]&=& \frac{f^2_p}{\left(1+f_p\right)}\sum^{\infty}_{n=0}\sum^{\infty}_{r=0}|\Gamma_{p,n}|^{2r}= \frac{f_p}{\left(1+f_p\right)},~~~~~\eea
    where we have used two following facts by assuming $\gamma_p\ll 1$ and $\Gamma_{p,n}\ll 1$:
    \bea \sum^{\infty}_{k=0}|\gamma_{p}|^{2k}&=&\frac{1}{\left(1-|\gamma_{p}|^2\right)},\\
    \sum^{\infty}_{n=0}\sum^{\infty}_{r=0}|\Gamma_{p,n}|^{2r}&=&\displaystyle\left(\sum^{\infty}_{n=0}\frac{1}{1-|\Gamma_{p,n}|^2}\right)=\frac{1}{f_p}.\eea 
    As a result finally we have:
    \bea {\bf Tr}\left(\rho_{\bf L}\right)=\Bigg\{\frac{1}{\left(1+f_p\right)}+\frac{f_p}{\left(1+f_p\right)}\Bigg\}=\frac{\left(1+f_p\right)}{\left(1+f_p\right)}=1,\eea    
It suggests that the reduced density operator utilised in this paper has the proper normalisation in its structure. The remainder of the computation carried out in this study will benefit greatly from the current derived structure of the reduced density operator for the region {\bf L}.

     \section{Entanglement negativity and logarithmic negativity in open chart}\label{ka3}              

Our main goal in this part is to formulate equations for the entanglement negativity and the logarithmic negativity between the region of {\bf R} and {\bf L} in the open chart of global de Sitter space time. Both of the aforementioned sections will be treated as causally unrelated during this computation.

The Bunch Davies quantum vacuum state can be factored according to the contributions made by the complementary and particular integral parts of the solution in terms of the quantum numbers $p$, $l$, and $m$ as follows:
\bea |{\bf BD}\rangle &=&\Bigg\{\sqrt{\frac{\left(1-|\gamma_p|^2\right)}{\left(1+f_p\right)}}\sum^{\infty}_{k=0}|\gamma_{p}|^{k}\bigg(|k;p,l,m\rangle_{{\bf R}^{'}}\otimes|k;p,l,m\rangle_{{\bf L}^{'}}\bigg)\nonumber\\
        &&\quad\quad\quad\quad\quad\quad+\frac{f_{p}}{\sqrt{\left(1+f_p\right)}}\sum^{\infty}_{n=0}\sum^{\infty}_{r=0}|\Gamma_{p,n}|^{r}\bigg(|n,r;p,l,m\rangle_{{\bf R}^{'}}\otimes| n,r;p,l,m\rangle_{{\bf L}^{'}}\bigg)\Bigg\},~~~~~~~~~\quad\eea  
     One can directly compute the expression for the eigenvalues, which is given by the following expression, by further employing the fundamental physical idea of Schmidt decomposition for a pure quantum state as mentioned in the earlier section of this paper:
                    \bea \sqrt{\lambda_k}&=&\Bigg\{\sqrt{\frac{\left(1-|\gamma_p|^2\right)}{\left(1+f_p\right)}}|\gamma_{p}|^{k}+\frac{f_{p}}{\sqrt{\left(1+f_p\right)}}\sum^{\infty}_{n=0}|\Gamma_{p,n}|^{k}\Bigg\}\quad\quad\forall k=[0,\infty].\eea
 Then the logarithmic negativity from the present theoretical set up can be computed as:
 \bea {\cal L}{\cal N}(p,\nu)&=&2\ln\Bigg(\sum^{\infty}_{k=0}\lambda_k\Bigg)\nonumber\\
 &=&2\ln\Bigg(\sum^{\infty}_{k=0}\Bigg\{\sqrt{\frac{\left(1-|\gamma_p|^2\right)}{\left(1+f_p\right)}}|\gamma_{p}|^{k}+\frac{f_{p}}{\sqrt{\left(1+f_p\right)}}\sum^{\infty}_{n=0}|\Gamma_{p,n}|^{k}\Bigg\}\Bigg)\nonumber\\
 &=&2\ln\Bigg(\sqrt{\frac{\left(1-|\gamma_p|^2\right)}{\left(1+f_p\right)}}\sum^{\infty}_{k=0}|\gamma_{p}|^{k}+\frac{f_{p}}{\sqrt{\left(1+f_p\right)}}\sum^{\infty}_{n=0}\sum^{\infty}_{k=0}|\Gamma_{p,n}|^{k}\Bigg)\nonumber\\
 &=&\ln\Bigg(\frac{1}{\left(1+f_p\right)}\Bigg\{\sqrt{\frac{\left(1+|\gamma_p|\right)}{\left(1-|\gamma_p|\right)}}+\frac{f_p}{\overline{f}_p}\Bigg\}^{2}\Bigg).\eea                    
        where we introduce a new notation $\overline{f}_p$,  which is defined as:
        \bea \overline{f}_p&=&\frac{1}{\left(\displaystyle \sum^{\infty}_{n=0}\frac{1}{1-|\Gamma_{p,n}|}\right)}\eea                                    
In the last step we have used the following results to compute the summations:
\bea \sum^{\infty}_{k=0}|\gamma_{p}|^{k}&=&\frac{1}{\left(1-|\gamma_p|\right)},\\
\sum^{\infty}_{k=0}|\Gamma_{p,n}|^{k}&=&\frac{1}{\left(1-|\Gamma_{p,n}|\right)}.\eea
Hence,  the entanglement negativity can be further computed in terms of the expression derived for the logarithmic negativity as:
\bea {\cal N}(p,\nu)&=&\frac{1}{2}\Bigg(\exp\left({\cal L}{\cal N}(p,\nu)\right)-1\Bigg)\nonumber\\
&=&\frac{1}{2}\Bigg(\frac{1}{\left(1+f_p\right)}\Bigg\{\sqrt{\frac{\left(1+|\gamma_p|\right)}{\left(1-|\gamma_p|\right)}}+\frac{f_p}{\overline{f}_p}\Bigg\}^{2}-1\Bigg)\eea
  Now one would anticipate that the two causally unrelated areas,  {\bf R} and {\bf L},  are quantum mechanically entangled with one another under the current framework for any finite values of $p$,  which is basically the direct outcome of having non vanishing contribution from both $|\gamma_{p}|$ and $|\Gamma_{p,n}|$. 
  
Now after integrating over $p$ in presence of appropriate contribution from the density of quantum mechanical states under consideration in open chart we finally obtain the following expression for the logarithmic negativity in the volume of a hyperboloid:
  \bea\label{eeq1}  {\cal L}{\cal N}(\nu)&=&V^{\bf reg}_{\bf H^3}\int^{\infty}_{0}~dp~{\cal D}(p)~ {\cal L}{\cal N}(p,\nu),\eea   
  where the quantity ${\cal D}(p)$ represents the density of quantum states corresponding to the radial contribution on the hyperboloid ${\bf H^3}$,  which is given by:
  \bea {\cal D}(p)=\frac{1}{2\pi^2}p^2.\eea
    \begin{figure*}[htb]
    \centering
    \subfigure[For $f_p=0$.]{
        \includegraphics[width=14.2cm,height=8.5cm] {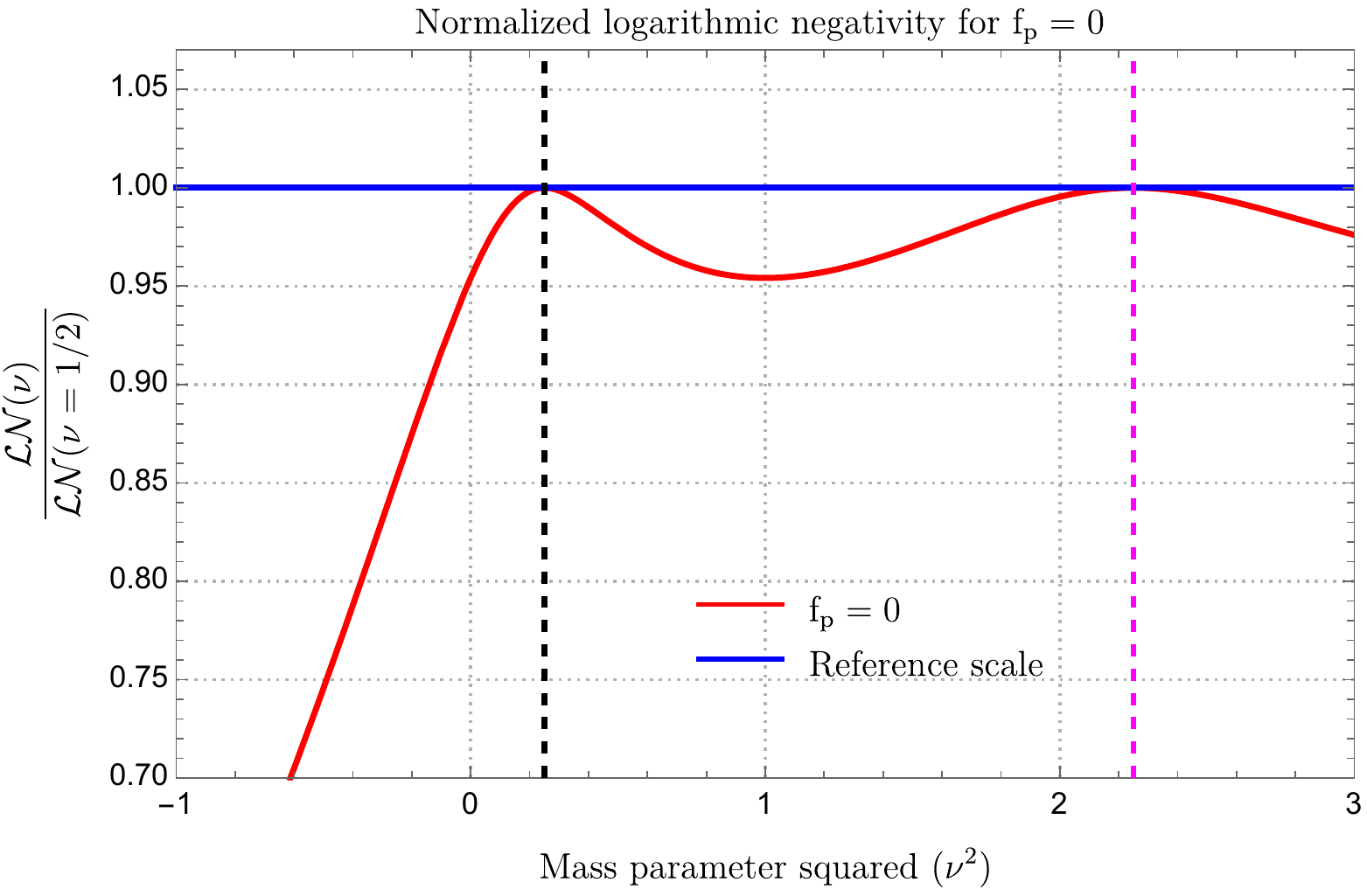}
        \label{L1}
    }
    \subfigure[For small $f_p\neq 0$.]{
        \includegraphics[width=14.2cm,height=8.5cm] {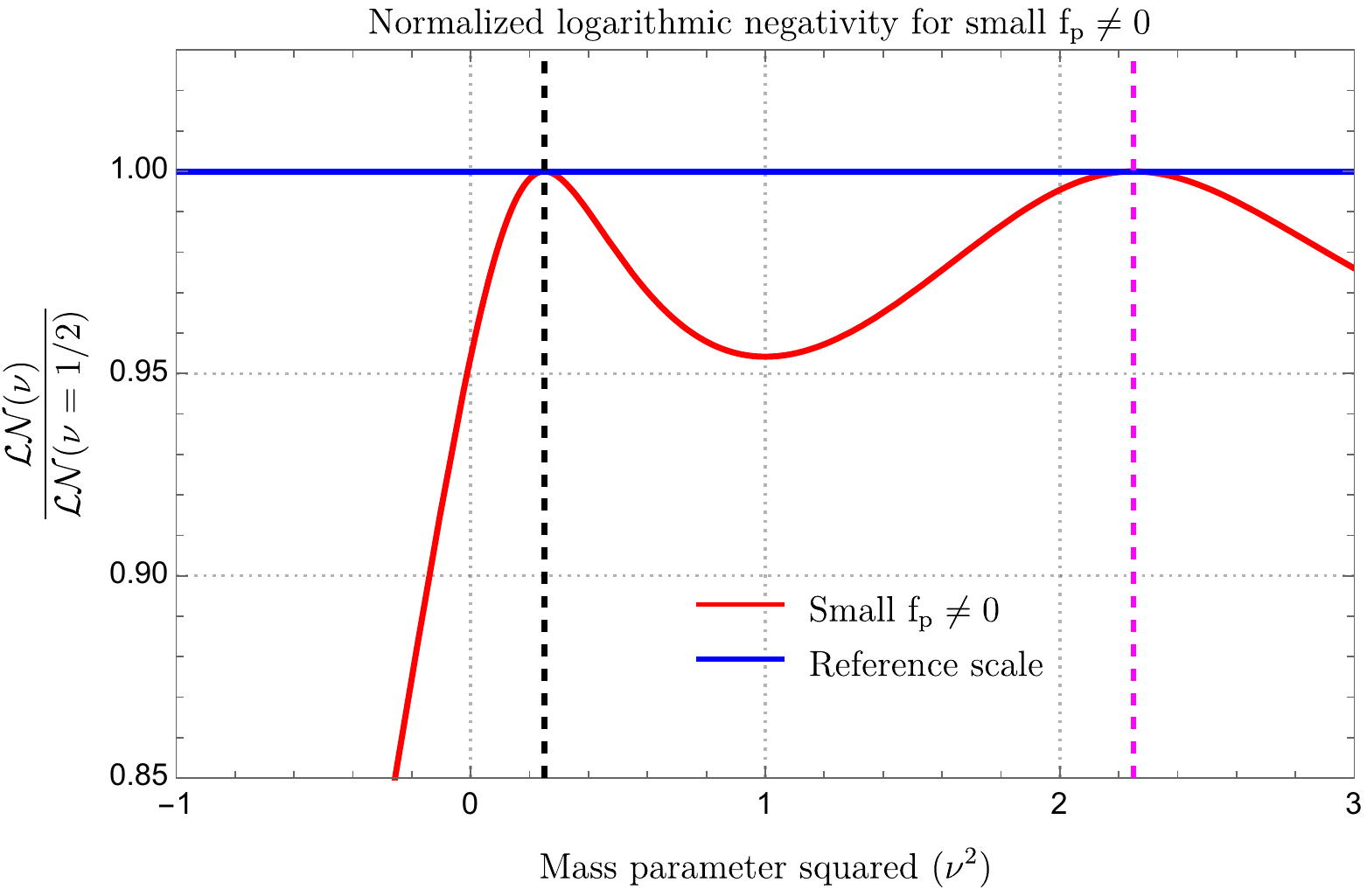}
        \label{L2}
       }
    \caption[Optional caption for list of figures]{Graphical behaviour of the normalized logarithmic negativity (${\cal LN}(\nu)/{\cal LN}(\nu=1/2)$) with mass parameter squared ($\nu^2$) for both $f_p=0$ and small $f_p\neq 0$.  Vertical dashed lines are drawn for $\nu=1/2$ (conformally coupled) and $\nu=3/2$ (massless case).  In both the plots reference scale corresponds to the scale at which ${\cal LN}(\nu)/{\cal LN}(\nu=1/2)=1$.} 
    \label{LN}
    \end{figure*}  
    \begin{figure*}[htb]
    \centering
    \subfigure[For $f_p=0$.]{
        \includegraphics[width=14.2cm,height=8.5cm] {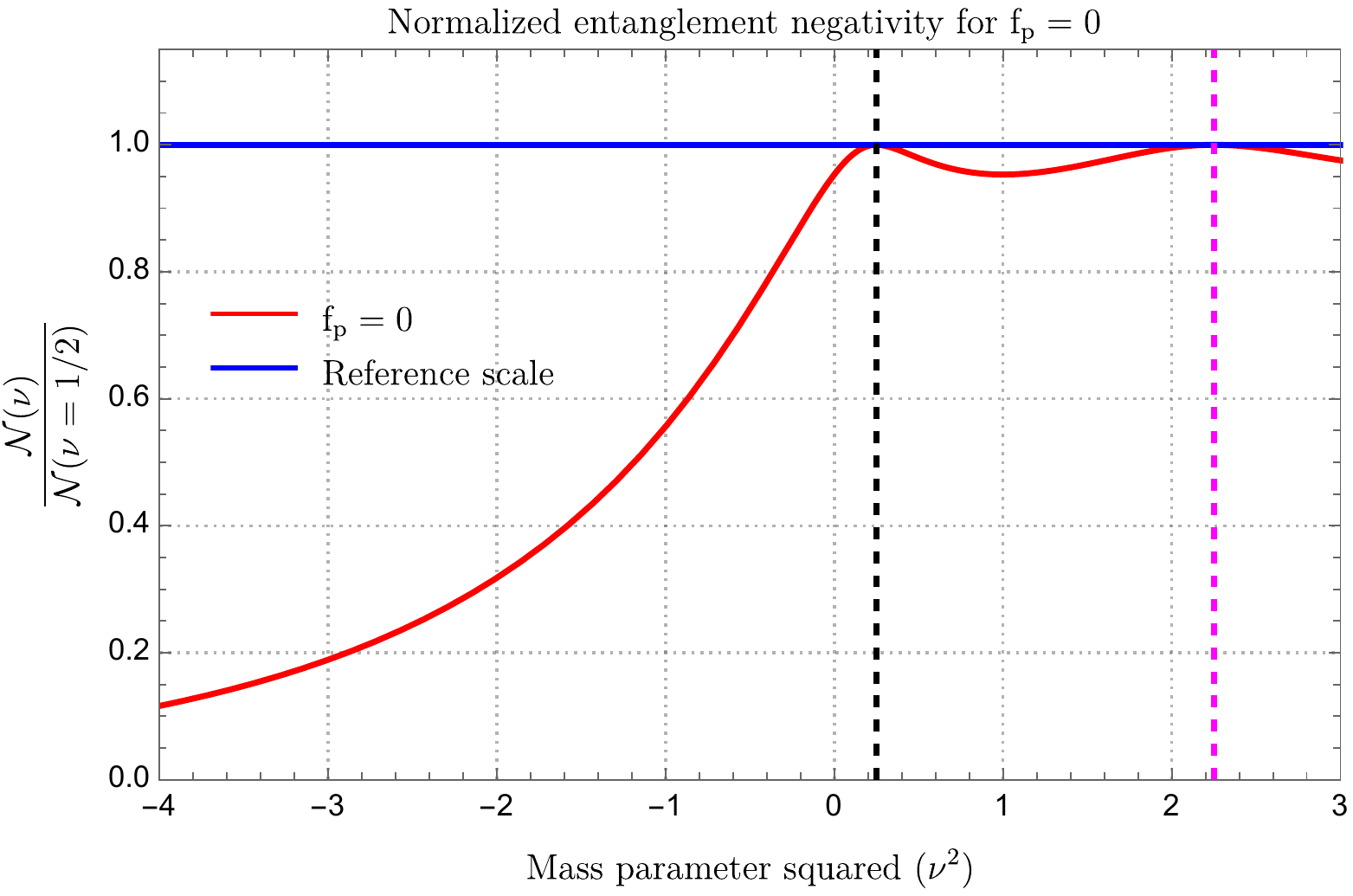}
        \label{N1}
    }
    \subfigure[For small $f_p\neq 0$.]{
        \includegraphics[width=14.2cm,height=8.5cm] {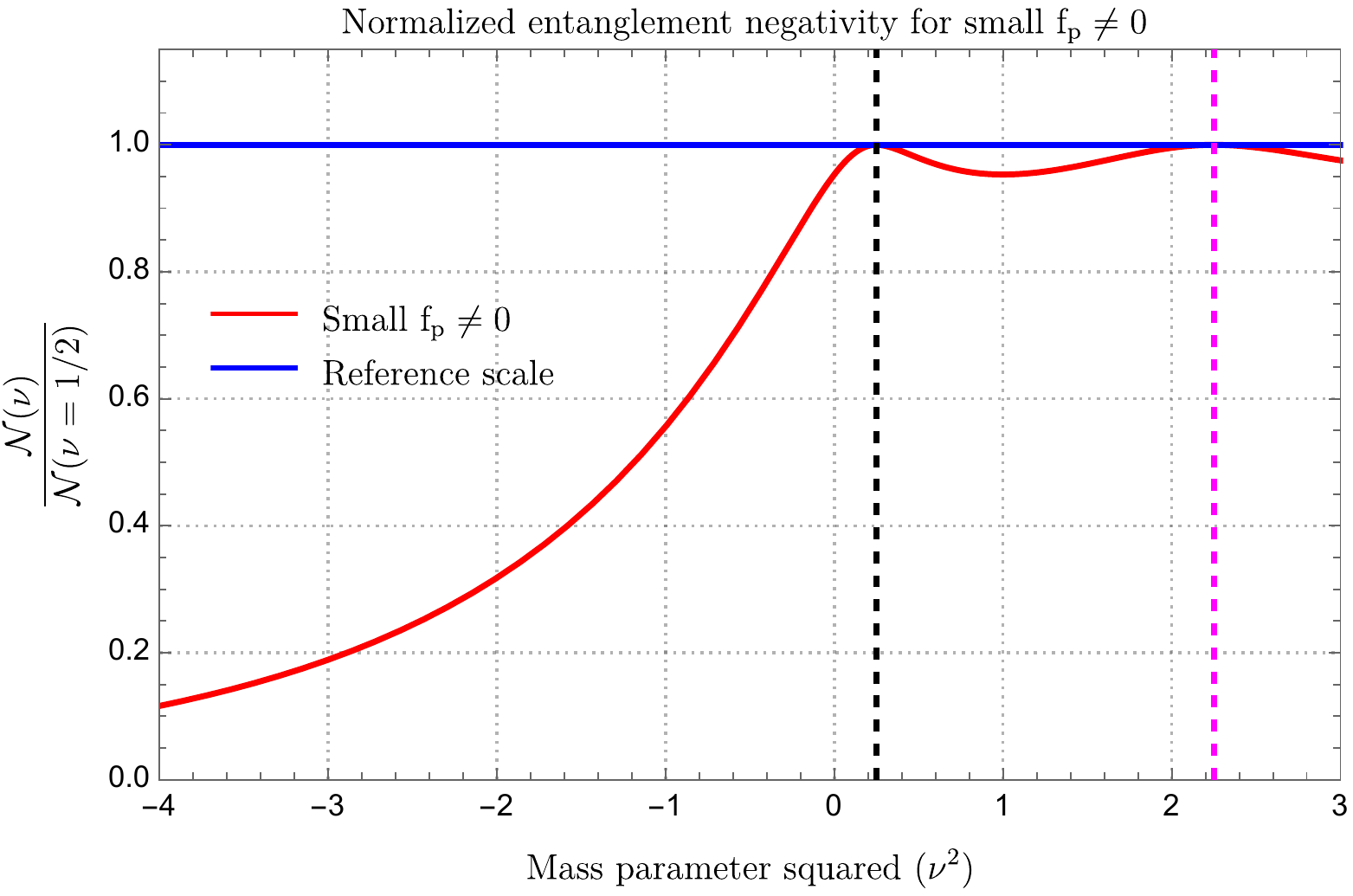}
        \label{N2}
       }
    \caption[Optional caption for list of figures]{Graphical behaviour of the normalized entanglement negativity (${\cal N}(\nu)/{\cal N}(\nu=1/2)$) with mass parameter squared ($\nu^2$) for both $f_p=0$ and small $f_p\neq 0$.  Vertical dashed lines are drawn for $\nu=1/2$ (conformally coupled) and $\nu=3/2$ (massless case).  In both the plots reference scale corresponds to the scale at which ${\cal N}(\nu)/{\cal N}(\nu=1/2)=1$.} 
    \label{N}
    \end{figure*}  
  Additionally,  in the above mentioned expression the regularized finite part of the volume of the hyperboloid ${\bf H^3}$ is given by:
  \bea V^{\bf reg}_{\bf H^3}&=&\frac{1}{2}V_{\bf S^2}=\frac{1}{2}\times 4\pi=2\pi.\eea
  Consequently,  equation (\ref{eeq1}) can be further expressed in terms of the following simplified form:     
  \bea\label{eexq1}  {\cal L}{\cal N}(\nu)&=&\frac{1}{\pi}\int^{\infty}_{0}~dp~p^2~ {\cal L}{\cal N}(p,\nu).\eea
  However,  in most of the physical problem in the upper limit of the above mentioned integration the integrand becomes divergent.  For this reason,  one need to introduce a regulator $\Lambda$ is the upper limit of the integration instead of strictly putting this to be infinity.  However,   for the computational purpose we fix the value of $\Lambda$ to be very large number.  In this specific problem this cut-off is physically treated to be the Ultra Violet (UV) cut-off.  In quantum field theoretic prescription sometimes this UV cut-off is physically interpreted as the manifestation of lattice regulator for the type of computation we are performing in this paper.  On the other hand,  it is important to note that,  in the lower limit of integration the integrand becomes convergent in most of the interesting physical situation. In the technical language this lower limit corresponds to the Infra Red (IR) which is safe for the particular problem we are doing in this paper.  Considering all of these facts stated above one can further recast equation (\ref{eexq1}) to quantify the regularized version of the logarithmic negativity in the following modified format:
      \bea\label{eexq2}  {\cal L}{\cal N}(\nu)&=&\frac{1}{\pi}\int^{\Lambda}_{0}~dp~p^2~ {\cal L}{\cal N}(p,\nu)\nonumber\\
      &=&\frac{1}{\pi}\int^{\Lambda}_{0}~dp~p^2~\ln\Bigg(\frac{1}{\left(1+f_p\right)}\Bigg\{\sqrt{\frac{\left(1+|\gamma_p|\right)}{\left(1-|\gamma_p|\right)}}+\frac{f_p}{\overline{f}_p}\Bigg\}^{2}\Bigg).\eea
      Also,   the regularized version of the entanglement negativity can be further computed using equation (\ref{eexq2}) as:
      \bea {\cal N}(\nu)&=&\frac{1}{2}\Bigg(\exp\left({\cal L}{\cal N}(\nu)\right)-1\Bigg)\nonumber\\
&=&\frac{1}{2}\Bigg[\exp\Bigg(\frac{1}{\pi}\int^{\Lambda}_{0}~dp~p^2~\ln\Bigg(\frac{1}{\left(1+f_p\right)}\Bigg\{\sqrt{\frac{\left(1+|\gamma_p|\right)}{\left(1-|\gamma_p|\right)}}+\frac{f_p}{\overline{f}_p}\Bigg\}^{2}\Bigg)\Bigg)-1\Bigg].\quad\eea
  Further,  if we substitute each of the components of the above equation from what we have derived in the previous section,  then one can clearly see from the complicated structure of this equation that the final result is not analytically computable.  For this specific reason we have analysed the above expression for the $f_p=0$ and small $f_p\neq 0$ numerically in this paper.  It basically covers both {\bf Case A} and {\bf Case B} solutions in the present context.

In figure (\ref{LN}(a)) and figure (\ref{LN}(b)),   we have explicitly depicted the behaviour of normalized logarithmic negativity (${\cal LN}(\nu)/{\cal LN}(\nu=1/2)$) with mass parameter squared ($\nu^2$) for both $f_p=0$ and small $f_p\neq 0$.  Similarly,  in figure (\ref{N}(a)) and figure (\ref{N}(b)),   we have explicitly plotted the behaviour of normalized entanglement negativity (${\cal N}(\nu)/{\cal N}(\nu=1/2)$) with mass parameter squared ($\nu^2$) for both $f_p=0$ and small $f_p\neq 0$.  Vertical dashed lines are drawn for $\nu=1/2$ and $\nu=3/2$ ({\bf Case A}) cases.  The outcomes and the physical interpretation of these plots are very interesting which we are writing point-wise in the following:
\begin{enumerate}
\item From both of these plots is is clearly observed that the normalized logarithmic negativity obtained from the small $f_p\neq 0$ is larger than the $f_p=0$.  The normalization has been been done with respect to value of the logarithmic negativity at $\nu=1/2$,   which is the conformal coupled result for the theory under consideration for the present computational framework.  

\item From both of these plots we found that when the mass parameter squared $\nu^2<0$ i.e.  $\nu=-i|\nu|$ then we are dealing with heavy effective mass where we get falling behaviour of normalized logarithmic negativity ${\cal LN}(\nu)/{\cal LN}(\nu=1/2)$ and normalized entanglement negativity ${\cal N}(\nu)/{\cal N}(\nu=1/2)$.  This implies that in the heavy region we have less quantum correlation and the effect of quantum mechanical entanglement is falling as as increase the mass in the definition of mass parameter.  Here it is important to note that we have not considered the possibility from $\nu=i|\nu|$,  because this will give rise to exponentially divergent contribution as a Boltzmann contribution which cannot be possible to handle in the present computation.  So to get a correct measure of quantum mechanical entanglement here in this analysis we have restricted our discussion to $\nu=-i|\nu|$ branch of solutions only.  This possibility corresponds to {\bf Case B} in the present context.

\item We also found that at the conformal coupling limit $\nu=1/2$ the normalized logarithmic negativity ${\cal LN}(\nu)/{\cal LN}(\nu=1/2)=1$ and normalized entanglement negativity ${\cal N}(\nu)/{\cal N}(\nu=1/2)=1$.  The same thing happened in the massless limit $\nu=3/2$ where from the numerical plots we found that the normalized logarithmic negativity and entanglement negativity again ${\cal LN}(\nu)/{\cal LN}(\nu=1/2)=1$ and ${\cal N}(\nu)/{\cal N}(\nu=1/2)=1$.  At both this points,  $\nu=1/2$ and $\nu=3/2$ ({\bf Case A}) we get the maximum effect from quantum mechanical entanglement in the present computation.  This further physically implies that at these points we get the maximum contribution from quantum correlations.  In both the plots we have drawn a horizon line at ${\cal LN}(\nu)/{\cal LN}(\nu=1/2)=1$ and ${\cal N}(\nu)/{\cal N}(\nu=1/2)=1$ to indicate the reference level of our computation.

\item Additionally we have found that in the small mass region $\nu^2>0$ we get oscillatory type of feature where the period of oscillation is increasing as we increase the value of $\nu^2$ in the positive axis.  This possibility correspond to {\bf Case A} and {\bf Case B} in the present context.

\item Also from both of these plots we infer that effect of quantum mechanical entanglement is larger in the small mass region $\nu^2>0$ compared to the contribution from heavy mass region,  where we have $\nu^2<0$.  So small mass or the massless cases are more favourable than the heavy mass profile if we want to achieve more quantum mechanical effects from the entanglement measure.

\item  Last but not the least,  we now comment on the comparison among the outcomes obtained from the plots for normalized logarithmic negativity ${\cal LN}(\nu)/{\cal LN}(\nu=1/2)$ and normalized entanglement negativity ${\cal N}(\nu)/{\cal N}(\nu=1/2)$ for both $f_p=0$ and small $f_p\neq 0$.  Normalized entanglement negativity gives better understanding regarding the information content compared to the normalized logarithmic negativity for both the cases $f_p=0$ and small $f_p\neq 0$.  It seems like overall features obtained from normalized logarithmic negativity and normalized entanglement negativity are almost same.  But due to having differences in the definitions of the corresponding quantum information theoretic measures the outcomes are more prominent in normalized entanglement negativity.  Though it is important to note that both of them are connected to each other mathematically.  The effect of heavy mass for $\nu^2<0$ is more prominently observed in the case of normalized entanglement negativity compared to the normalized logarithmic negativity.  In the asymptotic limit of $\nu^2<0$ we see that for normalized entanglement negativity the quantum entanglement and the corresponding correlation saturates to a constant value.  On the other hand,  we found that there is sharp fall in the case of normalized logarithmic negativity which further implies there is no asymptotic value at which it becomes constant.  However,  apart from having this significant difference in the large mass limit,  in the small mass and massless limiting cases we have found out exactly same behaviour from the presented plots.
\end{enumerate}
   
    \section{Comparison with entanglement entropy in open chart}   
    \label{ka4}
    \begin{figure*}[htb]
    \centering
    \subfigure[For $f_p=0$.]{
        \includegraphics[width=14.2cm,height=8.5cm] {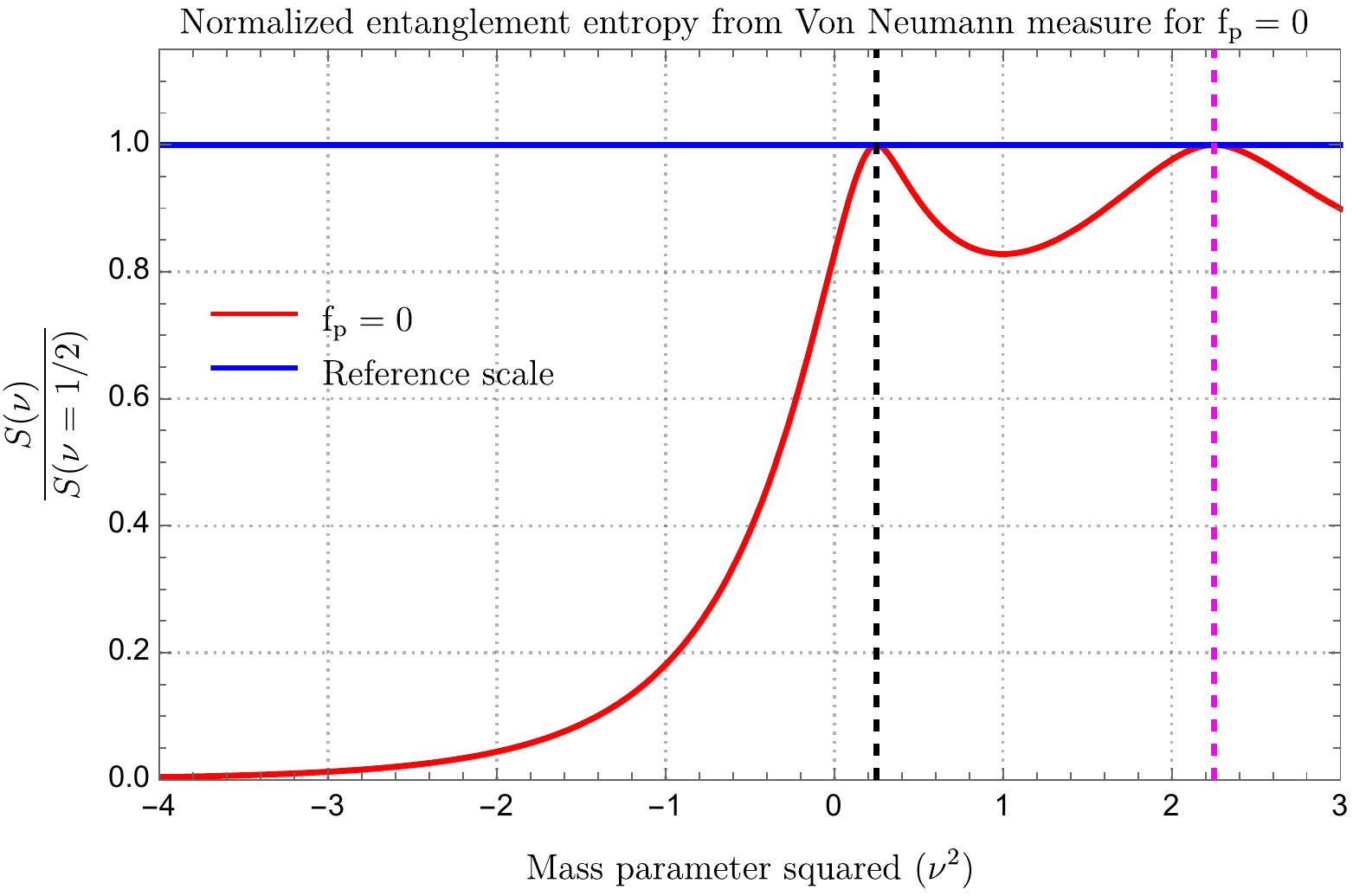}
        \label{VN1}
    }
    \subfigure[For small $f_p\neq 0$.]{
        \includegraphics[width=14.2cm,height=8.5cm] {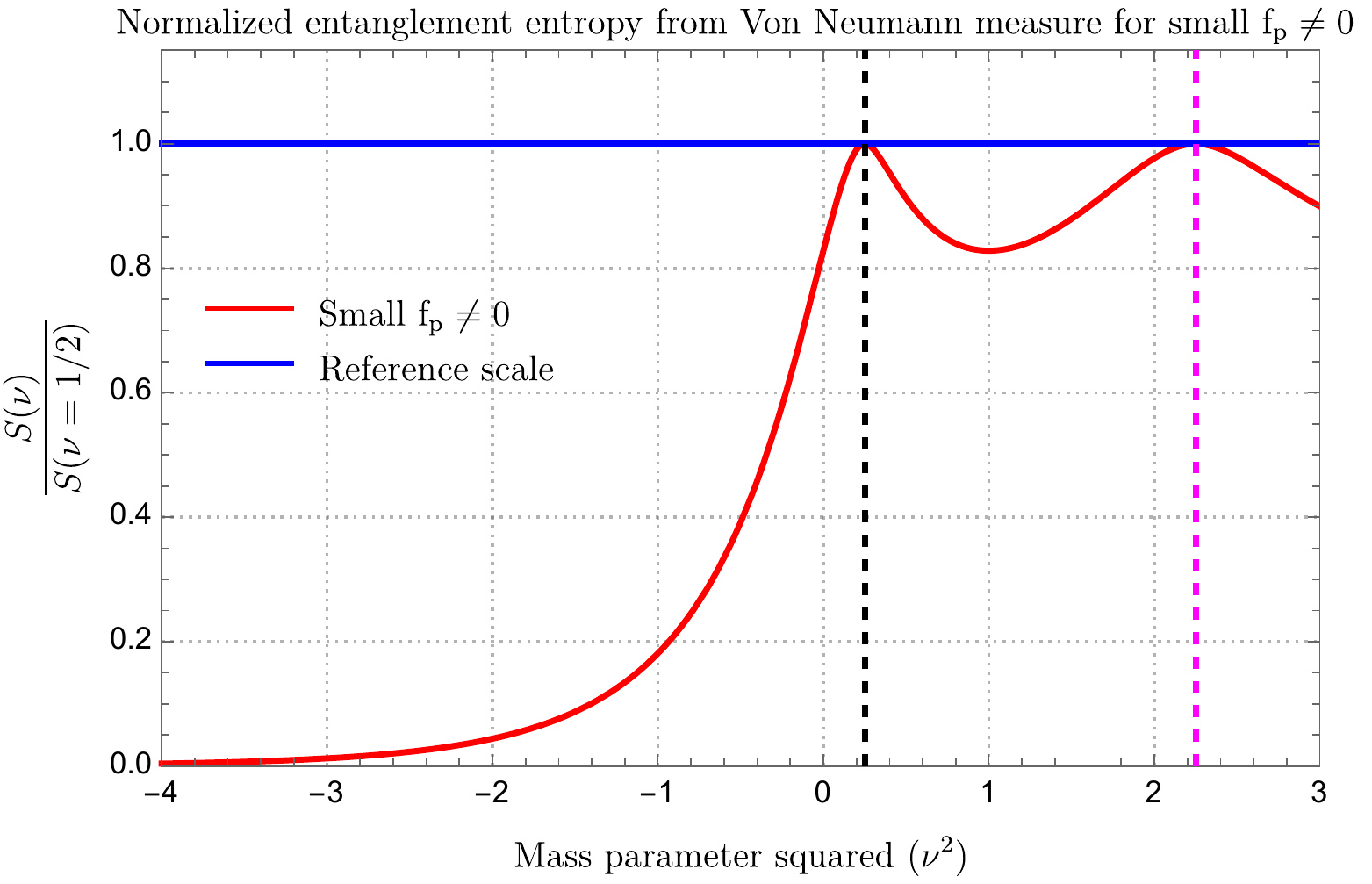}
        \label{VN2}
       }
    \caption[Optional caption for list of figures]{Graphical behaviour of the normalized entanglement entropy from Von Neumann measure ($S(\nu)/S(\nu=1/2)$) with mass parameter squared ($\nu^2$) for both $f_p=0$ and small $f_p\neq 0$.  Vertical dashed lines are drawn for $\nu=1/2$ (conformally coupled) and $\nu=3/2$ (massless case).  In both the plots reference scale corresponds to the scale at which $S(\nu)/S(\nu=1/2)=1$.} 
    \label{VN}
    \end{figure*}  
    \begin{figure*}[htb]
    \centering
    \subfigure[For $f_p=0$.]{
        \includegraphics[width=14.2cm,height=8.2cm] {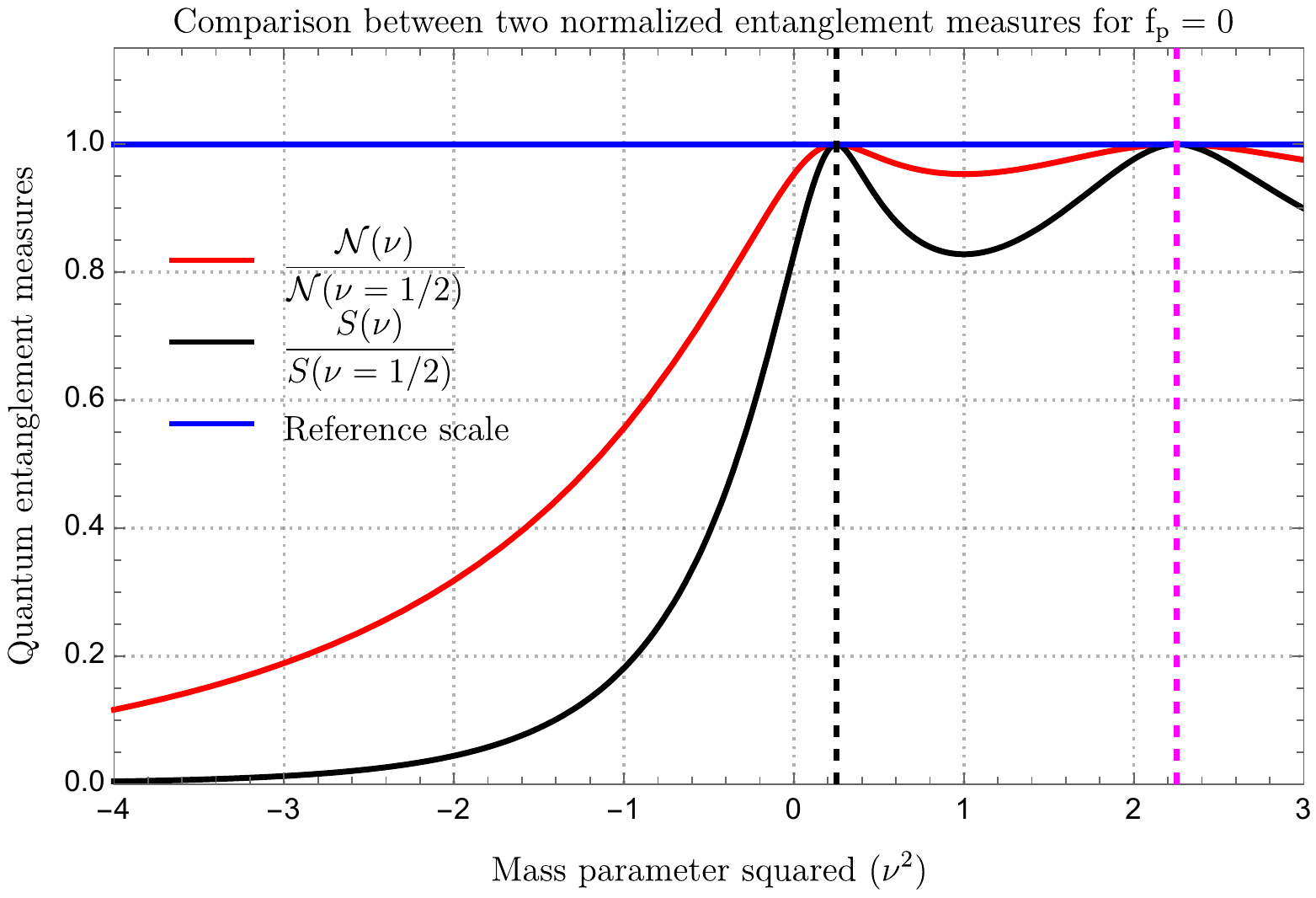}
        \label{COM1}
    }
    \subfigure[For small $f_p\neq 0$.]{
        \includegraphics[width=14.2cm,height=8.2cm] {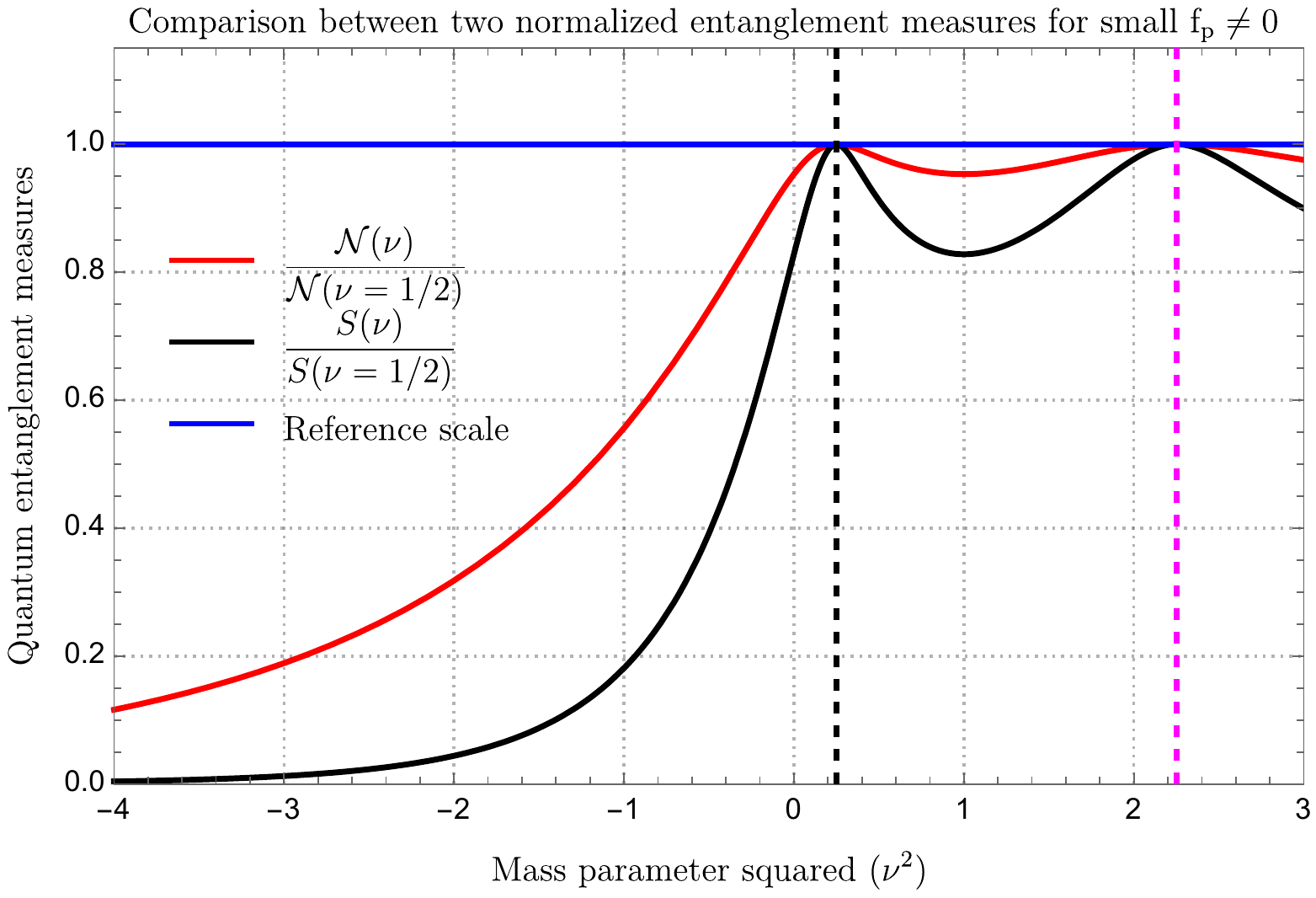}
        \label{COM2}
       }
    \caption[Optional caption for list of figures]{Graphical comparison between the behaviour of the normalized entanglement entropy from Von Neumann measure ($S(\nu)/S(\nu=1/2)$) and normalized entanglement negativity (${\cal N}(\nu)/{\cal N}(\nu=1/2)$) with mass parameter squared ($\nu^2$) for both $f_p=0$ and small $f_p\neq 0$.  Vertical dashed lines are drawn for $\nu=1/2$ (conformally coupled) and $\nu=3/2$ (massless case).  In both the plots reference scale corresponds to the scale at which ${\cal N}(\nu)/{\cal N}(\nu=1/2)=1$ and $S(\nu)/S(\nu=1/2)=1$.} 
    \label{COM}
    \end{figure*}  
    In this section we give a clear comparison between entanglement entropy and logarithmic negativity,  where we know both of them is used to describe the long range quantum correlation and entanglement effects in the present information theory motivated quantum field theoretic picture.
    
    Let us start our discussion with the entanglement entropy which is described by Von Neumann measure by the following expression \cite{Choudhury:2017bou}:
 \bea S(p,\nu)&=& -{\rm \bf Tr }\left[\rho_{\bf L}\ln \rho_{\bf L}\right]\nonumber\\
 &=&\Bigg\{- \left(1+\frac{f_p}{1+f_p}\right)\left[\ln\left(1-|\gamma_{p}|^2\right)+\frac{|\gamma_{p}|^2}{\left(1-|\gamma_{p}|^2\right)}\ln\left(|\gamma_{p}|^2\right)\right]\nonumber\\
 &&\quad\quad\quad\quad\quad\quad\quad\quad\quad\quad\quad\quad\quad\quad\quad\quad\quad\quad\quad -\left(1-f_p\right)\ln\left(1+f_p\right)\Bigg\},\eea
where we have computed each explicit parameter in the earlier section of this study, together with the explicit expression for the reduced density matrix.

 Now by following the same prescription stated before the regularized part of the entanglement entropy can be further expressed by the following simplified expression:
       \bea S(\nu)&=& V^{\bf reg}_{\bf H^3}\int^{\infty}_{0}~dp~{\cal D}(p)~S(p,\nu)= \frac{1}{\pi}\int^{\infty}_{0}~dp~p^2~S(p,\nu),\eea 
       where ${\cal D}(p)$ and $V^{\bf reg}_{\bf H^3}$ are defined earlier.   Further applying the same logical argument with the UV cut off for the computational purpose we further use the following UV cut off regulated version of Von Neumann entropy:
   \bea  S(\nu)&=& \frac{1}{\pi}\int^{\Lambda}_{0}~dp~p^2~S(p,\nu).\eea 
   Our next job is to numerically compute the expressions for entanglement entropy from the above mentioned UV regulated Von Neumann measure which will going to be extremely useful for the further comparison purpose with the results obtained from logarithmic negativity and entanglement negativity measures.
   
 In figure (\ref{VN}(a)) and figure (\ref{VN}(b)),  we have explicitly depicted the behaviour of normalized entanglement entropy from Von Neumann measure ($S(\nu)/S(\nu=1/2)$) with mass parameter squared ($\nu^2$) for both $f_p=0$ and small $f_p\neq 0$.  Similarly,  in figure (\ref{COM}(a)) and figure (\ref{COM}(b)),  we have explicitly plotted comparative behaviour between normalized entanglement negativity (${\cal N}(\nu)/{\cal N}(\nu=1/2)$) and  normalized entanglement entropy from Von Neumann measure ($S(\nu)/S(\nu=1/2)$) with mass parameter squared ($\nu^2$) for both $f_p=0$ and small $f_p\neq 0$.  Vertical dashed lines are drawn for $\nu=1/2$ and $\nu=3/2$ ({\bf Case A}) cases.  In both the plots we have drawn a horizon line at ${\cal N}(\nu)/{\cal N}(\nu=1/2)=1$ and $S(\nu)/S(\nu=1/2)=1$ to indicate the reference level of our computation.  The outcomes and the physical interpretation of these plots are very interesting which we are writing point-wise in the following:
 \begin{enumerate}
 \item From all of these plots we found that compared entanglement entropy computed from Von Neumann measure entanglement negativity give more information from the same system under consideration.  The reason behind this statement is both in the heavy mass $\nu^2<0$  and small mass $\nu^2>0$ regions we get more amplitude from the normalized version of the information content. 
 
 \item  In the heavy mass $\nu^2<0$ region we found that in the asymptotic limit entanglement entropy gives vanishing contribution,  but in the same limit entanglement negativity gives small but significantly non vanishing contribution.  On the other hand,  in the small mass $\nu^2>0$ region the amplitude of fluctuation of normalized entanglement entropy is larger amplitude than the normalized entanglement negativity.  
 
 \item Only at $\nu=1/2$ (conformally coupled) and $\nu=3/2$ (massless) we get exactly same contribution from both ${\cal N}(\nu)/{\cal N}(\nu=1/2)$ and $S(\nu)/S(\nu=1/2)$,  which is exactly ${\cal N}(\nu)/{\cal N}(\nu=1/2)=S(\nu)/S(\nu=1/2)=1$.

\end{enumerate}    
This is obviously a very interesting finding from our analysis that in case of the present quantum field theoretic set up entanglement negativity captures better information regarding quantum mechanical correlations and quantum entanglement than the Von Neumann measure of entanglement entropy.  

 \section{Logarithmic negativity between two causally unrelated patches of open chart}   
    \label{ka5}
  
    Our main goal in this part is to calculate and estimate the logarithmic negativity between two patches of the open chart of the global de Sitter space that are not causally related. We provide information on two physical observers whose role it is to explicitly determine and estimate the quantum mechanical entanglement in the current theoretical setup in order to serve the goal. For the purpose of computational simplification, we additionally assume that the quantum states corresponding to these two observers constitute an initial pure state within the multiverse and are maximally entangled to one another.Since we are dealing with two observers,  the corresponding framework in a more simpler language can be interpreted as biverse which can be easily further generalise to a general multiverse scenario.   From the starting point of the present construction we have explicitly mentioned that the sub regions {\bf R} and {\bf L} in the penrose diagram as well as in the Hilbert space construction is taken to be completely symmetric.  For this reason to extend this theoretical tool to compute the entanglement negativity from the corresponding biverse set up we consider that the region {\bf L} is in the inside of the de Sitter bubble.  One can do the similar thing by making use of the region {\bf R} as well.   Another important assumption we have considered during this computation is that there is no bubble wall exists in this framework for the open chart of the global de Sitter space.  Further in this biverse set up we place another observer in the other open chart of the global de Sitter bubble.  From the present theoretical computation our objective is to explicitly find the role of the inside observer to detect the quantum mechanical signatures of entanglement between two de Sitter bubbles using the well known Bunch Davies initial quantum states for the biverse.  Now in the present quantum field theoretic set up causality demands that the region {\bf R} has to be causally unrelated from the other region {\bf L}.  This is because of the fact that there is no access to the region {\bf R} during this computation.  For this specific reason one needs to consider the partial trace operation over the inaccessible region {\bf R},  which further give rise to the information loss regarding this specific region.  As a consequence the corresponding quantum mechanical state which describe the observer would be a mixed quantum state for this computation.  In this set up the quantum mechanical state associated with the other observer becomes a pure quantum state which belongs to the other causally unrelated patch of the open chart of the global de Sitter bubble.  This implies in the present computation we need to consider the entanglement between a mixed and pure quantum mechanical states which belong to two causally unrelated de Sitter bubbles.  If in the computation of quantum entanglement from two subsystems only pure quantum states are involved then in such a system Von Neumann measure is the best measure to quantify entanglement entropy.  But if any of the subsystem is described by mixed quantum state then for such a situation Von Neumann measure is not the appropriate quantum information theoretic measure.  In that case entanglement negativity or the logarithmic negativity can give better measure of quantum entanglement,  which is somewhat clear to us from the comparison that we have already drawn from our previous analysis.  He strongly believe that this new measure will going to explain various unexplored underlying physics involved in the present theoretical set up.

    \subsection{Computational set up and construction of maximally entangled states}

Starting with the total quantum vacuum state, which is actually represented by the product of the quantum vacuum states for each oscillator that we found computed explicitly in the previous section of this paper, we begin to study the effects from entanglement negativity between two causally unrelated patches of open chart. It's crucial to remember that each oscillator's quantum mechanical state is identified by one of the three quantum numbers $p$, $l$, or $m$ for future calculation purposes. For this reason, one must take product over $p$ in the final expression of total quantum. The final Bunch Davies quantum vacuum state in this configuration is stated as:
\bea  |0\rangle_{\bf BD} &=&\prod_{p}|0_p\rangle_{\bf BD},\eea
where we define the Bunch Davies states for each mode as:
    \bea |0_p\rangle_{\bf BD} &=&\Bigg\{\sqrt{\frac{\left(1-|\gamma_p|^2\right)}{\left(1+f_p\right)}}\sum^{\infty}_{k=0}|\gamma_{p}|^{k}\bigg(|k_p\rangle_{{\bf R}^{'}}\otimes|k_p\rangle_{{\bf L}^{'}}\bigg)\nonumber\\
        &&\quad\quad\quad\quad\quad\quad+\frac{f_{p}}{\sqrt{\left(1+f_p\right)}}\sum^{\infty}_{n=0}\sum^{\infty}_{r=0}|\Gamma_{p,n}|^{r}\bigg(|n,r_p\rangle_{{\bf R}^{'}}\otimes| n,r_p\rangle_{{\bf L}^{'}}\bigg)\Bigg\}.~~~~~~~~~\quad\eea 
   Here it is important to note that,  in the above expression for the simplification in the writing purpose we have removed the tags of $l$ and $m$ on the individual direct product states defined in the region {\bf R} and {\bf L} in the Bogoliubov transformed basis.  This simplification will going to further help us to deal with cumbersome expressions.
   
   In this specific computation we are actually considering smaller version of the multiverse where many causally unrelated patches in the open chart of the de Sitter bubbles forms a maximally entangled state which is necessary ingredient for the present calculation.  In this calculation it is assumed that each of the causally unrelated patches corresponds to a Bunch Davies quantum vacuum states which will finally form a maximally entangled state out of all possible Bunch Davies states.  However this is not a strict assumption to extract the required outcome from the present computational set up.  One can think about a scenario in which some quantum vacuum states are distinct from Bunch Davies vacuum and are entangled with one another in a more general version of the computation. This is quite an interesting possibility in this context.  Now we don't want to complicate our understanding at this level of computation.  For this reason we are going to restrict our analysis by considering the two Bunch Davies quantum vacuum states which belong to two causally unrelated patches of the open chart of the de Sitter bubbles. Now to theoretically model the present computational set up let us first consider two momentum modes having momenta $p=p_{\bf in}$ and $p=p_{\bf out}$ of the axionic field theory described by {\bf Case A} (massless) and {\bf Case B} (massive) respectively.   With this knowledge, it is possible to further express the maximally entangled state in terms of the unique contributions of Bunch Davies vacuum to the two causally independent regions of the de Sitter bubbles as follows:
   \bea |\Psi\rangle_{\bf ME}:&=&\frac{1}{\sqrt{2}}\sum_{i=0,1}\Bigg(|i_{p_{\bf out}}\rangle_{\bf BD_1}\otimes|i_{p_{\bf in}}\rangle_{\bf BD_2}\Bigg)\nonumber\\
   &=&\frac{1}{\sqrt{2}}\Bigg(|0_{p_{\bf out}}\rangle_{\bf BD_1}\otimes|0_{p_{\bf in}}\rangle_{\bf BD_2}+|1_{p_{\bf out}}\rangle_{\bf BD_1}\otimes|1_{p_{\bf in}}\rangle_{\bf BD_2}\Bigg).\eea   
    \begin{figure*}[htb]
    	\centering
    	{
    		\includegraphics[width=15.5cm,height=12.5cm] {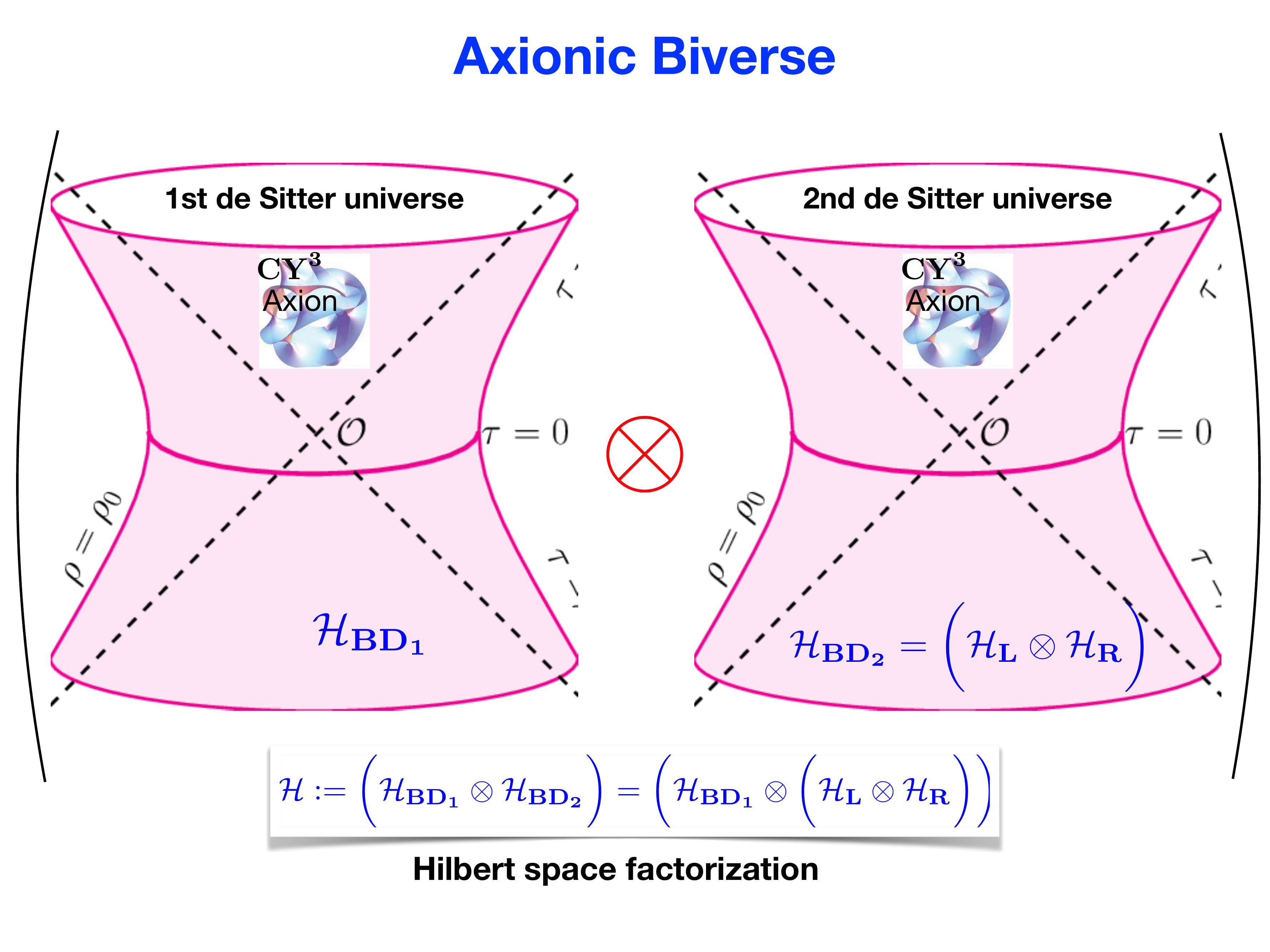}
    	}
    	\caption[Optional caption for list of figures]{Representative diagram of the Hilbert space factorization of the axionic biverse which is constructed out of two de Sitter vacua.  Each of the vacua is spanned in terms of the complete set of Bunch Davies states.} 
    	\label{biverse}
    \end{figure*}
   In the above equation,  $|0_{p_{\bf out}}\rangle_{\bf BD_1}$ and $|1_{p_{\bf out}}\rangle_{\bf BD_1}$ represent the ground and first single particle excited quantum state characterized by the momentum mode $p_{\bf out}$ in the first Bunch Davies vacuum obtained from the first patch of the open chart of global the de Sitter space.  Similarly,       
   $|0_{p_{\bf in}}\rangle_{\bf BD_2}$ and $|1_{p_{\bf in}}\rangle_{\bf BD_2}$ represent the ground and first single particle excited quantum state characterized by the momentum mode $p_{\bf in}$ in the second Bunch Davies vacuum obtained from the second patch of the open chart of global the de Sitter space.  For the further computational simplification purpose we assume that outside and inside observers are associated with two detectors having mode momenta $p_{\bf out}$ and $p_{\bf in}$ respectively.  It is also crucial to note that, given the current quantum field theoretic framework, the complete Hilbert space can be factored as follows:
   \bea {\cal H}:=\bigg({\cal H}_{\bf {BD}_1}\otimes{\cal H}_{\bf {BD}_2}\bigg)=\Bigg(\underbrace{\bigg({\cal H}_{\bf L}\otimes {\bf H}_{\bf R}\bigg)_{\bf {BD}_1}}_{\bf For ~first ~subspace}\otimes\underbrace{\bigg({\cal H}_{\bf L}\otimes {\bf H}_{\bf R}\bigg)_{\bf {BD}_2}}_{\bf For ~second ~subspace}\Bigg),\eea  
where the first and the second Bunch Davies vacuum states are associated with two subspace,  whose corresponding product Hilbert space can be further factorized in terms Hilbert spaces associated with region {\bf R} and {\bf L} as:
\bea    {\cal H}_{\bf {BD}_1}&=&\bigg({\cal H}_{\bf L}\otimes {\bf H}_{\bf R}\bigg)_{\bf {BD}_1},\\
{\cal H}_{\bf {BD}_2}&=&\bigg({\cal H}_{\bf L}\otimes {\bf H}_{\bf R}\bigg)_{\bf {BD}_2}.\eea
   In the multiverse picture one can further generalize this result y considering many Bunch Davies states and different quantum vacua than Bunch Davies,  such as $\alpha$ vacua,  Motta Allen vacua etc.  In figure (\ref{biverse}) representative diagram of the Hilbert space factorization of the axionic biverse which is constructed out of two de Sitter vacua.  Each of the vacua is spanned in terms of the complete set of Bunch Davies states which we have pointed clearly in this diagram.

    \subsection{Construction of excited quantum state for single oscillator}

   Our main task in this paragraph is to create the excited quantum state for a single oscillator, which will also be helpful to create the overall maximally entangled state required for the specific computation. Let's start with the characteristic matrix equation for the oscillators in the recently announced Bogoliubov transformed basis to demonstrate this issue in more detail:  
      \bea c_{J}&=& b_{I}{\cal G}^{I}_{J},~\quad
            C_{J(n)}=\bar{b}_{J(n)}\left({\cal G}_{(n)}\right)^{I}_{J}\quad\quad {\rm where}\quad\quad c_J=(c_q,c^{\dagger}_q), C_{J(n)}=(C_{q(n)},C^{\dagger}_{q(n)}),\quad\quad\quad\eea
  where we define the two square matrices as appearing in the above expressions by the following equations:
         \bea {\cal G}^{I}_{J}=\left(\begin{array}{ccc} U_q &~~~ V^{*}_q \\ V_q &~~~ U^{*}_q  \end{array}\right)
                    ,~\quad\quad
                    \left({\cal G}_{(n)}\right)^{I}_{J}=\left(\begin{array}{ccc} \overline{U}_{ q,n} &~~~ \overline{V}^{*}_{\sigma q,n} \\ \overline{V}_{ q,n} &~~~ \overline{U}^{*}_{ q,n}  \end{array}\right),\quad\quad\quad\quad\eea
where the components of the new matrices are given by:
                    \bea U_q &\equiv& {\rm \bf diag}\left(u,\overline{u}\right),\\
                    V_q &\equiv& {\rm \bf diag}\left(v,\overline{v}\right),\\
                     \overline{U}_{q,n} &\equiv& {\rm \bf diag}\left(U_n,\overline{U}_n\right),\\
                     \overline{V}_{q,n} &\equiv& {\rm \bf diag}\left(V_n,\overline{V}_n\right).\eea
Hence one can able to find the relationship between the $a$-type of oscillators with c-type of oscillators,  which are given by:

 \bea a^{(c)}_{J} &=& b_{J}\left({\cal M}^{-1}\right)^{I}_{J}=c_K\left({\cal G}^{-1}\right)^{K}_{I} \left({\cal M}^{-1}\right)^{I}_{J},\\
  a^{(p)}_{J(n)} &=& b_{J(n)}\left({\cal M}^{-1}_{(n)}\right)^{I}_{J}=C_{K(n)}\left({\cal G}^{-1}_{(n)}\right)^{K}_{I} \left({\cal M}^{-1}_{(n)}\right)^{I}_{J},\eea 
 which further form the following quantity:
 \bea a_{I}&=& \left[a^{(c)}_{I}+\sum^{\infty}_{n=0}a^{(p)}_{I(n)}\right]=\bigg[\underbrace{c_K\left({\cal G}^{-1}\right)^{K}_{I} \left({\cal M}^{-1}\right)^{I}_{J}}_{\bf Complementary~part}+\underbrace{\sum^{\infty}_{n=0}C_{K(n)}\left({\cal G}^{-1}_{(n)}\right)^{K}_{I} \left({\cal M}^{-1}_{(n)}\right)^{I}_{J}}_{\bf Particular~integral~part}\bigg].\quad\quad\eea 
Here the the product of two inverse matrices are parametrized by another $(4\times 4)$ square matrix,  which is given by:
   \bea \left({\cal G}^{-1}\right)^{K}_{I} \left({\cal M}^{-1}\right)^{I}_{J}&=&\left(\begin{array}{ccc} Q_{\sigma q} &~~~R^{*}_{\sigma q} \\ R_{\sigma q} &~~~ Q^{*}_{\sigma q}  \end{array}\right),
         \\
\left({\cal G}^{-1}_{(n)}\right)^{K}_{I} \left({\cal M}^{-1}_{(n)}\right)^{I}_{J}&=&\left(\begin{array}{ccc}\overline{Q}_{\sigma q,n} &~~~ \overline{R}^{*}_{\sigma q,n} \\ R_{\sigma q,n} &~~~\overline{Q}^{*}_{\sigma q,n}  \end{array}\right),\eea   
where the elements of both the matrices are itself $(2\times 2)$ matrices,  which are provided by the subsequent expressions:
\bea Q_{\sigma q}&=&\left(\begin{array}{ccc}  \widetilde{A}u~~~ &~~~-\widetilde{B}u+\widetilde{D}^{*}v\\ -\widetilde{B}u+\widetilde{D}^{*}v &~~~  \widetilde{A}u  \end{array}\right),\\
R_{\sigma q}&=&\left(\begin{array}{ccc} - \widetilde{A}v~~~ &\widetilde{B}v-\widetilde{D}^{*}u\\ \widetilde{B}v-\widetilde{D}^{*}u &~~~ - \widetilde{A}v  \end{array}\right),\\
Q_{\sigma q,n}&=&\left(\begin{array}{ccc} \widetilde{A}_nU_n~~~ &~~~-\widetilde{B}_nU_n+\widetilde{D}^{*}_nV_n\\ -\widetilde{B}_nU_n+\widetilde{D}^{*}_nV_n &~~~ \widetilde{A}_nU_n  \end{array}\right),\\
R_{\sigma q,n}&=&\left(\begin{array}{ccc} -\widetilde{A}_nV_n~~~ &\widetilde{B}_nV_n-\widetilde{D}^{*}_nU_n\\ \widetilde{B}_nV_n-\widetilde{D}^{*}_nU_n &~~~ -\widetilde{A}_nV_n  \end{array}\right).
         \eea
         Here the coefficients ($ \widetilde{A}$, $\widetilde{B}$,$\widetilde{D}$) and ($\widetilde{A}_n$, $\widetilde{B}_n$,$\widetilde{D}_n$) aare provided by the subsequent expressions:
         \bea  \widetilde{A}&=&\frac{\sqrt{\pi p}}{\left|\Gamma\left(\nu+\frac{1}{2}+ip\right)\right|}\frac{\exp\left(\frac{\pi p}{2}\right)}{\sqrt{\cosh 2\pi p+\cos 2\pi\nu}},\\ 
         \widetilde{B}&=& \widetilde{A}~\frac{\cos \pi\nu}{i \sinh \pi p}\nonumber\\
         &=&\frac{\sqrt{\pi p}}{\left|\Gamma\left(\nu+\frac{1}{2}+ip\right)\right|}\frac{\exp\left(\frac{\pi p}{2}\right)}{\sqrt{\cosh 2\pi p+\cos 2\pi\nu}}\frac{\cos \pi\nu}{i \sinh \pi p},\\
         \widetilde{D}&=&- \widetilde{A}~\frac{\cos (ip+\nu)\pi}{i \sinh \pi p}\exp\left(-\pi p\right)\frac{\Gamma\left(\nu+\frac{1}{2}+ip\right)}{\Gamma\left(\nu+\frac{1}{2}-ip\right)}\nonumber\\
         &=&-\frac{\sqrt{\pi p}}{\left|\Gamma\left(\nu+\frac{1}{2}+ip\right)\right|}\frac{\exp\left(-\frac{\pi p}{2}\right)}{\sqrt{\cosh 2\pi p+\cos 2\pi\nu}}~\frac{\cos (ip+\nu)\pi}{i \sinh \pi p}\frac{\Gamma\left(\nu+\frac{1}{2}+ip\right)}{\Gamma\left(\nu+\frac{1}{2}-ip\right)},\eea
         and 
         \bea
        \widetilde{A}_n &=&\frac{\sqrt{\pi p_n}}{\left|\Gamma\left(\nu+\frac{1}{2}+ip_n\right)\right|}\frac{\exp\left(\frac{\pi p_n}{2}\right)}{\sqrt{\cosh 2\pi p_n+\cos 2\pi\nu}},\\
       \widetilde{B}_n &=&\widetilde{A}_n~\frac{\cos \pi\nu}{i \sinh \pi p_n}\nonumber\\
         &=&\frac{\sqrt{\pi p_n}}{\left|\Gamma\left(\nu+\frac{1}{2}+ip_n\right)\right|}\frac{\exp\left(\frac{\pi p_n}{2}\right)}{\sqrt{\cosh 2\pi p_n+\cos 2\pi\nu}}\frac{\cos \pi\nu}{i \sinh \pi p_n},\\ 
       \widetilde{D}_n &=&-\widetilde{A}_n~\frac{\cos (ip_n+\nu)\pi}{i \sinh \pi p_n}\exp\left(-\pi p_n\right)\frac{\Gamma\left(\nu+\frac{1}{2}+ip_n\right)}{\Gamma\left(\nu+\frac{1}{2}-ip_n\right)}\nonumber\\
         &=&-\frac{\sqrt{\pi p_n}}{\left|\Gamma\left(\nu+\frac{1}{2}+ip_n\right)\right|}\frac{\exp\left(-\frac{\pi p_n}{2}\right)}{\sqrt{\cosh 2\pi p_n+\cos 2\pi\nu}}~\frac{\cos (ip_n+\nu)\pi}{i \sinh \pi p_n}\frac{\Gamma\left(\nu+\frac{1}{2}+ip_n\right)}{\Gamma\left(\nu+\frac{1}{2}-ip_n\right)}. \eea  
   From the structure of the above mentioned matrices we have found the following characteristics:
   \begin{enumerate}
   \item First of all we have found that for each of the above mentioned matrices the diagonal and off-diagonal elements are same,  which implies each of them are symmetric matrices under the matrix transpose operation i.e.
   \bea &&Q_{q\sigma}=Q^{T}_{\sigma q}=Q_{\sigma q},\\
   &&R_{q\sigma}=R^{T}_{\sigma q,n}=R_{\sigma q,n},\\
   &&Q_{q\sigma,n}=Q^{T}_{\sigma q,n}=Q_{\sigma q,n},\\
   &&R_{q\sigma,n}=R^{T}_{\sigma q,n}=R_{\sigma q,n}.\eea
   
   \item In the above mentioned expressions we have also used the following relationships to write down the final result in a more simplified and compact format:
   \bea && \widetilde{A}^{*}= \widetilde{A},\quad\quad \widetilde{B}^{*}=-\widetilde{B},\quad\quad u^{*}=u=\overline{u},\quad\quad\quad\quad  v^{*}=v=\overline{v}, \\
   &&\widetilde{A}^{*}_n=\widetilde{A}_n,\quad\quad \widetilde{B}^{*}_n=-\widetilde{B}_n,\quad\quad U^{*}_n=U_n=\overline{U}_n,\quad\quad  V^{*}_n=V_n=\overline{V}_n.\eea  
   
   \item If we set $B=0$ and $\widetilde{B}=0$ along with $v=0$ and $V_n=0$ correspond to case of conformal coupling ($\nu=1/2$) and the massless theory ($\nu=3/2$). 
   
   \end{enumerate}
  
   The following formulas in the region {\bf L} provide the creation and annihilation operators for oscillators of the $a$ type, and fixing this will fix their explicit mathematical structures:
   \bea a^{\dagger}_{\bf L} :&=&\underbrace{\Bigg( \widetilde{A}uc^{\dagger}_{\bf L}- \widetilde{A}vc_{\bf L}+\left(\widetilde{B}u+\widetilde{D}v\right)c^{\dagger}_{\bf R} -\left(\widetilde{B}v+\widetilde{D}u\right)c_{\bf R}   \Bigg)}_{\bf Complementary~part}\nonumber\\
   &&\quad\quad+\underbrace{\sum^{\infty}_{n=0}\Bigg(\widetilde{A}_nU_nC^{\dagger}_{{\bf L}(n)}-\widetilde{A}_nV_nC_{{\bf L}(n)}+\left(\widetilde{B}_nU_n+\widetilde{D}_nV_n\right)C^{\dagger}_{{\bf R}(n)} -\left(\widetilde{B}_nV_n+\widetilde{D}_nU_n\right)C_{{\bf R}(n)}   \Bigg)}_{\bf Particular~integral~part},\nonumber\\
   && \\
  a_{\bf L} :&=&\underbrace{\Bigg( \widetilde{A}uc_{\bf L}- \widetilde{A}vc^{\dagger}_{\bf L}+\left(-\widetilde{B}u+\widetilde{D}^{*}v\right)c_{\bf R} -\left(-\widetilde{B}v+\widetilde{D}^{*}u\right)c^{\dagger}_{\bf R}   \Bigg)}_{\bf Complementary~part}\nonumber\\
   &&\quad\quad+\underbrace{\sum^{\infty}_{n=0}\Bigg(\widetilde{A}_nU_nC_{{\bf L}(n)}-\widetilde{A}_nV_nC^{\dagger}_{{\bf L}(n)}+\left(-\widetilde{B}_nU_n+\widetilde{D}^{\dagger}_nV_n\right)C_{{\bf R}(n)} -\left(-\widetilde{B}_nV_n+\widetilde{D}^{\dagger}_nU_n\right)C^{\dagger}_{{\bf R}(n)}   \Bigg)}_{\bf Particular~integral~part},\nonumber\\
   && \eea
    The following expression represents the excited quantum state for a single oscillator and corresponds to the quantum state of the inside observer:
   \bea |1_{p_{\bf in}}\rangle_{{\bf BD_2}}&=& a^{\dagger}_{\bf L} |0_{p_{\bf in}}\rangle_{{\bf BD_2}}\nonumber\\
   &=&\Bigg\{\underbrace{\Bigg( \widetilde{A}uc^{\dagger}_{\bf L}- \widetilde{A}vc_{\bf L}+\left(\widetilde{B}u+\widetilde{D}v\right)c^{\dagger}_{\bf R} -\left(\widetilde{B}v+\widetilde{D}u\right)c_{\bf R}   \Bigg)}_{\bf Complementary~part}\nonumber\\
   &&+\underbrace{\sum^{\infty}_{n=0}\Bigg(\widetilde{A}_nU_nC^{\dagger}_{{\bf L}(n)}-\widetilde{A}_nV_nC_{{\bf L}(n)}+\left(\widetilde{B}_nU_n+\widetilde{D}_nV_n\right)C^{\dagger}_{{\bf R}(n)} -\left(\widetilde{B}_nV_n+\widetilde{D}_nU_n\right)C_{{\bf R}(n)}   \Bigg)}_{\bf Particular~integral~part}\nonumber\\ 
   &&\quad\quad\quad\quad\quad\Bigg\}\Bigg\{\sqrt{\frac{\left(1-|\gamma_{p_{\bf in}}|^2\right)}{\left(1+f_{p_{\bf in}}\right)}}\sum^{\infty}_{k=0}|\gamma_{p_{\bf in}}|^{k}\bigg(|k_{p_{\bf in}}\rangle_{{\bf R}^{'}}\otimes|k_{p_{\bf in}}\rangle_{{\bf L}^{'}}\bigg)\nonumber\\
        &&\quad\quad\quad\quad\quad\quad+\frac{f_{{p_{\bf in}}}}{\sqrt{\left(1+f_{p_{\bf in}}\right)}}\sum^{\infty}_{n=0}\sum^{\infty}_{r=0}|\Gamma_{{p_{\bf in}},n}|^{r}\bigg(|n,r_{p_{\bf in}}\rangle_{{\bf R}^{'}}\otimes| n,r_{p_{\bf in}}\rangle_{{\bf L}^{'}}\bigg)\Bigg\}~~~~~~~~~\nonumber\\
   &=&\Bigg\{\sqrt{\frac{\left(1-|\gamma_{p_{\bf in}}|^2\right)}{\left(1+f_{p_{\bf in}}\right)}}\Bigg[\Delta_1\sum^{\infty}_{k=0}|\gamma_{p_{\bf in}}|^{k}\sqrt{k+1}\bigg(|k_{p_{\bf in}}\rangle_{{\bf R}^{'}}\otimes|(k+1)_{p_{\bf in}}\rangle_{{\bf L}^{'}}\bigg)\nonumber\\
   &&\quad\quad\quad\quad+\Delta_2\sum^{\infty}_{k=0}|\gamma_{p_{\bf in}}|^{k}\sqrt{k+1}\bigg(|(k+1)_{p_{\bf in}}\rangle_{{\bf R}^{'}}\otimes|k_{p_{\bf in}}\rangle_{{\bf L}^{'}}\bigg)\Bigg]\nonumber\\
        &&+\frac{f_{{p_{\bf in}}}}{\sqrt{\left(1+f_{p_{\bf in}}\right)}}\Bigg[\sum^{\infty}_{n=0} \Delta_{3,n}\sum^{\infty}_{r=0}|\Gamma_{{p_{\bf in}},n}|^{r}\sqrt{r+1}\bigg(|n,r_{p_{\bf in}}\rangle_{{\bf R}^{'}}\otimes| n,(r+1)_{p_{\bf in}}\rangle_{{\bf L}^{'}}\bigg)\nonumber\\
        &&\quad\quad\quad\quad+\sum^{\infty}_{n=0}\Delta_{4,n}\sum^{\infty}_{r=0} |\Gamma_{{p_{\bf in}},n}|^{r}\sqrt{r+1}\bigg(|n,(r+1)_{p_{\bf in}}\rangle_{{\bf R}^{'}}\otimes| n,r_{p_{\bf in}}\rangle_{{\bf L}^{'}}\bigg)\Bigg]\Bigg\}.\eea  
   In the above expression we introduce four symbols $\Delta_1$,  $\Delta_2$,  $\Delta_{3,n}$  and  $\Delta_{4,n}$ which are defined by the following expressions:
   \bea \Delta_1 &=& \Bigg( \widetilde{A}u-(\widetilde{B}v+\widetilde{D}u)\gamma_{p_{\bf in}}\Bigg),\\
   \Delta_2 &=&\Bigg(- \widetilde{A}v\gamma_{p_{\bf in}}+(\widetilde{B}u+\widetilde{D}v)\Bigg),\\
   \Delta_{3,n} &=&\Bigg(\widetilde{A}_nU_n -\left(\widetilde{B}_nV_n+\widetilde{D}_nU_n\right)\Bigg),\\
   \Delta_{4,n} &=&\Bigg(-\widetilde{A}_nV_n \Gamma_{{p_{\bf in}},n}+\left(\widetilde{B}_nU_n+\widetilde{D}_nV_n\right)\Bigg).\eea   
   To compute the expression for the excited quantum state for the single oscillator we have used the usual harmonic oscillator algebra in terms of the quantum states.

   We must perform a partial trace operation over all of the degrees of freedom in the region ${\bf R}$ since the building of the current theoretical framework requires that the inside observer be located at the region ${\bf L}$ of one of the open charts of the global de Sitter space. This will ultimately result in a density matrix, which we have precisely computed in the following subsection. Due to this, we suggest the following ansatz for factorising the entire Hilbert space used in the computation here:
   \bea {\cal H}:=\Bigg({\cal H}_{\bf BD_1}\otimes{\cal H}_{\bf BD_2}\Bigg)=\Bigg({\cal H}_{\bf BD_1}\otimes\bigg({\cal H}_{\bf L}\otimes{\cal H}_{\bf R}\bigg)_{\bf BD_2}\Bigg).\eea
   Though the similar factorization exists in the first Bunch Davies quantum vacuum state,  for the time being to perform the present computation we don't need the factorization details of this subspace.  The prime for this particular choice is because to construct the density matrix the details of the subspace which belongs to the first Bunch Davies quantum vacuum state is not explicitly required. This will help us to compute the rest of the computations of this paper in a very simplified fashion.  In the next subsection we will discuss the technical details of this construction to formulate the maximal entangled state and hence the reduced density matrix.
    \subsection{Construction of reduced density matrix at the inside observer}

    Our primary objective in this section is to derive the equation for the reduced density matrix in the region {\bf L} by performing a partial trace operation over the contributions from the region {\bf R}. The matching maximally entangled state, denoted by the following phrase, must first be created:
   \bea |\Psi\rangle_{\bf ME}:
   &=&\frac{1}{\sqrt{2}}\Bigg(|0_{p_{\bf out}}\rangle_{\bf BD_1}\otimes\Bigg\{\sqrt{\frac{\left(1-|\gamma_{p_{\bf in}}|^2\right)}{\left(1+f_{p_{\bf in}}\right)}}\sum^{\infty}_{k=0}|\gamma_{p_{\bf in}}|^{k}\bigg(|k_{p_{\bf in}}\rangle_{{\bf R}^{'}}\otimes|k_{p_{\bf in}}\rangle_{{\bf L}^{'}}\bigg)\nonumber\\
        &&\quad\quad\quad\quad\quad\quad+\frac{f_{{p_{\bf in}}}}{\sqrt{\left(1+f_{p_{\bf in}}\right)}}\sum^{\infty}_{n=0}\sum^{\infty}_{r=0}|\Gamma_{{p_{\bf in}},n}|^{r}\bigg(|n,r_{p_{\bf in}}\rangle_{{\bf R}^{'}}\otimes| n,r_{p_{\bf in}}\rangle_{{\bf L}^{'}}\bigg)\Bigg\}\nonumber\\
        &&+|1_{p_{\bf out}}\rangle_{\bf BD_1}\otimes\Bigg\{\sqrt{\frac{\left(1-|\gamma_{p_{\bf in}}|^2\right)}{\left(1+f_{p_{\bf in}}\right)}}\Bigg[\Delta_1\sum^{\infty}_{k=0}|\gamma_{p_{\bf in}}|^{k}\sqrt{k+1}\bigg(|k_{p_{\bf in}}\rangle_{{\bf R}^{'}}\otimes|(k+1)_{p_{\bf in}}\rangle_{{\bf L}^{'}}\bigg)\nonumber\\
   &&\quad\quad\quad\quad+\Delta_2\sum^{\infty}_{k=0}|\gamma_{p_{\bf in}}|^{k}\sqrt{k+1}\bigg(|(k+1)_{p_{\bf in}}\rangle_{{\bf R}^{'}}\otimes|k_{p_{\bf in}}\rangle_{{\bf L}^{'}}\bigg)\Bigg]\nonumber\\
        &&+\frac{f_{{p_{\bf in}}}}{\sqrt{\left(1+f_{p_{\bf in}}\right)}}\Bigg[\sum^{\infty}_{n=0} \Delta_{3,n}\sum^{\infty}_{r=0}|\Gamma_{{p_{\bf in}},n}|^{r}\sqrt{r+1}\bigg(|n,r_{p_{\bf in}}\rangle_{{\bf R}^{'}}\otimes| n,(r+1)_{p_{\bf in}}\rangle_{{\bf L}^{'}}\bigg)\nonumber\\
        &&\quad\quad\quad\quad+\sum^{\infty}_{n=0}\Delta_{4,n}\sum^{\infty}_{r=0} |\Gamma_{{p_{\bf in}},n}|^{r}\sqrt{r+1}\bigg(|n,(r+1)_{p_{\bf in}}\rangle_{{\bf R}^{'}}\otimes| n,r_{p_{\bf in}}\rangle_{{\bf L}^{'}}\bigg)\Bigg]\Bigg\}\Bigg).\quad\quad\quad\eea   
    From the above mentioned detailed structure of the maximal entangled state constructed in the present set up it is observed that the scale dependence in this state comes through the quantities,  $\gamma_{\bf p_{in}}$,  $\Gamma_{{p_{\bf in}},n}$,  $\Delta_1$,  $\Delta_2$,  $\Delta_{3,n}$  and  $\Delta_{4,n}$ appearing in the computation.  This particular fact is the direct outcome of the factorization of the inside observer's subspace in to two symmetric subspaces {\bf R} and {\bf L}.  Our further job is to to study the imprints of this scale dependence on the physical outcomes of the systems to explore the unknown facts from the theoretical set up under consideration.  
    
    We now need to take a partial trace over the degrees of freedom of region {\bf R}, because we already know that the inside observer's subspace does not get any information content from this region. This will allow us to create the reduced density matrix out of this configuration. Due to the fact that the above-mentioned newly built maximally entangled quantum state, which is actually a mixed state in the current prescription, will be taken into account throughout this computation, we must be mindful of this fact. As a result, the reduced density matrix can be expressed simply as follows: 
    \bea \rho_{\bf reduced}:&=&{\bf Tr}_{{\bf R}^{'}}\left[|\Psi\rangle_{\bf ME}\quad{}_{\bf ME}\langle\Psi|\right]\nonumber\\
    &=&\sum^{\infty}_{m_{\bf p_{in}}=0}{}_{{\bf R}^{'}}\langle m_{\bf p_{in}}|\Psi\rangle_{\bf ME}{}_{\bf ME}\langle\Psi|m\rangle_{{\bf R}^{'}}+\sum^{\infty}_{s=0}\sum^{\infty}_{m_{\bf p_{in}}=0}{}_{{\bf R}^{'}}\langle s,m_{\bf p_{in}}|\Psi\rangle_{\bf ME}{}_{\bf ME}\langle\Psi|s,m_{\bf p_{in}}\rangle_{{\bf R}^{'}}\nonumber\\
    &=& \Bigg\{\sum^{\infty}_{m_{\bf p_{in}}=0}\rho_{m_{\bf p_{in}}}+\sum^{\infty}_{m_{\bf p_{in}}=0}\sum^{\infty}_{s=0}\rho_{m_{\bf p_{in}},s}\Bigg\},\eea
   where we define $\rho_{m_{\bf p_{in}}}$ and $\rho_{m_{\bf p_{in}},s}$ by the following expressions:   
    \bea \rho_m &=& \frac{\left(1-|\gamma_{p_{\bf in}}|^2\right)}{2\left(1+f_{p_{\bf in}}\right)}|\gamma_{p_{\bf in}}|^{2m_{\bf p_{in}}}\Bigg\{|0_{\bf p_{out}}\rangle_{\bf BD_1}|m_{\bf p_{in}}\rangle_{{\bf L}^{'}}~{}_{\bf BD_1}\langle 0|{}_{{\bf L}^{'}}\langle m_{\bf p_{in}}|\nonumber\\
    &&+\Delta^{*}_2\gamma_{p_{\bf in}}\sqrt{m_{\bf p_{in}}+1}|0_{\bf p_{out}}\rangle_{\bf BD_1}|m_{\bf p_{in}}+1\rangle_{{\bf L}^{'}}~{}_{\bf BD_1}\langle 1|{}_{{\bf L}^{'}}\langle m_{\bf p_{in}}|\nonumber\\
    &&+\Delta_2\gamma^{*}_{p_{\bf in}}\sqrt{m_{\bf p_{in}}+1}|1_{\bf p_{out}}\rangle_{\bf BD_1}|m_{\bf p_{in}}\rangle_{{\bf L}^{'}}~{}_{\bf BD_1}\langle 0|{}_{{\bf L}^{'}}\langle m_{\bf p_{in}}+1|\nonumber\\
    &&+|\Delta_2|^{2}(m_{\bf p_{in}}+1)|1_{\bf p_{out}}\rangle_{\bf BD_1}|m_{\bf p_{in}}\rangle_{{\bf L}^{'}}~{}_{\bf BD_1}\langle 1|{}_{{\bf L}^{'}}\langle m_{\bf p_{in}}|\nonumber\\
    &&+\Delta^{*}_1 \sqrt{m_{\bf p_{in}}+1}|0_{\bf p_{out}}\rangle_{\bf BD_1}|m_{\bf p_{in}}\rangle_{{\bf L}^{'}}~{}_{\bf BD_1}\langle 1|{}_{{\bf L}^{'}}\langle m_{\bf p_{in}}+1|\nonumber\\
    &&+\Delta_1 \sqrt{m_{\bf p_{in}}+1}|1_{\bf p_{out}}\rangle_{\bf BD_1}|m_{\bf p_{in}}+1\rangle_{{\bf L}^{'}}~{}_{\bf BD_1}\langle 0|{}_{{\bf L}^{'}}\langle m_{\bf p_{in}}|\nonumber\\
    &&+\Delta^{*}_1 \Delta_2 \gamma^{*}_{p_{\bf in}}\sqrt{(m_{\bf p_{in}}+1)(m_{\bf p_{in}}+2)}|1_{\bf p_{out}}\rangle_{\bf BD_1}|m_{\bf p_{in}}\rangle_{{\bf L}^{'}}~{}_{\bf BD_1}\langle 1|{}_{{\bf L}^{'}}\langle m_{\bf p_{in}}+2|\nonumber\\
    &&+\Delta_1 \Delta^{*}_2 \gamma_{p_{\bf in}}\sqrt{(m_{\bf p_{in}}+1)(m_{\bf p_{in}}+2)}|1_{\bf p_{out}}\rangle_{\bf BD_1}|m_{\bf p_{in}}+2\rangle_{{\bf L}^{'}}~{}_{\bf BD_1}\langle 1|{}_{{\bf L}^{'}}\langle m_{\bf p_{in}}|\nonumber\\
    &&+|\Delta_1|^{2}(m_{\bf p_{in}}+1)|1_{\bf p_{out}}\rangle_{\bf BD_1}|m_{\bf p_{in}}+1\rangle_{{\bf L}^{'}}~{}_{\bf BD_1}\langle 1|{}_{{\bf L}^{'}}\langle m_{\bf p_{in}}+1|\Bigg\},\eea
    and 
    \bea
   \rho_{m,s} &=&\frac{f^{2}_{p_{\bf in}}}{2\left(1+f_{p_{\bf in}}\right)}|\Gamma_{{p_{\bf in}},s}|^{2m_{\bf p_{in}}}\Bigg\{|0_{\bf p_{out}}\rangle_{\bf BD_1}|s,m_{\bf p_{in}}\rangle_{{\bf L}^{'}}~{}_{\bf BD_1}\langle 0|{}_{{\bf L}^{'}}\langle s,m_{\bf p_{in}}|\nonumber\\
    &&+\Delta^{*}_{4,s}\Gamma_{{p_{\bf in}},s}\sqrt{m_{\bf p_{in}}+1}|0_{\bf p_{out}}\rangle_{\bf BD_1}|s,(m_{\bf p_{in}}+1)\rangle_{{\bf L}^{'}}~{}_{\bf BD_1}\langle 1|{}_{{\bf L}^{'}}\langle s,m_{\bf p_{in}}|\nonumber\\
    &&+\Delta_{4,s}\Gamma^{*}_{{p_{\bf in}},s}\sqrt{m_{\bf p_{in}}+1}|1_{\bf p_{out}}\rangle_{\bf BD_1}|s,m_{\bf p_{in}}\rangle_{{\bf L}^{'}}~{}_{\bf BD_1}\langle 0|{}_{{\bf L}^{'}}\langle s, (m_{\bf p_{in}}+1)|\nonumber\\
    &&+|\Delta_{4,s}|^{2}(m_{\bf p_{in}}+1)|1_{\bf p_{out}}\rangle_{\bf BD_1}|s,m_{\bf p_{in}}\rangle_{{\bf L}^{'}}~{}_{\bf BD_1}\langle 1|{}_{{\bf L}^{'}}\langle s,m_{\bf p_{in}}|\nonumber\\
    &&+\Delta^{*}_{3,s} \sqrt{m_{\bf p_{in}}+1}|0_{\bf p_{out}}\rangle_{\bf BD_1}|s,m_{\bf p_{in}}\rangle_{{\bf L}^{'}}~{}_{\bf BD_1}\langle 1|{}_{{\bf L}^{'}}\langle s,(m_{\bf p_{in}}+1)|\nonumber\\
    &&+\Delta_{3,s} \sqrt{m_{\bf p_{in}}+1}|1_{\bf p_{out}}\rangle_{\bf BD_1}|s,m_{\bf p_{in}}+1\rangle_{{\bf L}^{'}}~{}_{\bf BD_1}\langle 0|{}_{{\bf L}^{'}}\langle s,m_{\bf p_{in}}|\nonumber\\
    &&+\Delta^{*}_1 \Delta_2 \Gamma^{*}_{{p_{\bf in}},s}\sqrt{(m_{\bf p_{in}}+1)(m_{\bf p_{in}}+2)}|1_{\bf p_{out}}\rangle_{\bf BD_1}|s,m_{\bf p_{in}}\rangle_{{\bf L}^{'}}~{}_{\bf BD_1}\langle 1|{}_{{\bf L}^{'}}\langle s,(m_{\bf p_{in}}+2)|\nonumber\\
    &&+\Delta_{3,s} \Delta^{*}_{4,s} \Gamma_{{p_{\bf in}},s}\sqrt{(m_{\bf p_{in}}+1)(m_{\bf p_{in}}+2)}|1_{\bf p_{out}}\rangle_{\bf BD_1}|s,(m_{\bf p_{in}}+2)\rangle_{{\bf L}^{'}}~{}_{\bf BD_1}\langle 1|{}_{{\bf L}^{'}}\langle s,m_{\bf p_{in}}|\nonumber\\
    &&+|\Delta_{3,s}|^{2}(m_{\bf p_{in}}+1)|1_{\bf p_{out}}\rangle_{\bf BD_1}|s,(m_{\bf p_{in}}+1)\rangle_{{\bf L}^{'}}~{}_{\bf BD_1}\langle 1|{}_{{\bf L}^{'}}\langle s,(m_{\bf p_{in}}+1)|\Bigg\}.\eea
The internal observer is essentially described by the quantum mechanical state that emerges in this situation for both the complementary and specific integral parts, where the corresponding observer is positioned in one of the regions of the open chart of the global de Sitter space. Furthermore, it is significant to remember that all mode eigen values for the complementary and specific integral parts are identical and marked with the symbol $m_{p_{\bf in}}$.  This is because of the fact that in the particular integral part the index $s$ which is appearing due to putting source term is not going to effect the eigen values of the mode function at the end of the day.  Though one can tag the corresponding quantum states of the particular integral part with $s$ and  $m_{p_{\bf in}}$,  to make a distinction from the quantum modes of the complementary part which is tagged by only the quantum number $m_{p_{\bf in}}$.  This is a very crucial point which is very important to mention at this stage of computation to avoid all further unnecessary confusion.
      \subsection{Partial transpose operation and the comment on the negative eigenvalues}
      
      Identifying the negative eigenvalues of the obtained formula for the reduced density matrix discussed before is the main objective of this section. In order to perform this computation, we separate the contributions from the complementary section and the particular integral component. Here, we focus on the partial transpose operation with respect to the component that corresponds to the first quantum vacuum state of Bunch Davies, which is given by the following expressions:
         \bea \rho^{T,{\bf BD_1}}_m &=&\frac{\left(1-|\gamma_{p_{\bf in}}|^2\right)}{2\left(1+f_{p_{\bf in}}\right)}|\gamma_{p_{\bf in}}|^{2m_{\bf p_{in}}}\Bigg\{|0_{\bf p_{out}}\rangle_{\bf BD_1}|m_{\bf p_{in}}\rangle_{{\bf L}^{'}}~{}_{\bf BD_1}\langle 0|{}_{{\bf L}^{'}}\langle m_{\bf p_{in}}|\nonumber\\
    &&+\Delta^{*}_2\gamma_{p_{\bf in}}\sqrt{m_{\bf p_{in}}+1}|1_{\bf p_{out}}\rangle_{\bf BD_1}|m_{\bf p_{in}}+1\rangle_{{\bf L}^{'}}~{}_{\bf BD_1}\langle 0|{}_{{\bf L}^{'}}\langle m_{\bf p_{in}}|\nonumber\\
    &&+\Delta_2\gamma^{*}_{p_{\bf in}}\sqrt{m_{\bf p_{in}}+1}|0_{\bf p_{out}}\rangle_{\bf BD_1}|m_{\bf p_{in}}\rangle_{{\bf L}^{'}}~{}_{\bf BD_1}\langle 1|{}_{{\bf L}^{'}}\langle m_{\bf p_{in}}+1|\nonumber\\
    &&+|\Delta_2|^{2}(m_{\bf p_{in}}+1)|1_{\bf p_{out}}\rangle_{\bf BD_1}|m_{\bf p_{in}}\rangle_{{\bf L}^{'}}~{}_{\bf BD_1}\langle 1|{}_{{\bf L}^{'}}\langle m_{\bf p_{in}}|\nonumber\\
    &&+\Delta^{*}_1 \sqrt{m_{\bf p_{in}}+1}|1_{\bf p_{out}}\rangle_{\bf BD_1}|m_{\bf p_{in}}\rangle_{{\bf L}^{'}}~{}_{\bf BD_1}\langle 0|{}_{{\bf L}^{'}}\langle m_{\bf p_{in}}+1|\nonumber\\
    &&+\Delta_1 \sqrt{m_{\bf p_{in}}+1}|0_{\bf p_{out}}\rangle_{\bf BD_1}|m_{\bf p_{in}}+1\rangle_{{\bf L}^{'}}~{}_{\bf BD_1}\langle 1|{}_{{\bf L}^{'}}\langle m_{\bf p_{in}}|\nonumber\\
    &&+\Delta^{*}_1 \Delta_2 \gamma^{*}_{p_{\bf in}}\sqrt{(m_{\bf p_{in}}+1)(m_{\bf p_{in}}+2)}|1_{\bf p_{out}}\rangle_{\bf BD_1}|m_{\bf p_{in}}\rangle_{{\bf L}^{'}}~{}_{\bf BD_1}\langle 1|{}_{{\bf L}^{'}}\langle m_{\bf p_{in}}+2|\nonumber\\
    &&+\Delta_1 \Delta^{*}_2 \gamma_{p_{\bf in}}\sqrt{(m_{\bf p_{in}}+1)(m_{\bf p_{in}}+2)}|1_{\bf p_{out}}\rangle_{\bf BD_1}|m_{\bf p_{in}}+2\rangle_{{\bf L}^{'}}~{}_{\bf BD_1}\langle 1|{}_{{\bf L}^{'}}\langle m_{\bf p_{in}}|\nonumber\\
    &&+|\Delta_1|^{2}(m_{\bf p_{in}}+1)|1_{\bf p_{out}}\rangle_{\bf BD_1}|m_{\bf p_{in}}+1\rangle_{{\bf L}^{'}}~{}_{\bf BD_1}\langle 1|{}_{{\bf L}^{'}}\langle m_{\bf p_{in}}+1|\Bigg\},\eea
    and 
    \bea
   \rho^{T,{\bf BD_1}}_{m,s} &=&\frac{f^{2}_{p_{\bf in}}}{2\left(1+f_{p_{\bf in}}\right)}|\Gamma_{{p_{\bf in}},s}|^{2m_{\bf p_{in}}}\Bigg\{|0_{\bf p_{out}}\rangle_{\bf BD_1}|s,m_{\bf p_{in}}\rangle_{{\bf L}^{'}}~{}_{\bf BD_1}\langle 0|{}_{{\bf L}^{'}}\langle s,m_{\bf p_{in}}|\nonumber\\
    &&+\Delta^{*}_{4,s}\Gamma_{{p_{\bf in}},s}\sqrt{m_{\bf p_{in}}+1}|1_{\bf p_{out}}\rangle_{\bf BD_1}|s,(m_{\bf p_{in}}+1)\rangle_{{\bf L}^{'}}~{}_{\bf BD_1}\langle 0|{}_{{\bf L}^{'}}\langle s,m_{\bf p_{in}}|\nonumber\\
    &&+\Delta_{4,s}\Gamma^{*}_{{p_{\bf in}},s}\sqrt{m_{\bf p_{in}}+1}|0_{\bf p_{out}}\rangle_{\bf BD_1}|s,m_{\bf p_{in}}\rangle_{{\bf L}^{'}}~{}_{\bf BD_1}\langle 1|{}_{{\bf L}^{'}}\langle s, (m_{\bf p_{in}}+1)|\nonumber\\
    &&+|\Delta_{4,s}|^{2}(m_{\bf p_{in}}+1)|1_{\bf p_{out}}\rangle_{\bf BD_1}|s,m_{\bf p_{in}}\rangle_{{\bf L}^{'}}~{}_{\bf BD_1}\langle 1|{}_{{\bf L}^{'}}\langle s,m_{\bf p_{in}}|\nonumber\\
    &&+\Delta^{*}_{3,s} \sqrt{m_{\bf p_{in}}+1}|1_{\bf p_{out}}\rangle_{\bf BD_1}|s,m_{\bf p_{in}}\rangle_{{\bf L}^{'}}~{}_{\bf BD_1}\langle 0|{}_{{\bf L}^{'}}\langle s,(m_{\bf p_{in}}+1)|\nonumber\\
    &&+\Delta_{3,s} \sqrt{m_{\bf p_{in}}+1}|0_{\bf p_{out}}\rangle_{\bf BD_1}|s,m_{\bf p_{in}}+1\rangle_{{\bf L}^{'}}~{}_{\bf BD_1}\langle 1|{}_{{\bf L}^{'}}\langle s,m_{\bf p_{in}}|\nonumber\\
    &&+\Delta^{*}_{3,s} \Delta_{4,s} \Gamma^{*}_{{p_{\bf in}},s}\sqrt{(m_{\bf p_{in}}+1)(m_{\bf p_{in}}+2)}|1_{\bf p_{out}}\rangle_{\bf BD_1}|s,m_{\bf p_{in}}\rangle_{{\bf L}^{'}}~{}_{\bf BD_1}\langle 1|{}_{{\bf L}^{'}}\langle s,(m_{\bf p_{in}}+2)|\nonumber\\
    &&+\Delta_{3,s} \Delta^{*}_{4,s} \Gamma_{{p_{\bf in}},s}\sqrt{(m_{\bf p_{in}}+1)(m_{\bf p_{in}}+2)}|1_{\bf p_{out}}\rangle_{\bf BD_1}|s,(m_{\bf p_{in}}+2)\rangle_{{\bf L}^{'}}~{}_{\bf BD_1}\langle 1|{}_{{\bf L}^{'}}\langle s,m_{\bf p_{in}}|\nonumber\\
    &&+|\Delta_{3,s}|^{2}(m_{\bf p_{in}}+1)|1_{\bf p_{out}}\rangle_{\bf BD_1}|s,(m_{\bf p_{in}}+1)\rangle_{{\bf L}^{'}}~{}_{\bf BD_1}\langle 1|{}_{{\bf L}^{'}}\langle s,(m_{\bf p_{in}}+1)|\Bigg\}.\eea
           Now if from the above mentioned partial transposed version of the reduced density matrices computed from the complementary and particular integral part after taking addition and summing over $s$ if we found that at least one eigenvalue is negative,  then we can conclude from our theoretical set up that quantum mechanical states corresponding to the inside and outside observers are entangled.

           Let's first express the transposed version of the reduced density matrices derived from the complementary and particular integral component in square matrix form before continuing with the computation, which is given by the following expressions:
     \bea \rho^{T,{\bf BD_1}}_{m}&=&\frac{\left(1-|\gamma_{p_{\bf in}}|^2\right)}{2\left(1+f_{p_{\bf in}}\right)}|\gamma_{p_{\bf in}}|^{2m_{\bf p_{in}}}\left(\begin{array}{ccc} A_{m_{\bf p_{in}}} &~~~B_{m_{\bf p_{in}}} &~~~C_{m_{\bf p_{in}}}\\ B^{*}_{m_{\bf p_{in}}} &~~~ D_{m_{\bf p_{in}}} &~~~ 0\\ C^{*}_{m_{\bf p_{in}}} &~~~ 0 &~~~ 0  \end{array}\right),\\\rho^{T,{\bf BD_1}}_{m_{\bf p_{in}},s}&=&\frac{f^{2}_{p_{\bf in}}}{2\left(1+f_{p_{\bf in}}\right)}|\Gamma_{{p_{\bf in}},s}|^{2m_{\bf p_{in}}}\left(\begin{array}{ccc} A_{m_{\bf p_{in}},s} &~~~B_{m_{\bf p_{in}},s} &~~~C_{m_{\bf p_{in}},s}\\ B^{*}_{m_{\bf p_{in}},s} &~~~ D_{m_{\bf p_{in}},s} &~~~ 0\\ C^{*}_{m_{\bf p_{in}},s} &~~~ 0 &~~~ 0  \end{array}\right),\eea
           where we define each of entries of the above mentioned square matrices by the following expressions:
           \bea A_{m_{\bf p_{in}}} &=& 1+|\Delta_2|^{2}(m_{\bf p_{in}}+1),\\
           B_{m_{\bf p_{in}}} &=& \sqrt{m_{\bf p_{in}}+1}\left(\Delta_2 \gamma^{*}_{p_{\bf in}}+\Delta^{*}_1\right),\\
           C_{m_{\bf p_{in}}} &=&  \sqrt{(m_{\bf p_{in}}+1)(m_{\bf p_{in}}+2)}~\Delta^{*}_1\Delta_2 \gamma^{*}_{p_{\bf in}},\\
          D_{m_{\bf p_{in}}} &=& |\Delta_1|^{2}(m_{\bf p_{in}}+1),\eea
          and 
           \bea A_{m_{\bf p_{in}},s} &=& 1+|\Delta_{4,s}|^{2}(m_{\bf p_{in}}+1),\\
           B_{m_{\bf p_{in}},s} &=& \sqrt{m_{\bf p_{in}}+1}\left(\Delta_{4,s} \Gamma^{*}_{{p_{\bf in}},s}+\Delta^{*}_{3,s}\right),\\
           C_{m_{\bf p_{in}},s} &=&  \sqrt{(m_{\bf p_{in}}+1)(m_{\bf p_{in}}+2)}~\Delta^{*}_{3,s}\Delta_{4,s} \Gamma^{*}_{{p_{\bf in}},s},\\
          D_{m_{\bf p_{in}},s} &=& |\Delta_{3,s}|^{2}(m_{\bf p_{in}}+1).\eea
          Our next job is to compute the eigenvalue equations from the total partial transposed  matrix,  which is given by the following expression:
          \bea && \widetilde{\lambda}^{3}_{m_{\bf p_{in}}}-\overline{A}_{m_{\bf p_{in}}}\widetilde{\lambda}^{2}_{m_{\bf p_{in}}}+\overline{B}_{m_{\bf p_{in}}}\widetilde{\lambda}_{m_{\bf p_{in}}}+\overline{C}_{m_{\bf p_{in}}}=0.\eea
         where in the above mentioned two expressions we have introduced some shorthand redefined symbols which are given by the following expressions:
         \bea \overline{A}_{m_{\bf p_{in}}} &=&\frac{1}{2\left(1+f_{p_{\bf in}}\right)}\Bigg\{|\gamma_{p_{\bf in}}|^{2m_{\bf p_{in}}}\left(1-|\gamma_{p_{\bf in}}|^2\right) \left(A_{m_{\bf p_{in}}}+D_{m_{\bf p_{in}}}\right)\nonumber\\
         &&\quad\quad\quad\quad\quad\quad\quad\quad\quad\quad +f^{2}_{p_{\bf in}}\sum^{\infty}_{s=0}|\Gamma_{{p_{\bf in}},s}|^{2m}\left(A_{m_{\bf p_{in}},s}+D_{m_{\bf p_{in}},s}\right)\Bigg\},\quad\\
        \overline{B}_{m_{\bf p_{in}}} &=& \frac{1}{4\left(1+f_{p_{\bf in}}\right)^2}\Bigg\{|\gamma_{p_{\bf in}}|^{4m_{\bf p_{in}}}\left(1-|\gamma_{p_{\bf in}}|^2\right)^2\left(A_{m_{\bf p_{in}}} D_{m_{\bf p_{in}}}-\left(|B_{m_{\bf p_{in}}}|^{2}+|C_{m_{\bf p_{in}}}|^{2}\right)\right)\nonumber\\
        &&\quad\quad +f^{4}_{p_{\bf in}}\sum^{\infty}_{s=0}|\Gamma_{{p_{\bf in}},s}|^{4m_{\bf p_{in}}} \left(A_{m_{\bf p_{in}},s} D_{m_{\bf p_{in}},s}-\left(|B_{m_{\bf p_{in}},s}|^{2}+|C_{m_{\bf p_{in}},s}|^{2}\right)\right)\Bigg\},\\
        \overline{C}_{m_{\bf p_{in}}} &=&\frac{1}{8\left(1+f_{p_{\bf in}}\right)^3}\Bigg\{|\gamma_{p_{\bf in}}|^{6m_{\bf p_{in}}} \left(1-|\gamma_{p_{\bf in}}|^2\right)^3|C_{m_{\bf p_{in}}}|^{2}D_{m_{\bf p_{in}}}\nonumber\\
        &&\quad\quad\quad\quad\quad\quad\quad\quad\quad\quad+f^{6}_{p_{\bf in}}\sum^{\infty}_{s=0}|\Gamma_{{p_{\bf in}},s}|^{6m_{\bf p_{in}}}|C_{m_{\bf p_{in}},s}|^{2}D_{m_{\bf p_{in}},s}\Bigg\}.\eea
        The real root of the eigenvalue computed from the $(m_{\bf p_{in}},m_{\bf p_{in}}+1)$ block is given by:
        \bea \widetilde{\lambda}_{m_{\bf p_{in}}} &=&\frac{1}{3}\Bigg[\overline{A}_{m_{\bf p_{in}}}+\frac{f(\overline{A}_{m_{\bf p_{in}}},\overline{B}_{m_{\bf p_{in}}},\overline{C}_{m_{\bf p_{in}}})}{\sqrt[3]{2}} -\frac{\sqrt[3]{2} \left(3 \overline{B}_{m_{\bf p_{in}}}-\overline{A}^2_{m_{\bf p_{in}}}\right)}{f(\overline{A}_{m_{\bf p_{in}}},\overline{B}_{m_{\bf p_{in}}},\overline{C}_{m_{\bf p_{in}}})}\Bigg]. \eea
        where we define the newly defined function $f(\overline{A}_{m_{\bf p_{in}}},\overline{B}_{m_{\bf p_{in}}},\overline{C}_{m_{\bf p_{in}}})$ which is defined as:
        \bea f(\overline{A}_{m_{\bf p_{in}}},\overline{B}_{m_{\bf p_{in}}},\overline{C}_{m_{\bf p_{in}}}):&=&\Bigg[2 \overline{A}^3_{m_{\bf p_{in}}}-9 \overline{A}_{m_{\bf p_{in}}} \overline{B}_{m_{\bf p_{in}}}-27 \overline{C}_{m_{\bf p_{in}}}\nonumber\\
 &&+3 \sqrt{3} \Bigg\{18 \overline{A}_{m_{\bf p_{in}}}\overline{ B}_{m_{\bf p_{in}}}\overline{ C}_{m_{\bf p_{in}}}+4 \overline{B}^3_{m_{\bf p_{in}}}+27 \overline{C}^2_{m_{\bf p_{in}}}\nonumber\\
 &&-4 \overline{A}^3_{m_{\bf p_{in}}} \overline{C}_{m_{\bf p_{in}}}-\overline{A}^2_{m_{\bf p_{in}}} \overline{B}^2_{m_{\bf p_{in}}}\Bigg\}^{\displaystyle \frac{1}{2}} \Bigg]^{\displaystyle \frac{1}{3}}.\quad\eea      
      Then the logarithmic negativity from the present set up can be further computed as:
      \bea {\cal LN}&=&\ln\left(2\sum_{\widetilde{\lambda}_{m_{\bf p_{in}}}<0}\widetilde{\lambda}_{m_{\bf p_{in}}}+1\right)\nonumber\\
      &=&\ln\Bigg(\frac{2}{3}\Bigg[\overline{A}_{m_{\bf p_{in}}}+\frac{f(\overline{A}_{m_{\bf p_{in}}},\overline{B}_{m_{\bf p_{in}}},\overline{C}_{m_{\bf p_{in}}})}{\sqrt[3]{2}} -\frac{\sqrt[3]{2} \left(3 \overline{B}_{m_{\bf p_{in}}}-\overline{A}^2_{m_{\bf p_{in}}}\right)}{f(\overline{A}_{m_{\bf p_{in}}},\overline{B}_{m_{\bf p_{in}}},\overline{C}_{m_{\bf p_{in}}})}\Bigg]+1\Bigg).\quad\quad\quad\eea
 In the next subsection we have studied this possibility numerically to extract the unknown facts from the present set up.
  \subsection{Computation of logarithmic negativity: Numerical study}
    \begin{figure*}[htb]
    \centering
    \subfigure[For $f_{p_{\bf in}}=0$.]{
        \includegraphics[width=14.2cm,height=8.5cm] {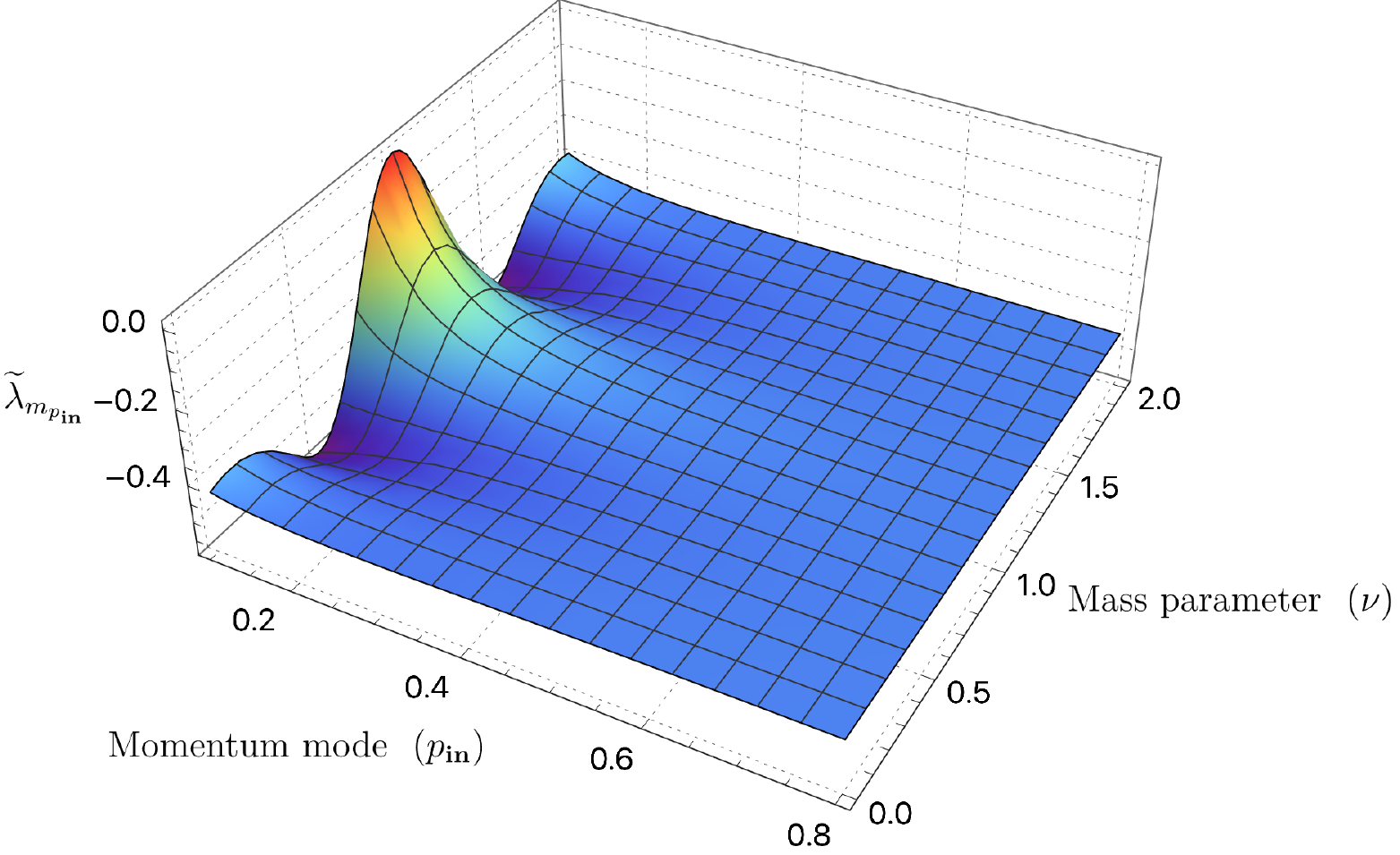}
        \label{EN1}
    }
    \subfigure[For small $f_{p_{\bf in}}\neq 0$.]{
        \includegraphics[width=14.2cm,height=8.5cm] {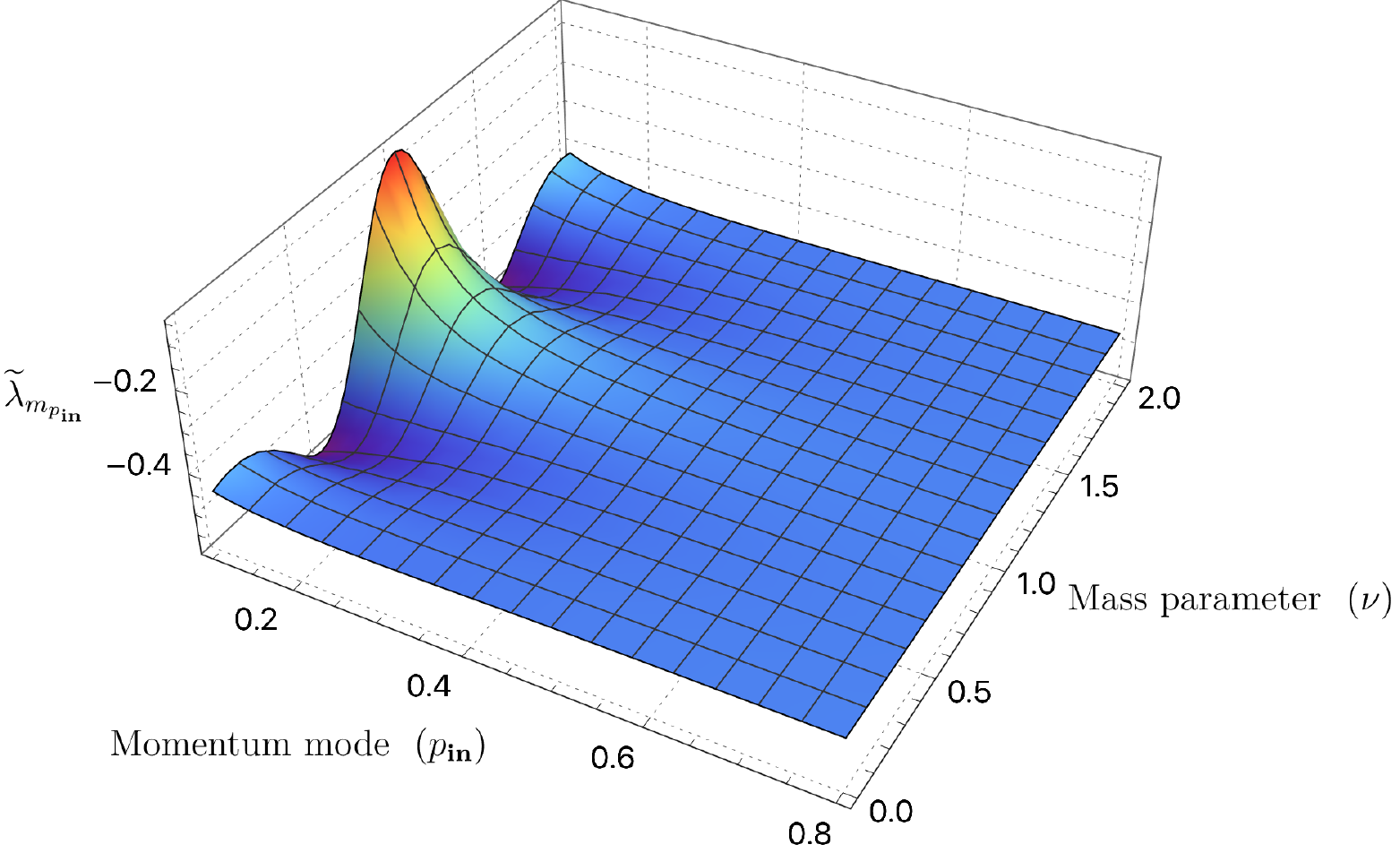}
        \label{EN2}
       }
    \caption[Optional caption for list of figures]{Representative 3D plot of the eignenvalue of the partial transposed matrix with the mass parameter and the corresponding momentum mode associated with the computation.  Here the partial transpose operation is taken with respect to the quantum vacuum state of the first Bunch Davies state which is characterizing the corresponding open chart of the global de Sitter space. } 
    \label{3D}
    \end{figure*}  

    \begin{figure*}[htb]
    \centering
    \subfigure[For $f_{p_{\bf in}}=0$.]{
        \includegraphics[width=14.2cm,height=8.9cm] {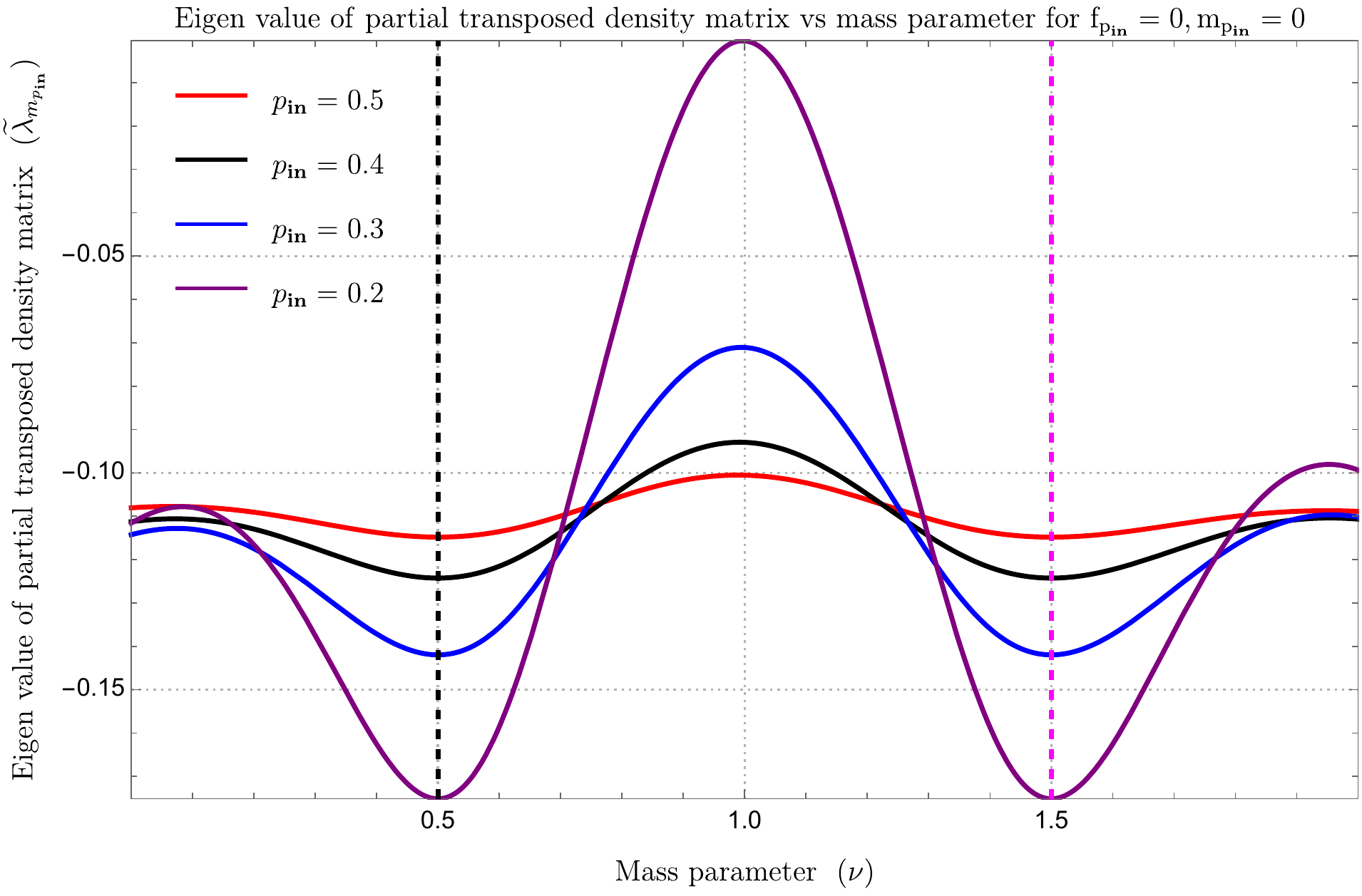}
        \label{EN1}
    }
    \subfigure[For small $f_{p_{\bf in}}\neq 0$.]{
        \includegraphics[width=14.2cm,height=8.9cm] {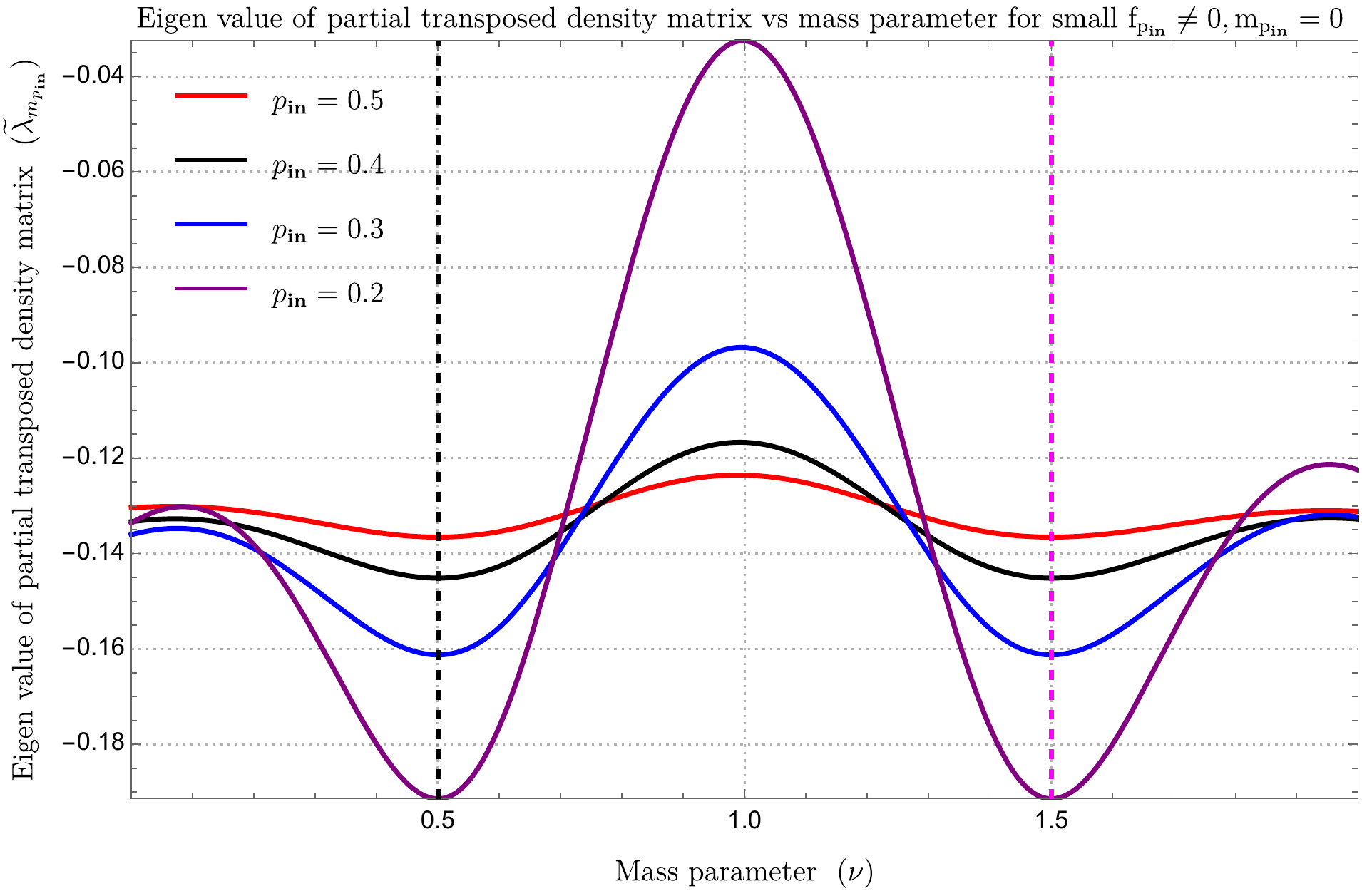} 
        \label{EN2}
       }
    \caption[Optional caption for list of figures]{Graphical behaviour of the of the eignenvalue of the partial transposed matrix with the mass parameter for given value of the momentum mode associated with the computation.  } 
    \label{EN}
    \end{figure*}  
    \begin{figure*}[htb]
    \centering
    \subfigure[For $f_{p_{\bf in}}=0$.]{
        \includegraphics[width=14.2cm,height=8.9cm] {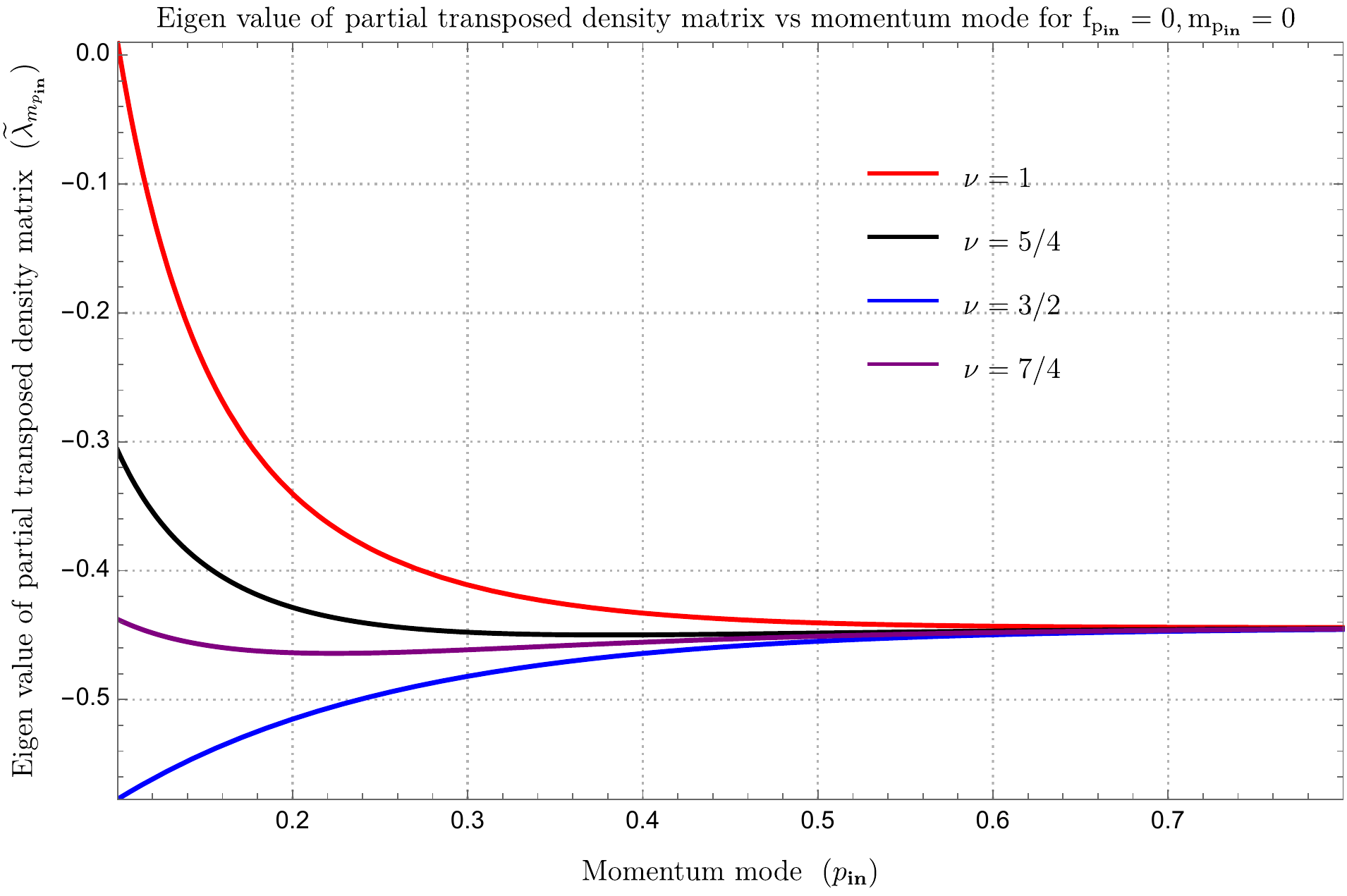}
        \label{Ep1}
    }
    \subfigure[For small $f_{p_{\bf in}}\neq 0$.]{
        \includegraphics[width=14.2cm,height=8.9cm] {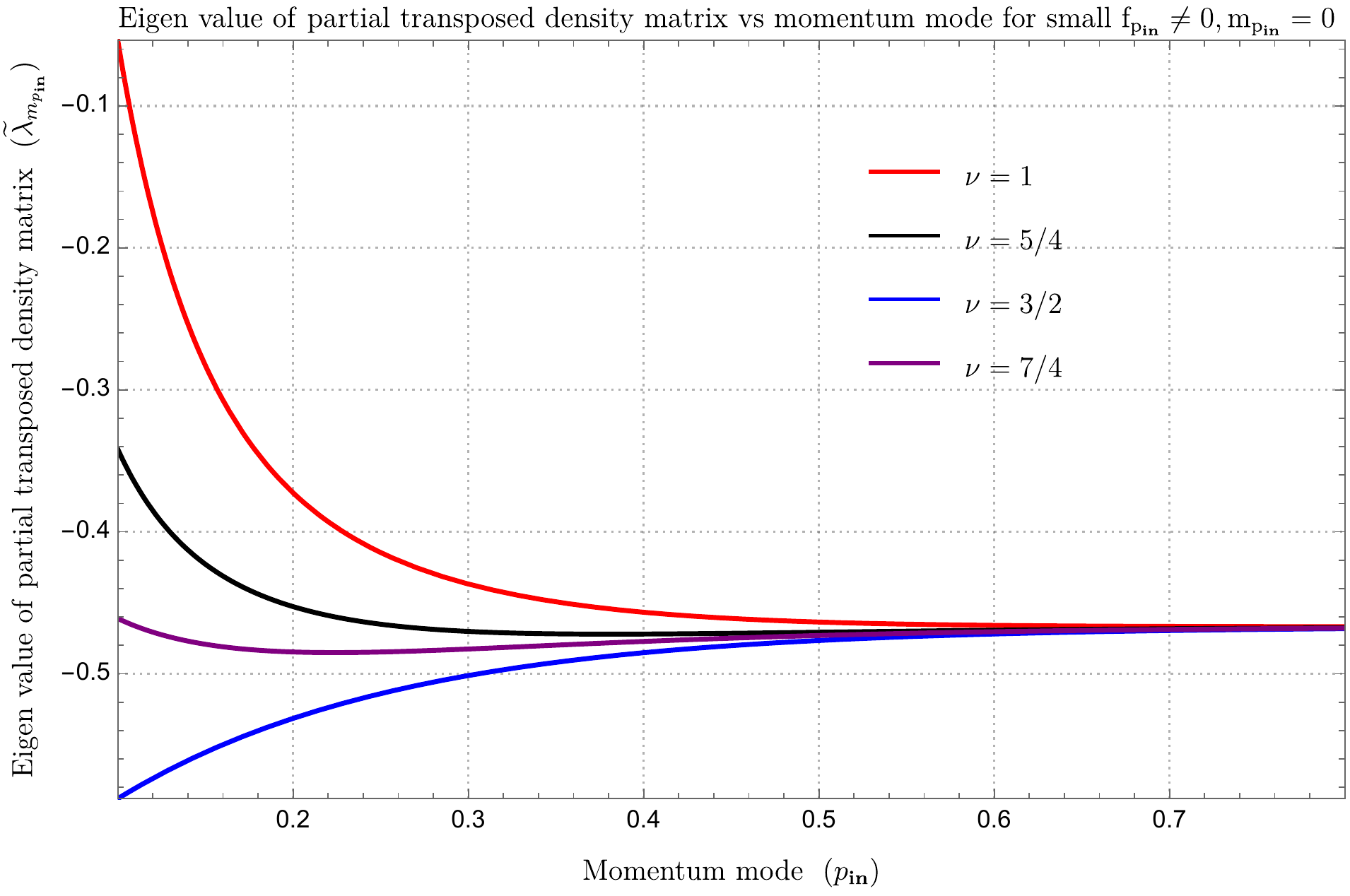}
        \label{Ep2}
       }
    \caption[Optional caption for list of figures]{Graphical behaviour of the of the eignenvalue of the partial transposed matrix with the momentum mode associated with the computation for the given value of mass parameter.  }  
    \label{EP}
    \end{figure*}  
     
    \begin{figure*}[htb]
    \centering
    \subfigure[For $f_{p_{\bf in}}=0$.]{
        \includegraphics[width=14.2cm,height=8.9cm] {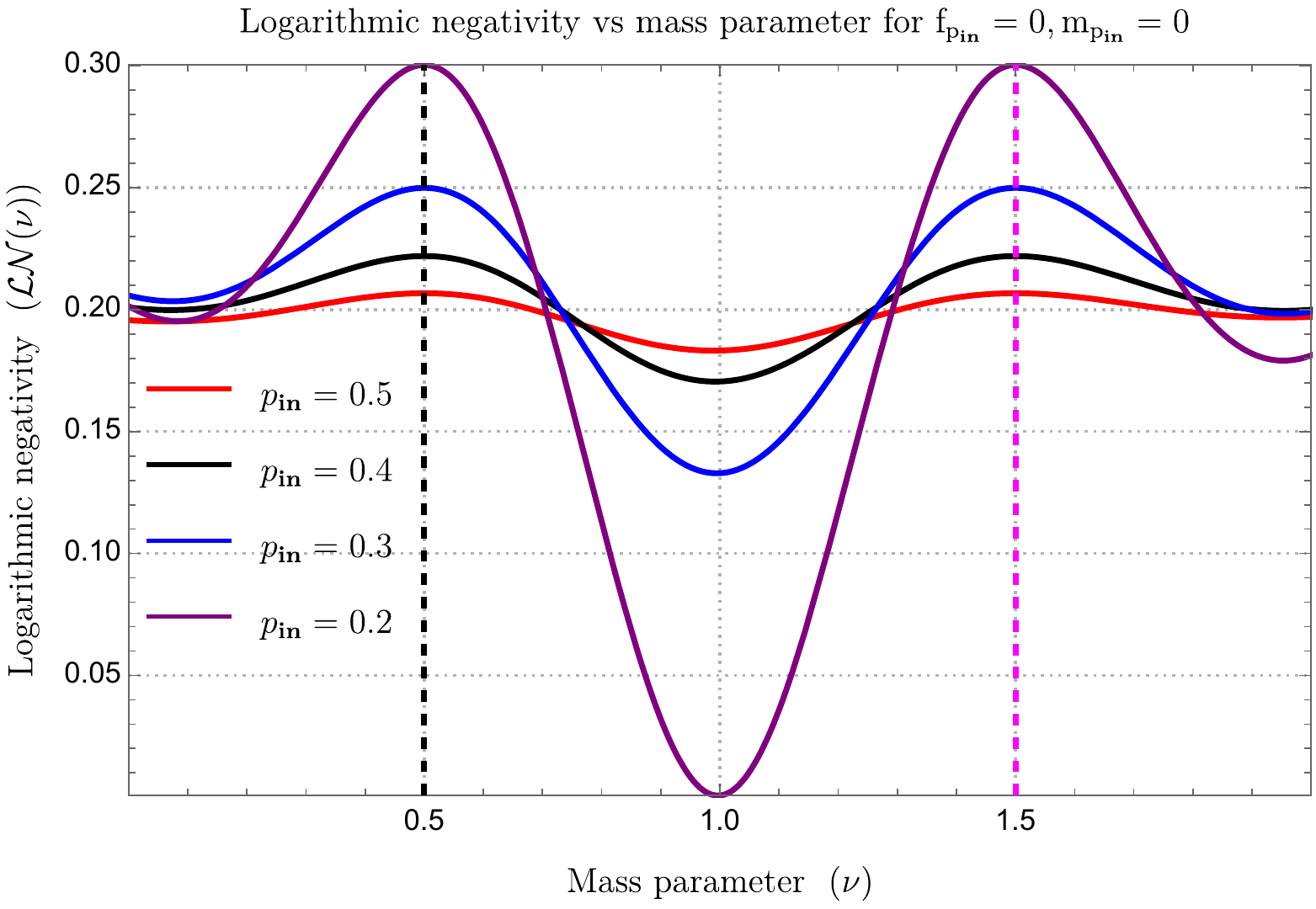}
        \label{LEN1}
    }
    \subfigure[For small $f_{p_{\bf in}}\neq 0$.]{
        \includegraphics[width=14.2cm,height=8.9cm] {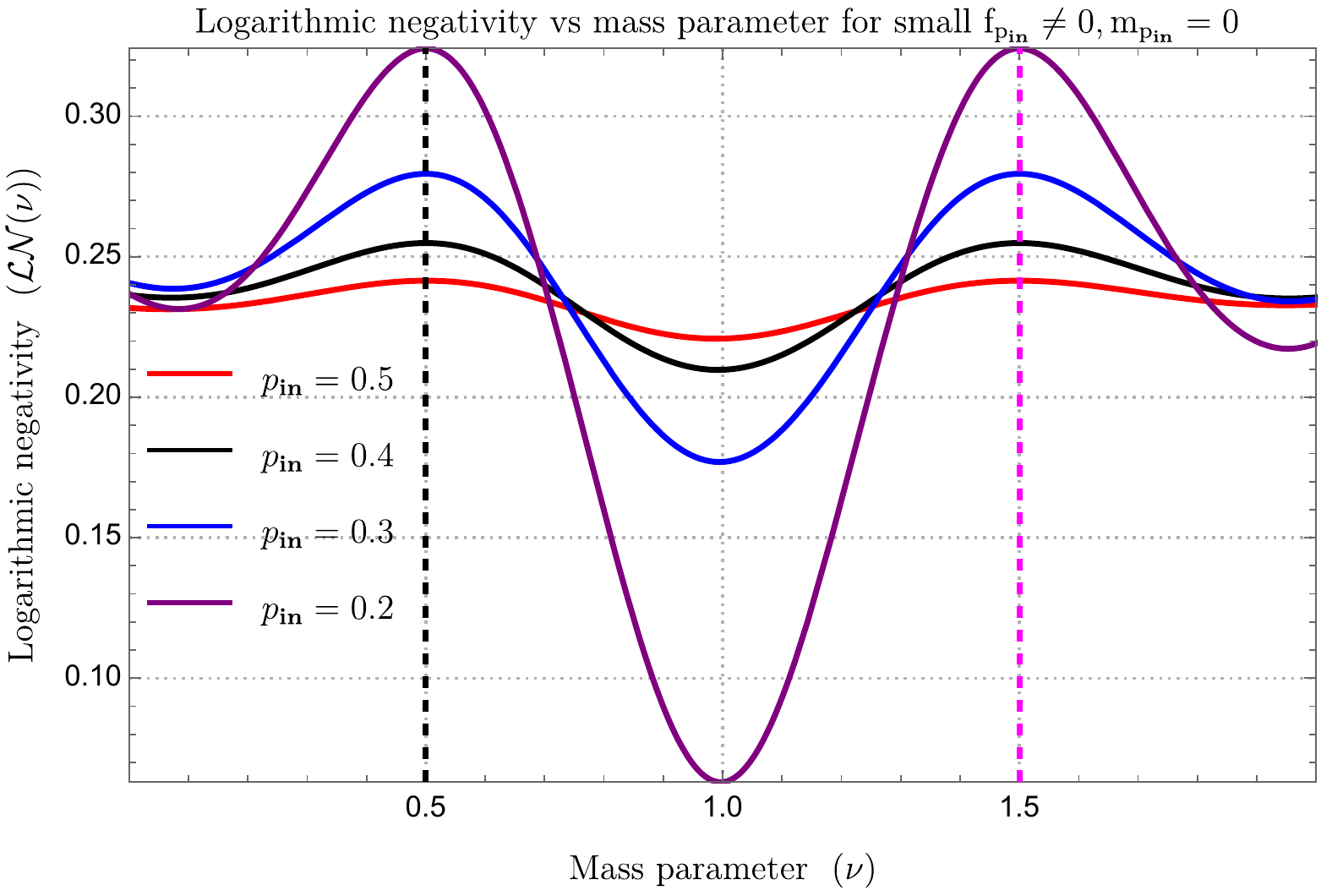}
        \label{LEN2}
       }
    \caption[Optional caption for list of figures]{Graphical behaviour of the of the logarithmic negativity computed from the negative eigenvalues of the partial transposed reduced density matrix with the mass parameter associated with the computation for the given value of momentum mode.   } 
    \label{LEN}
    \end{figure*}  
    \begin{figure*}[htb]
    \centering
    \subfigure[For $f_{p_{\bf in}}=0$.]{
        \includegraphics[width=14.2cm,height=8.9cm] {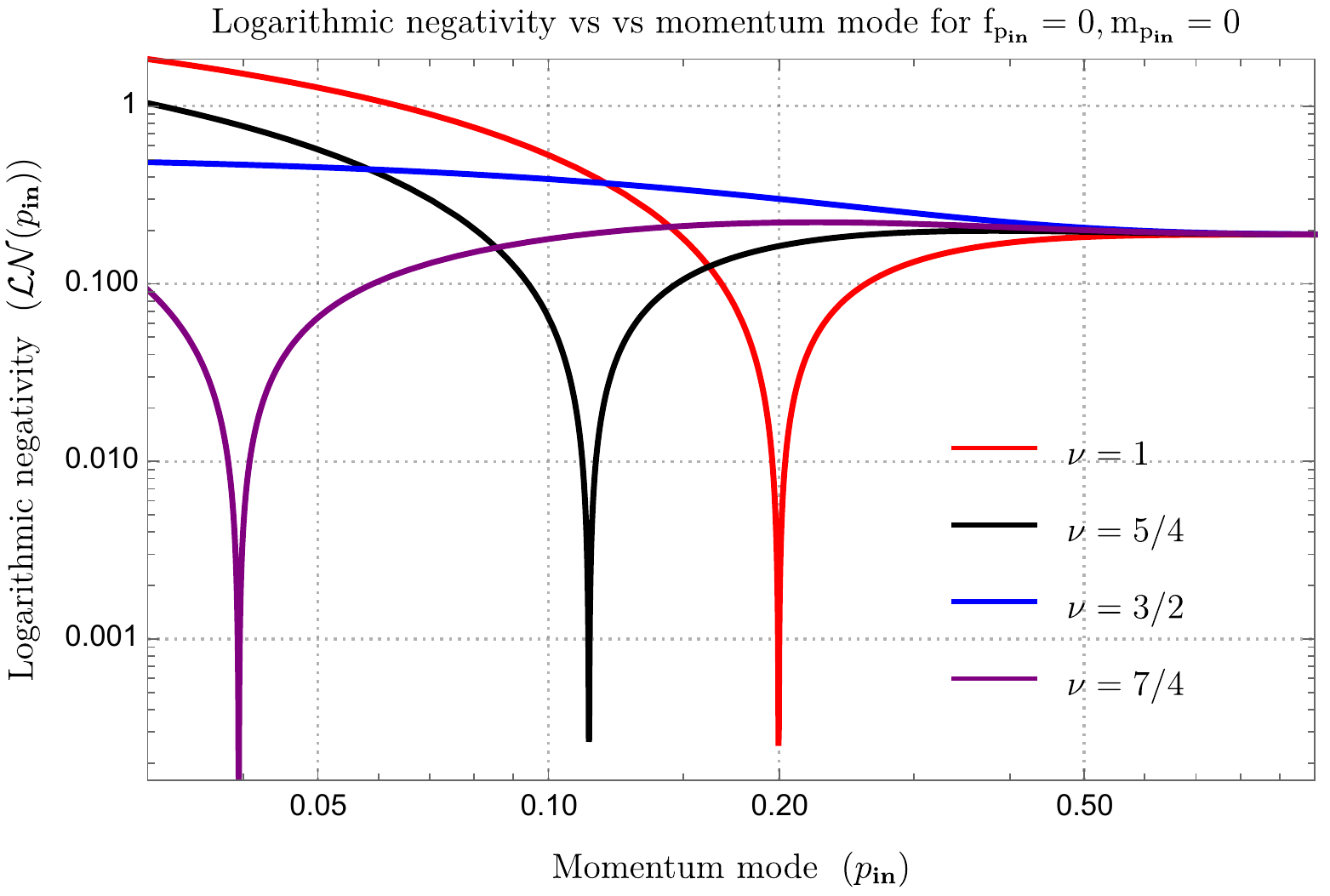}
        \label{LNp1}
    }
    \subfigure[For small $f_{p_{\bf in}}\neq 0$.]{
        \includegraphics[width=14.2cm,height=8.9cm] {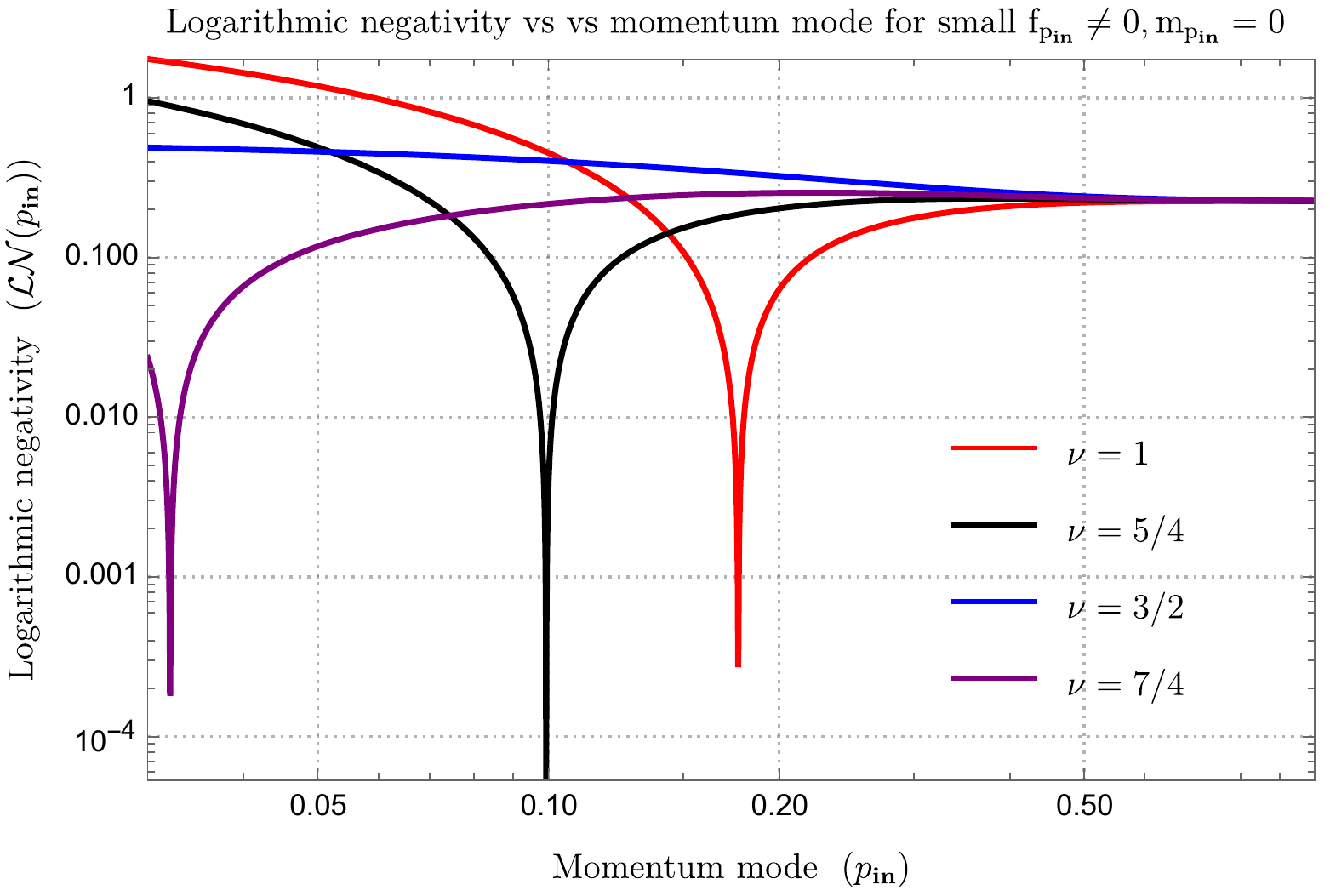}
        \label{LNp2}
       }
    \caption[Optional caption for list of figures]{Graphical behaviour of the of the logarithmic negativity computed from the negative eigenvalues of the partial transposed reduced density matrix with the momentum mode associated with the computation for the given value of mass parameter. } 
    \label{LNp}
    \end{figure*}  
      In figure (\ref{3D}(a)) and (\ref{3D}(b)),  we have depicted the representative 3D plot of the eignenvalue of the partial transposed matrix with the mass parameter and the corresponding momentum mode associated with the computation.  In these plots we have considered two possibilities,  vanishing $f_{p_{\bf in}}=0$ and a very small value but $f_{p_{\bf in}}\neq 0$.  Here the partial transpose operation is taken with respect to the quantum vacuum state of the first Bunch Davies state which is characterizing the corresponding open chart of the global de Sitter space.  The outcomes and the physical interpretation of these plots are very interesting which we are writing point-wise in the following:
     \begin{enumerate}
     \item   In these plots we can clearly observe that the corresponding eignevalue that we have plotted for the mode having $m_{\bf p_{in}}=0$ gives negative contributions for both figure (\ref{3D}(a)) and (\ref{3D}(b)).  This is a clear signature of having quantum mechanical entanglement in the biverse as well as well as the multiverse picture that we have theoretically constructed in this paper. 
     
     \item   The other higher modes $m_{\bf p_{in}}>0$ gives positive contribution to the eigenvalue for which we have not incorporated those plots in this paper.  Most importantly such contributions will not be in support of quantum entanglement in the present theoretical picture,  so those solutions are not interesting in the present context.  
     
     \item  Comparing both of these plots we can also observe that for $f_{p_{\bf in}}=0$ we get less quantum entanglement compared to the case that we have studied for small $f_{p_{\bf in}}\neq 0$. 
     
     \item Also we have found that for heavy mass field which is governed by $\nu=-i|\nu|$ or $\nu^2<0$ the corresponding eigenvalue of the partial transposed version of the reduced density matrix is highly positive.  Since this fact is not the desirable one,  we have not included such plots in this paper.  But it is strictly confirmed that to get high amount of entanglement heavy field is not desirable in the present prescribed framework.  For this reason we have only shown the plots for massless or partially massless fields which can be studied using the solution for the branch $\nu>0$. 
     
     \item From both of these plots we can clearly observe that for the values of the mass parameter $\nu=1/2$ (conformal coupling) and $\nu=3/2$ (massless case) there are two dips in the eigenvalue spectrum,   which correspond to maximum entanglement from the theoretical set up under consideration.  On the other hand,  in both of these plots we see that at $\nu=1$ there is minimum contribution from the quantum entanglement compared to the previously mentioned two values of the mass parameter $\nu$.
     
     \item For any other values of the mass parameter lying within the window $0<\nu<2$ we get the intermediate amount of quantum mechanical entanglement.  Particularly,  it is important to note that for $\nu=0$ the amount of entanglement is larger than the amount of entanglement obtained for the value $\nu=2$.
     
     \item It seems like that the eigenvalue spectrum is distributed symmetrically around the value of the mass parameter $\nu=1$.  But actually this is not the case.  Crucial observation suggests that the hight of the spectrum is lesser at $\nu=0$ compared to $\nu=2$.
     
     \item In both of these plots we have restricted our parameter space within the region $0<p_{\bf in}<0.8$ and $0<\nu<2$,   which we found the most suitable parameter space to get the negative contribution from the eigenvalue spectrum for the mode $m_{\bf p_{in}}=0$.
     \end{enumerate}
     
    In figure (\ref{EN}(a)) and (\ref{EN}(b)),  we have depicted the representative graphical behaviour of the of the eignenvalue of the partial transposed matrix with the mass parameter for given value of the momentum mode associated with the computation.  In these plots we have considered two possibilities,  vanishing $f_{p_{\bf in}}=0$ and a very small value but $f_{p_{\bf in}}\neq 0$.  In both of these plots we have fixed the value of the momentum mode within the region $0.2<p_{\bf in}<0.5$ and have studied the behaviour of the eigenvalue spectrum with respect to mass parameter within the range $0<\nu<2$.  In our computation we found that this is the most desirable window of parameters within which one can get maximum contribution from the quantum entanglement in terms of getting maximum negative contribution from the eigenvalue.
    
    Further,   in figure (\ref{EP}(a)) and (\ref{EP}(b)),  we have depicted the representative graphical graphical behaviour of the of the eignenvalue of the partial transposed matrix with the momentum mode associated with the computation for the given value of mass parameter.   In these plots we have considered two possibilities,  vanishing $f_{p_{\bf in}}=0$ and a very small value but $f_{p_{\bf in}}\neq 0$.  In both of these plots we have fixed the value of the mass parameter within the region $1<\nu<7/4$ and have studied the behaviour of the eigenvalue spectrum with respect to momentum mode within the range $0.1<p_{\bf in}<0.8$.  We have found that for very small value of momentum mode with the mass parameter in the region $\nu<1$ is not desirable in the present context as it gives very large positive eigenvalue.  Also from this analysis we have found that for very large value of the momentum mode for any values of the mass parameter $\nu$ the corresponding eigenvalue spectrum will saturate to a negative value.  After comparing figure (\ref{EP}(a)) and (\ref{EP}(b)),  we also have found that this asymptotic saturation value for small $f_{p_{\bf in}}\neq 0$ is larger compared to result obtained from $f_{p_{\bf in}}=0$ case.
    
Next,  in figure (\ref{LEN}(a)) and (\ref{LEN}(b)),  we have depicted the representative graphical graphical behaviour of the the logarithmic negativity computed from the negative eigenvalues of the partial transposed reduced density matrix with the mass parameter associated with the computation for the given value of momentum mode.  Finally,  in figure (\ref{LNp}(a)) and (\ref{LNp}(b)),  we have depicted the representative graphical graphical behaviour of the the logarithmic negativity with the momentum mode for the given value of mass parameter.  In these plots we have considered two possibilities,  vanishing $f_{p_{\bf in}}=0$ and a very small value but $f_{p_{\bf in}}\neq 0$.  The outcomes and the physical interpretation of these plots are very interesting which we are writing point-wise in the following:
     \begin{enumerate}
     \item From the figure (\ref{LEN}(a)) and (\ref{LEN}(b)),  we have found that for $f_{p_{\bf in}}=0$  logarithmic negativity vanishes at $\nu=1$,  but for small $f_{p_{\bf in}}\neq 0$ it is non zero but very small.  It further implies that at the value of the mass parameter $\nu=1$ we have almost negligible contribution from the quantum mechanical entanglement on large scales. Additionally it is important to note that,  this particular outcome is obtained for a specific value of the momentum mode,  $p_{\bf in}=0.2$.   
     
     \item On the other hand,  we have found that at $\nu=1/2$ and $\nu=3/2$ the obtained value of the logarithmic negativity from the present theoretical set up reach the maximum value,   which further correspond to the maximum quantum mechanical entanglement or maximum correlation.  This outcome is true for the momentum mode,  $p_{\bf in}=0.2$.   One important thing here to mention that the large scale limit corresponds to the small value of the momentum mode $p_{\bf in}$ in our present computation.
     
     \item  For the other values of momentum mode lying within the window $0.2<p_{\bf in}<0.5$ we have found that the variation with respect to the mass parameter $\nu$ is less compared to the case that we have studied for the momentum mode $p_{\bf in}=0.2$.   Comparing all the outcomes obtained for the different momentum modes within the mentioned range we have found that if we increase the value of $p_{\bf in}$ then the corresponding variation is reduced and we get intermediate values of negativity,  which corresponds to the intermediate amount of quantum mechanical entanglement for the system under consideration.  This further implies the fact that lower values of the momentum mode for the prescribed analysis is more desirable as it is giving higher amplitude of logarithmic negativity,  hence higher amount of quantum mechanical entanglement.
     
     \item We have also found that the underlying quantum mechanical entanglement directly put impact on the shape of the spectrum of logarithmic negativity at the large scales which nearly equals or exceeds the mass parameter's value $\nu\sim 3/2$.

     \item Further,  in figure (\ref{LNp}(a)) and (\ref{LNp}(b)),  we have found out that for large values of the momentum mode $p_{\bf in}>0.5$ the corresponding logarithmic negativity computed from the prescribed theoretical set up saturates to a constant non zero positive non negligible value.  This further implies constant amount of quantum entanglement for any arbitrary positive real value of the mass parameter $\nu$.  However,  in this asymptotic limit one cannot distinguish the individual effect of the mass parameter in the present computation. This also implies that the low momentum modes are more desirable for the present analysis to distinguish the individual effects of the mass parameter $\nu$.
     
     \item Last but not least,  we have also found that except $\nu=3/2$ for all other values of the mass parameter $\nu$ we get a sharp changing behaviour in the specific value of the momentum mode.
     \end{enumerate}
     
      \subsection{Small scale limit of logarithmic negativity: Analytical study}
      
    In the small scale limit one needs to take the limit $p_{\bf in}\rightarrow \infty$ to compute the logarithmic negativity from the present computational set up.  In this limit we have:
    \bea \Delta_1=\Delta^{*}_1\rightarrow 1,  \quad\Delta_2=\Delta^{*}_2\rightarrow 0, \quad\Delta_{3,s}=\Delta^{*}_{3,s}\rightarrow 1,\quad \Delta_{4,s}=\Delta^{*}_{4,s}\rightarrow 0.\eea
    The following simplified formulas in the small scale limit are obtained after taking partial transposition with regard to the subsystem corresponding to the first Bunch Davies quantum vacuum state:
       \bea \rho^{T,{\bf BD_1}}_{m_{\bf p_{in}}} &=&\frac{\left(1-|\gamma_{p_{\bf in}}|^2\right)}{2\left(1+f_{p_{\bf in}}\right)}|\gamma_{p_{\bf in}}|^{2m_{\bf p_{in}}}\Bigg\{|0_{\bf p_{out}}\rangle_{\bf BD_1}|m_{\bf p_{in}}\rangle_{{\bf L}^{'}}~{}_{\bf BD_1}\langle 0|{}_{{\bf L}^{'}}\langle m_{\bf p_{in}}|\nonumber\\
    &&+\sqrt{m_{\bf p_{in}}+1}|1_{\bf p_{out}}\rangle_{\bf BD_1}|m_{\bf p_{in}}\rangle_{{\bf L}^{'}}~{}_{\bf BD_1}\langle 0|{}_{{\bf L}^{'}}\langle m_{\bf p_{in}}+1|\nonumber\\
    &&+\sqrt{m_{\bf p_{in}}+1}|0_{\bf p_{out}}\rangle_{\bf BD_1}|m_{\bf p_{in}}+1\rangle_{{\bf L}^{'}}~{}_{\bf BD_1}\langle 1|{}_{{\bf L}^{'}}\langle m_{\bf p_{in}}|\nonumber\\
    &&+(m_{\bf p_{in}}+1)|1_{\bf p_{out}}\rangle_{\bf BD_1}|m_{\bf p_{in}}+1\rangle_{{\bf L}^{'}}~{}_{\bf BD_1}\langle 1|{}_{{\bf L}^{'}}\langle m_{\bf p_{in}}+1|\Bigg\},\eea  
    and 
    \bea
   \rho^{T,{\bf BD_1}}_{m_{\bf p_{in}},s} &=&\frac{f^{2}_{p_{\bf in}}}{2\left(1+f_{p_{\bf in}}\right)}|\Gamma_{{p_{\bf in}},s}|^{2m_{\bf p_{in}}}\Bigg\{|0_{\bf p_{out}}\rangle_{\bf BD_1}|s,m_{\bf p_{in}}\rangle_{{\bf L}^{'}}~{}_{\bf BD_1}\langle 0|{}_{{\bf L}^{'}}\langle s,m_{\bf p_{in}}|\nonumber\\
    &&+\sqrt{m_{\bf p_{in}}+1}|1_{\bf p_{out}}\rangle_{\bf BD_1}|s,m_{\bf p_{in}}\rangle_{{\bf L}^{'}}~{}_{\bf BD_1}\langle 0|{}_{{\bf L}^{'}}\langle s,(m_{\bf p_{in}}+1)|\nonumber\\
    &&+\sqrt{m_{\bf p_{in}}+1}|0_{\bf p_{out}}\rangle_{\bf BD_1}|s,m_{\bf p_{in}}+1\rangle_{{\bf L}^{'}}~{}_{\bf BD_1}\langle 1|{}_{{\bf L}^{'}}\langle s,m_{\bf p_{in}}|\nonumber\\
    &&+(m_{\bf p_{in}}+1)|1_{\bf p_{out}}\rangle_{\bf BD_1}|s,(m_{\bf p_{in}}+1)\rangle_{{\bf L}^{'}}~{}_{\bf BD_1}\langle 1|{}_{{\bf L}^{'}}\langle s,(m_{\bf p_{in}}+1)|\Bigg\}.\eea
           Let us express the above mentioned transposed version of the reduced density matrices computed from the complementary and particular integral part in square matrix form,  which are given by the following expressions:
           \bea \rho^{T,{\bf BD_1}}_{m_{\bf p_{in}}}&=&\frac{\left(1-|\gamma_{p_{\bf in}}|^2\right)}{2\left(1+f_{p_{\bf in}}\right)}|\gamma_{p_{\bf in}}|^{2m_{\bf p_{in}}}\left(\begin{array}{ccc} 1 &~~~\sqrt{m_{\bf p_{in}}+1} &~~~0\\ \sqrt{m_{\bf p_{in}}+1} &~~~ (m_{\bf p_{in}}+1) &~~~ 0\\ 0 &~~~ 0 &~~~ 0  \end{array}\right),\\
    \rho^{T,{\bf BD_1}}_{m_{\bf p_{in}},s}&=&\frac{f^{2}_{p_{\bf in}}}{2\left(1+f_{p_{\bf in}}\right)}|\Gamma_{{p_{\bf in}},s}|^{2m_{\bf p_{in}}}\left(\begin{array}{ccc} 1 &~~~\sqrt{m_{\bf p_{in}}+1} &~~~0\\ \sqrt{m_{\bf p_{in}}+1} &~~~ (m_{\bf p_{in}}+1) &~~~ 0\\ 0 &~~~ 0 &~~~ 0  \end{array}\right),\eea
        Next we compute the eigenvalue equation from the total partial transposed matrix after summing over source mode $s$,  which is given by the following expression: 
          \bea && \widetilde{\lambda}^{2}_{m_{\bf p_{in}}}\left(\widetilde{\lambda}_{m_{\bf p_{in}}}-\frac{\left(1-|\gamma_{p_{\bf in}}|^2\right)}{2\left(1+f_{p_{\bf in}}\right)}|\gamma_{p_{\bf in}}|^{2m_{\bf p_{in}}}(m_{\bf p_{in}}+2)-\frac{f^{2}_{p_{\bf in}}}{2\left(1+f_{p_{\bf in}}\right)}(m_{\bf p_{in}}+2)g_{p_{\bf in}}\right)=0.\quad\quad\quad\eea   
           The non trivial root of the eigenvalue computed from the $(m_{\bf p_{in}},m_{\bf p_{in}}+1)$ block in the small scale limit is given by: 
        \bea \widetilde{\lambda}_{m_{\bf p_{in}}} &=&\frac{(m_{\bf p_{in}}+2)}{2\left(1+f_{p_{\bf in}}\right)}\Bigg[\left(1-|\gamma_{p_{\bf in}}|^2\right)|\gamma_{p_{\bf in}}|^{2m_{\bf p_{in}}}+f^{2}_{p_{\bf in}} g_{p_{\bf in}}\Bigg],\eea
        where we define:
        \bea g_{p_{\bf in}}:=\sum^{\infty}_{s=0}|\Gamma_{{p_{\bf in}},s}|^{2m_{\bf p_{in}}}.\eea
       Then the logarithmic negativity from the present set up can be further computed as:
      \bea {\cal LN}&=&\ln\left(2\sum_{\widetilde{\lambda}_{m_{\bf p_{in}}}<0}\widetilde{\lambda}_{m_{\bf p_{in}}}+1\right)\nonumber\\
      &=&\ln\Bigg(
   \frac{(m_{\bf p_{in}}+2)}{\left(1+f_{p_{\bf in}}\right)}\Bigg[\left(1-|\gamma_{p_{\bf in}}|^2\right)|\gamma_{p_{\bf in}}|^{2m_{\bf p_{in}}}+f^{2}_{p_{\bf in}} g_{p_{\bf in}}\Bigg]+1\Bigg).\eea
 We get a negative contribution from the eigenvalue if the following condition is satisfied for the small scale limiting situation:
    \bea \Bigg[|\gamma_{p_{\bf in}}|^{2m_{\bf p_{in}}}\left(1-|\gamma_{p_{\bf in}}|^{2}\right)+f^{2}_{p_{\bf in}}g_{p_{\bf in}}\Bigg]<0.\eea 
        Now from the previous analysis we have already found that $m_{\bf p_{in}}=0$ mode is most desirable one to obtain the negative contribution from the eigenvalue spectrum.  In such a case using the well known Riemann zeta function regularization we can write the following result for $m_{\bf p_{in}}=0$:
          \bea g_{p_{\bf in}}:=\sum^{\infty}_{s=0}1=1+\sum^{\infty}_{s=1}1=1+\left(1+1+1+\cdots\right)=1+\zeta(0)=1-\frac{1}{2}=\frac{1}{2}.\eea
          which also suggests the following restriction on the small scale limiting circumstance:
        \bea |\gamma_{p_{\bf in}}|>\sqrt{\Bigg(1+\frac{f^{2}_{p_{\bf in}}}{2}\Bigg)}.\eea

    \subsection{Massless limit of logarithmic negativity: Analytical study}
       In the massless limit one needs to take the limit $\nu=3/2$ (exact masslessness) or $\nu=1/2$ (conformal invariance) to compute the logarithmic negativity from the present computational set up.  In this limit we have:
        \bea \Delta_1\rightarrow  \bigg( \widetilde{A}-\widetilde{D}\gamma_{p_{\bf in}}\bigg)u,  \quad\Delta_2\rightarrow 0, \quad\Delta_{3,s}\rightarrow  \bigg( \widetilde{A}_n-\widetilde{D}_n\Gamma_{{p_{\bf in}},n}\bigg)U_n,\quad \Delta_{4,s}\rightarrow 0.\quad\quad\quad\eea
        The partial transposition operation for the subsystem corresponding to the first Bunch Davies quantum vacuum state is given by the following formulas for the massless limiting situation:
       \bea \rho^{T,{\bf BD_1}}_{m_{\bf p_{in}}} &=&\frac{\left(1-|\gamma_{p_{\bf in}}|^2\right)}{2\left(1+f_{p_{\bf in}}\right)}|\gamma_{p_{\bf in}}|^{2m_{\bf p_{in}}}\Bigg\{|0_{\bf p_{out}}\rangle_{\bf BD_1}|m_{\bf p_{in}}\rangle_{{\bf L}^{'}}~{}_{\bf BD_1}\langle 0|{}_{{\bf L}^{'}}\langle m_{\bf p_{in}}|\nonumber\\
    &&+\Delta^{*}_1 \sqrt{m_{\bf p_{in}}+1}|1_{\bf p_{out}}\rangle_{\bf BD_1}|m_{\bf p_{in}}\rangle_{{\bf L}^{'}}~{}_{\bf BD_1}\langle 0|{}_{{\bf L}^{'}}\langle m_{\bf p_{in}}+1|\nonumber\\
    &&+\Delta_1 \sqrt{m_{\bf p_{in}}+1}|0_{\bf p_{out}}\rangle_{\bf BD_1}|m_{\bf p_{in}}+1\rangle_{{\bf L}^{'}}~{}_{\bf BD_1}\langle 1|{}_{{\bf L}^{'}}\langle m_{\bf p_{in}}|\nonumber\\
    &&+|\Delta_1|^{2}(m_{\bf p_{in}}+1)|1_{\bf p_{out}}\rangle_{\bf BD_1}|m_{\bf p_{in}}+1\rangle_{{\bf L}^{'}}~{}_{\bf BD_1}\langle 1|{}_{{\bf L}^{'}}\langle m_{\bf p_{in}}+1|\Bigg\},\eea
    and 
    \bea
   \rho^{T,{\bf BD_1}}_{m_{\bf p_{in}},s} &=&\frac{f^{2}_{p_{\bf in}}}{2\left(1+f_{p_{\bf in}}\right)}|\Gamma_{{p_{\bf in}},s}|^{2m_{\bf p_{in}}}\Bigg\{|0_{\bf p_{out}}\rangle_{\bf BD_1}|s,m_{\bf p_{in}}\rangle_{{\bf L}^{'}}~{}_{\bf BD_1}\langle 0|{}_{{\bf L}^{'}}\langle s,m_{\bf p_{in}}|\nonumber\\
    &&+\Delta^{*}_{3,s} \sqrt{m_{\bf p_{in}}+1}|1_{\bf p_{out}}\rangle_{\bf BD_1}|s,m_{\bf p_{in}}\rangle_{{\bf L}^{'}}~{}_{\bf BD_1}\langle 0|{}_{{\bf L}^{'}}\langle s,(m_{\bf p_{in}}+1)|\nonumber\\
    &&+\Delta_{3,s} \sqrt{m_{\bf p_{in}}+1}|0_{\bf p_{out}}\rangle_{\bf BD_1}|s,m_{\bf p_{in}}+1\rangle_{{\bf L}^{'}}~{}_{\bf BD_1}\langle 1|{}_{{\bf L}^{'}}\langle s,m_{\bf p_{in}}|\nonumber\\
    &&+|\Delta_{3,s}|^{2}(m_{\bf p_{in}}+1)|1_{\bf p_{out}}\rangle_{\bf BD_1}|s,(m_{\bf p_{in}}+1)\rangle_{{\bf L}^{'}}~{}_{\bf BD_1}\langle 1|{}_{{\bf L}^{'}}\langle s,(m_{\bf p_{in}}+1)|\Bigg\}.\quad\eea
    In this special case the factors $\Delta_1$ and $\Delta_{3,s}$ can be further simplified as:
    \bea \Delta_1 &=& \bigg( \widetilde{A}-\widetilde{D}\gamma_{p_{\bf in}}\bigg)u\nonumber\\
    &=&\frac{1}{\sinh \pi p_{\bf in}}\Bigg[\exp(\pi p_{\bf in})-i\exp(-\pi p_{\bf in})\frac{(1+p_{\bf in})}{(1-p_{\bf in})}\frac{\Gamma(ip_{\bf in})}{\Gamma(-ip_{\bf in})}\Bigg],\\
   \Delta_{3,s} &=& \bigg( \widetilde{A}_s-\widetilde{D}_s\Gamma_{p_{{\bf in},s}}\bigg)U_s\nonumber\\
    &=&\frac{1}{\sinh \pi p_{{\bf in},s}}\Bigg[\exp(\pi p_{{\bf in},s})-i\exp(-\pi p_{{\bf in},s})\frac{(1+p_{{\bf in},s})}{(1-p_{{\bf in},s})}\frac{\Gamma(ip_{{\bf in},s})}{\Gamma(-ip_{{\bf in},s})}\Bigg]. \eea
           
       Let us express the above mentioned transposed version of the reduced density matrices computed from the complementary and particular integral part in square matrix form,  which are given by the following expressions for the massless case:
           \bea \rho^{T,{\bf BD_1}}_{m_{\bf p_{in}}}&=&\frac{\left(1-|\gamma_{p_{\bf in}}|^2\right)}{2\left(1+f_{p_{\bf in}}\right)}|\gamma_{p_{\bf in}}|^{2m_{\bf p_{in}}}\left(\begin{array}{ccc} 1 &~~~ \sqrt{m_{\bf p_{in}}+1}\Delta^{*}_1 &~~~0\\  \sqrt{m_{\bf p_{in}}+1}\Delta_1 &~~~ 0 &~~~ 0\\ 0 &~~~ 0 &~~~ 0  \end{array}\right),\\
           \rho^{T,{\bf BD_1}}_{m_{\bf p_{in}},s}&=&\frac{f^{2}_{p_{\bf in}}}{2\left(1+f_{p_{\bf in}}\right)}|\Gamma_{{p_{\bf in}},s}|^{2m_{\bf p_{in}}}\left(\begin{array}{ccc} 1 &~~~\sqrt{m_{\bf p_{in}}+1}\Delta^{*}_{3,s} &~~~0\\ \sqrt{m_{\bf p_{in}}+1}\Delta_{3,s} &~~~ 0 &~~~ 0\\ 0 &~~~ 0 &~~~ 0  \end{array}\right).\eea
          Next we compute the eigenvalue equation from the total partial transposed matrix after summing over source mode $s$,  which is given by the following expression:
          \bea && \widetilde{\lambda}_{m_{\bf p_{in}}}\left(\widetilde{\lambda}^{2}_{m_{\bf p_{in}}}-\overline{A}_{m_{\bf p_{in}}}\widetilde{\lambda}_{m_{\bf p_{in}}}+\overline{B}_{m_{\bf p_{in}}}\right)=0.\eea
         where in the above mentioned expression we have introduced some shorthand redefined symbols which are given by the following expressions:
      \bea \overline{A}_m &=&\frac{1}{2\left(1+f_{p_{\bf in}}\right)}\Bigg[\left(1-|\gamma_{p_{\bf in}}|^2\right)|\gamma_{p_{\bf in}}|^{2m}+f^{2}_{p_{\bf in}}g_{p_{\bf in}}\Bigg],\quad\\
        \overline{B}_m &=& -\frac{ \left(m_{\bf p_{in}}+1\right)}{4\left(1+f_{p_{\bf in}}\right)^2}\Bigg[\left(1-|\gamma_{p_{\bf in}}|^2\right)^2|\gamma_{p_{\bf in}}|^{4m}|\Delta_1|^{2}+f^{4}_{p_{\bf in}}\sum^{\infty}_{s=0}|\Gamma_{{p_{\bf in}},s}|^{4m}|\Delta_{3,s}|^{2}\Bigg].\quad\quad\eea
        The non trivial roots of the eigenvalue computed from the $(m_{\bf p_{in}},m_{\bf p_{in}}+1)$ block are given by:
        \bea \widetilde{\lambda}^{\pm}_{m_{\bf p_{in}}} &=&\frac{1}{2}\Bigg[\overline{A}_{m_{\bf p_{in}}}\pm \sqrt{\overline{A}^{2}_{m_{\bf p_{in}}}-4 \overline{B}_{m_{\bf p_{in}}}}\Bigg]. \eea
 Then the logarithmic negativity computed from the negative eigenvalue from the present set up can be written for the massless limit as:
      \bea {\cal LN}&=&\ln\left(2\sum_{\widetilde{\lambda}_{m_{\bf p_{in}}}<0}\widetilde{\lambda}_{m_{\bf p_{in}}}+1\right)=\ln\Bigg(\overline{A}_{m_{\bf p_{in}}}+1- \sqrt{\overline{A}^{2}_{m_{\bf p_{in}}}-4 \overline{B}_{m_{\bf p_{in}}}}\Bigg).\eea 
       Here it is important to note that the eigenvalue and the associated logarithmic negativity computed in this particular massless limiting situation with $f_{p_{\bf in}}=0$ is similar to the case that is obtained for explaining the entanglement between an inertial and a non-inertial frame of reference for a free massless scalar degree of freedom in Minkowski flat space time as discussed in the reference \cite{Fuentes-Schuller:2004iaz}.  Small difference appearing due to having separate thermal behaviour of the Minkowski space time and global de Sitter space time.  Now this difference in the result is more prominent and significant once we consider the effect of source term in the effective axion potential with small $f_{p_{\bf in}}\neq 0$.  
\section{Conclusion}
\label{ka6}

We conclude our discussion with the following points which we have found from our analysis performed in this paper:
\begin{itemize}
\item  Firstly,  we have started with the basic discussion regarding the entanglement negativity and logarithmic negativity for a general quantum mechanical set up which is appearing in the context of quantum information theory.  We have provided the technical details for the related computations from a general quantum mechanical set up.  Further we have provided a proper physical justification that why the above mentioned two measures are physically relevant and significant for the computation we want to perform for the open chart of the global de Sitter space.

\item Then we have given a detailed justification of the factorization of the total Hilbert space in open chart to associate our computation in the region {\bf L} and {\bf R},   which is necessarily required to construct the reduced density matrix and to compute the above mentioned entanglement measures from the system under consideration.  

\item  Additionally, we have fully covered the specifics of the geometrical arrangement of the open chart of the de Sitter space, which is the platform on which we intend to carry out the remaining calculation. We have independently determined the metric's structure in the area between  {\bf L} and {\bf R}, which is a crucial piece of knowledge for determining how scalar modes would behave based on our calculations.

\item  Next, we computed the explicit equation for the mode function using the string theory-derived axionic effective interaction, with which we created the Bunch Davies vacuum state and subsequently the expression for the reduced density matrix.

\item Further,  we have computed the expressions for the entanglement negativity and logarithmic negativity for the mentioned axionic effective interaction in open chart.  We have found that,  the newly studied quantum information theoretic measures are more significant compared to the Von Neumann measure of entanglement entropy,  which is commonly used to describe the impact of quantum entanglement.  

\item The result that offers the most promise is that it enables us to compute and estimate the quantum entanglement between the inside and outside of a de Sitter bubble without the need for a boundary.

\item We also found that for the large mass the amount of the quantum entanglement decays exponential in the entanglement negativity vs mass parameter squared plots.  On the other hand this decaying behaviour is slightly different when we consider the logarithmic negativity measure in this context.

\item We also have found that for the case of conformal coupling $\nu=1/2$ and massless case $\nu=3/2$ in the logarithmic negativity and entanglement negativity spectrum there are two consecutive peaks appear of equal hights.  This is really an interesting feature we have found from the prescribed theoretical set up studied in this paper.  Additionally we have found that apart from having two prominent peaks for the mentioned values of the mass parameter we have oscillation in the spectrum due to having small mass parameter of the axion field during the de Sitter expansion in the global coordinates. This oscillation becomes more rapid if the mass parameter is very small.  On the other hand,  the period of oscillation is larger and less rapid if the mass parameter is large.

\item 

Then, using the Bunch Davies quantum vacuum state, we expanded our computation to compute the formula for the logarithmic negativity between two causally independent patches of the open chart of the global de Sitter space.  This is a scaled-down version of the well-known multiverse scenario and is known as the biverse picture. In order to perform this calculation, we presupposed a correct factorization of the Hilbert spaces in the two subspaces that are currently spanned by the modes of the Bunch Davies quantum vacuum states.  We don't need to be aware of the explicit content of the first subspace in order to apply our methodology to compute the quantum entanglement between two de Sitter spaces in the global coordinates. This is necessary because we need to take a partial trace over the first de Sitter space's entire subspace content.  However, the explicit sub factorization of the second subspace is crucial in this situation since it significantly affects the explicit content of the modes from the regions {\bf R} and {\bf L}.

\item 

The most important aspect of this particular computation is the maximally entangled state, which we used to build the reduced density matrix by extracting all the data from the initial Bunch Davies vacuum state.  We have further used this result to compute the expression for the partial transposed version of the reduced density matrix.  We have next found that the mode corresponding to $m_{\bf p_{in}}=0$ (ground state) correspond to the negative eigen value spectrum,  using which we have numerically studied the behaviour of logarithmic negativity from the prescribed theoretical set up. We have found from our analysis that the mode corresponding to $m_{\bf p_{in}}=0$ can produce large measure of quantum entanglement due to having negative eigen value spectrum.  For the other values of $m_{\bf p_{in}}$ we have found that the corresponding eigen values are largely positive which is not desirable to construct a biverse which can produce large effect of quantum mechanical entanglement at the end.

\item In the biverse construction we also have found conformally coupled case with $\nu=1/2$ and the massless case with $\nu=3/2$ are the two very special points in the entanglement spectrum where the amount of the quantum correlation is equal and high in amplitude.  On the other hand,   we have found that for the mass parameter $\nu=1$ the amount of quantum correlation estimated from the corresponding picture reaches its minimum value.  This is obviously a promising information which we have obtained after performing our analysis on the biverse picture and it is expected that it can be generalized to any multiverse scenario constructed out of the specific model that we have studied in this paper.

\item In the present context the global coordinates can be treated as the closed slicing from the point of view of FLRW cosmology.  By performing coordinate coordinate transformation one can transform the global to static and then static to the flat slicing in the planer patch of de Sitter space.  For this reason whatever results we have obtained in this paper for the global patch of the de Sitter space can be directly translated in the planer patch of the de Sitter space,   which further implies that our derived results hold good for primordial cosmology. 
\end{itemize}

Here we have some interesting immediate future direction on which one can extend our analysis:
\begin{itemize}

\item We have restricted our analysis for the estimation of entanglement negativity and logarithmic negativity by considering the Bunch Davies quantum vacuum state.  Immediately one can extend our analysis for a general non Bunch Davies vacua,  such as $\alpha$ vacua.  It is expected to have many interesting outcome if we extend our analysis for non Bunch Davies vacua because the quantum correlations and its various unknown applications will be known from this type of future analysis.

\item Since the global coordinates and planar coordinates of de Sitter space are connected via coordinate transformation,  it is good to explicitly know how the present results can be explained within the framework of primordial cosmology.   This possibility one can seriously think for future work to connect with observation.

\item One can further compute various other quantum information theoretic measures,  like quantum discord,  fidelity and many more interesting quantities from the present theoretical set up.

\item The direct connection between the higher point quantum correlations with the all of these possible quantum information theoretic measures,  imprints of quantum entanglement in quantum correlations computed in the quantum field theory of de Sitter space and the primordial cosmological set ups are another interesting possibilities which one can study in future from the present set up that we have constructed in this paper. 

\item Extending the present computation to study role quantum mechanical decoherence \cite{Burgess:2022nwu,Martin:2021znx,Martin:2018lin,Martin:2018zbe,Liu:2016aaf} and quantum diffusion \cite{Ezquiaga:2022qpw} might be very useful for the future study which can explain various unknown facts from the present set up in global as well as in the planer patch of de Sitter space.

\item The construction of squeezed quantum mechanical states and its consequences is a common area of research in cosmological set up \cite{Choudhury:2022btc,Choudhury:2021brg,Adhikari:2021ked,Adhikari:2021pvv,Choudhury:2020hil,Bhargava:2020fhl,Adhikari:2022oxr,Park:2022xyf,Ruan:2021wep,DiGiulio:2021fhf,Bhattacharyya:2020rpy,Li:2021kfq,Grishchuk:1993ds,Albrecht:1992kf}.  It would be really good if we can able to construct a squeezed quantum state out of the present theoretical set up that we are considering in this paper. This will going to help to figure out various quantum information theoretic measures and its applications in various contexts.  If the squeezed state construction is not possible then also one can study various other possibilities out of the present set up \cite{Adhikari:2022whf,Adhikari:2021ckk,Choudhury:2022xip,Choudhury:2022dox}.

\item Till now the computation is restricted to a quantum system which is completely adiabatic in nature,   because we are considering closed quantum system.  It would be really good if we can study the open quantum system version of the present set up within the framework of quantum field theory of de Sitter space \cite{Colas:2022kfu,Colas:2022hlq,Choudhury:2022ati,Banerjee:2021lqu,Choudhury:2020dpf,Banerjee:2020ljo,Akhtar:2019qdn,Choudhury:2018rjl,Choudhury:2018bcf,Chaykov:2022zro,Chaykov:2022pwd,Burgess:2021luo,Kaplanek:2021fnl,Burgess:2015ajz}.
\end{itemize}



	\subsection*{Acknowledgements}
SC would like to thank the work
friendly environment of The Thanu Padmanabhan Centre For Cosmology and Science Popularization (CCSP), 
Shree Guru Gobind Singh Tricentenary (SGT) University,  Gurugram,  Delhi-NCR for providing tremendous support in research and offer the Assistant Professor (Senior Grade) position.  SC also thanks
all the members of our newly formed virtual international
non-profit consortium Quantum Aspects of the SpaceTime \& Matter (QASTM) for elaborative discussions.  Last but not
least, we would like to acknowledge our debt to the people belonging to the various parts of the world for their
generous and steady support for research in natural sciences.

\clearpage

\addcontentsline{toc}{section}{References}
\bibliographystyle{utphys}
\bibliography{references2}

\end{document}